\documentclass[eqsecnum,twocolumn,tightenlines,floats,floatfix,prc,nofootinbib,preprintnumbers]{revtex4}

\def\bea{\begin{eqnarray}}
\def\eea{\end{eqnarray}}

\def\st#1{{\kern-4pt} \not\!#1}

\def\sp{\kern +3pt}
\def\sm{\kern -3pt}

\def\Unit{1\kern -4pt 1}

\usepackage[usenames]{color}

\def\be{\begin{equation}}
\def\ee{\end{equation}}
\def\ba{\begin{eqnarray}}
\def\ea{\end{eqnarray}}

\usepackage{graphics}
\usepackage{graphicx}
\usepackage{epsf} 
\usepackage{amsmath}
\usepackage{amssymb}
\usepackage{slashed}
\usepackage{bm}
\usepackage{bbm}

\usepackage{bbold}
\usepackage{mathbbol}

\def\sfrac#1#2{{\textstyle \frac{#1}{#2}}}

\begin{document}

\phantom{0}
\vspace{-0.2in}
\hspace{5.5in}

\preprint{}

\vspace{-1in}

\title
{\bf Improved empirical parametrizations of 
the $\gamma^\ast N \to \Delta(1232)$ and 
$\gamma^\ast N \to N(1520)$ transition amplitudes 
and the  Siegert's theorem}
\author{G.~Ramalho  
\vspace{-0.1in} }

\affiliation{
International Institute of Physics, Federal 
University of Rio Grande do Norte, 
Campus Lagoa Nova - Anel Vi\'ario da UFRN, 
Lagoa Nova, Natal-RN, 59070-405, Brazil}

\vspace{0.2in}
\date{\today}

\phantom{0}

\begin{abstract}
In the nucleon electroexcitation reactions, 
$\gamma^\ast N \to R$, where $R$ is a nucleon resonance ($N^\ast$),
the electric amplitude $E$, 
and the longitudinal amplitude $S_{1/2}$, 
are related by $E \propto \frac{\omega}{|{\bf q}|}S_{1/2}$,
in the pseudo-threshold limit ($|{\bf q}| \to 0$), 
where $\omega$ and $|{\bf q}|$ are 
respectively the energy and the magnitude of three-momentum of the photon. 
The previous relation is usually refereed to as the Siegert's theorem.
The form of the electric amplitude, 
defined in terms of the 
transverse amplitudes $A_{1/2}$ and $A_{3/2}$,
and the explicit coefficients of the relation, 
depend on the angular momentum and parity ($J^P$) of the resonance $R$.
The Siegert's theorem 
is the consequence of the structure 
of the electromagnetic transition current,
which induces constraints between the 
electromagnetic form factors in the pseudo-threshold limit.
In the present work, we study the implications 
of the Siegert's theorem for the $\gamma^\ast N \to \Delta(1232)$ and 
$\gamma^\ast N \to N(1520)$ transitions.
For the $\gamma^\ast N \to N(1520)$ transition,
in addition to the relation between electric amplitude
and longitudinal amplitude,
we also obtain a relation between 
the two transverse amplitudes: $A_{1/2}= A_{3/2} /\sqrt{3}$,
at the pseudo-threshold.
The constraints at the pseudo-threshold 
are tested for the MAID2007 parametrizations 
of the reactions under discussion.
New parametrizations for the amplitudes $A_{1/2}$, $A_{3/2}$
and $S_{1/2}$, for the  $\gamma^\ast N \to \Delta(1232)$ and 
$\gamma^\ast N \to N(1520)$ transitions, 
valid for small and large $Q^2$, are proposed.
The new parametrizations are consistent with 
both: the pseudo-threshold constraints (Siegert's theorem) 
and the empirical data.
\end{abstract}

\vspace*{0.9in}  
\maketitle

\section{Introduction}

The information relative to the structure of the electromagnetic transition 
between the nucleon and a nucleon excitation $R$ 
($\gamma^\ast N \to R$), can be 
parametrized in terms of helicity amplitudes dependent 
on the photon polarization states,
and the transfer momentum squared $q^2$,
which is restricted to the region $Q^2= -q^2 > 0$~\cite{NSTAR,Aznauryan12}.
For the transitions $\frac{1}{2}^+ \to \frac{1}{2}^\pm, \frac{3}{2}^\pm$,
where $J^P$ represents the state 
of angular momentum $J$, parity $P=\pm$,
and $\frac{1}{2}^+$ is the nucleon state,
one can define the transverse amplitudes 
$A_{1/2}, A_{3/2}$ and the longitudinal or scalar amplitude $S_{1/2}$
($A_{3/2}$ can be defined only for $J=\sfrac{3}{2}$).

Although those amplitudes are in principle 
independent functions, there are relations 
between them in the limit where the photon 
momentum $|{\bf q}|$  vanishes.
This limit is called the pseudo-threshold limit,
corresponding to the case where both the 
nucleon and the resonance $R$ are at rest.
At the pseudo-threshold limit ($|{\bf q}|=0$), 
one has $Q^2 = -\omega^2$, where 
$\omega = M_R-M$ is the photon energy,
and $M_R$, $M$ are respectively 
the resonance and the nucleon masses.
Since the pseudo-threshold limit is 
defined by $Q^2= Q_{PS}^2 < 0$, with $Q_{PS}^2= - (M_R-M)^2$,
it belongs to the unphysical region, 
where the helicity amplitudes cannot be measured from the 
$\gamma^\ast N \to R$ transition.
The extension of models for the $Q^2< 0$ region is 
important, however, for studies of reactions 
such as the $\Delta$ Dalitz decay ($\Delta \to e^+ e^- N$)
and Dalitz decays of other resonances~\cite{DeltaTL,WhitePaper}.

At the pseudo-threshold 
the matrix element of the electric 
multipole $E$, defined by 
the spatial current density  {\boldmath{$J$}},
and the matrix element of the Coulomb multipole 
$S_{1/2}$, defined by the charge density $\rho$, can be related by 
$E \propto \frac{\omega}{|{\bf q}|} S_{1/2}$~\cite{Buchmann98}.
This result, obtained in the limit of the 
long wavelengths ($|{\bf q}| \to 0$), 
is usually refereed as 
the Siegert's theorem~\cite{Buchmann98,Atti78,AmaldiBook,Drechsel92,Tiator16}.
Although defined below $Q^2 = 0$,
the relations between amplitudes can  be used to 
test analytic properties of theoretical models 
and to test the consistency of phenomenological parametrizations.

The exact proportionality between the electric
amplitude $E$ and the  scalar amplitude $S_{1/2}$ 
depends on the angular momentum-parity 
state ($J^P$) of the resonance.
The constraints for the helicity amplitudes
can in general be derived from 
the analysis of the transition currents, 
expressed in a covariant form 
in terms of the properties of the nucleon and the
resonance, which define a minimal number of independent 
structure form factors~\cite{Bjorken66,Devenish76,Jones73}. 

In Ref.~\cite{newPaper}, the implications 
of the pseudo-threshold limit for 
the $\gamma^\ast N \to N(1535)$ transition form factors 
and helicity amplitudes, and their implications 
in the parametrizations of the data are discussed in detail.
In the present work we discuss 
the consequences of the  pseudo-threshold limit 
for the $\gamma^\ast N \to \Delta(1232)$ 
and $\gamma^\ast N \to N(1520)$ transitions.

For the $\gamma^\ast N \to \Delta(1232)$ transition,
we will conclude that, in the pseudo-threshold limit, 
one has  $\frac{E}{|{\bf q}|} = \sqrt{2} \omega \frac{S_{1/2}}{|{\bf q}|^2}$,
where $\omega = M_R -M$, and $M_R$ is the $\Delta$ mass.
Note that the previous relation differs from 
the usual form
$E = \sqrt{2} \frac{\omega}{|{\bf q}|} S_{1/2}$~\cite{Tiator2006,Drechsel2007},
by a factor $1/|{\bf q}|$. 
This difference has implications in shape 
of the parametrizations of the data, 
as we will show.
[Along the paper, we will interpret the factors like 
$S_{1/2}/|{\bf q}|$ or $S_{1/2}/|{\bf q}|^2$, 
as functions defined also for $|{\bf q}|=0$,
with the result given by the limit $|{\bf q}| \to 0$,
when the limit exists.]

As for the  $\gamma^\ast N \to N(1520)$ transition,
the pseudo-threshold limit induces two constraints 
in the helicity amplitudes.
The trivial constraint is expressed as 
$\sfrac{1}{2}E= \sqrt{2} \frac{\omega}{|{\bf q}|}S_{1/2}$,
where $\omega$ is now defined in terms of 
the $N(1520)$ mass ($M_R$).
In addition, one has also the relation 
$A_{1/2}= A_{3/2} /\sqrt{3}$, at the pseudo-threshold.

The explicit form of the electric and 
the scalar amplitudes will be defined 
later for cases $\frac{3}{2}^+$ and $\frac{3}{2}^-$.
Defining $\lambda_R = \sqrt{2}(M_R -M)$,
we can express the correlation between 
the electric and scalar amplitudes
(Siegert's theorem)  
as $\frac{E}{|{\bf q}|} = \lambda_R \frac{S_{1/2}}{|{\bf q}|^2}$ 
for the  $\gamma^\ast N \to \Delta(1232)$ transition,
and $\sfrac{1}{2}E= \lambda_R \frac{S_{1/2}}{|{\bf q}|}$
for the  $\gamma^\ast N \to N(1520)$ transition.

In order to take into account 
the constraints from the pseudo-threshold limit,
in this work we present new parametrizations 
of the  $\gamma^\ast N \to \Delta(1232)$  and 
$\gamma^\ast N \to N(1520)$ helicity amplitudes.
We will conclude at the end, that, an overall 
description of the data for low $Q^2$ and large $Q^2$, 
including the pseudo-threshold,
is possible using smooth representations 
of the helicity amplitudes.
The presented parametrizations 
are compared with the MAID2007 
parametrizations~\cite{Tiator2006,Drechsel2007,MAID2009}.
Parametrizations for very large $Q^2$, 
that simulate the expected falloff 
from perturbative QCD (pQCD), will be proposed.

This article is organized as follows:
In Sec.~\ref{secGen}, we discuss the formalism 
associated with the electromagnetic transition current,
helicity amplitudes and transition form factors.
In Secs.~\ref{secDelta} and \ref{secN1520}, 
we study the $\gamma^\ast N \to \Delta(1232)$ 
and $\gamma^\ast N \to N(1520)$ transitions, respectively.
Parametrizations of the data appropriate for very large $Q^2$
are discussed In Sec.~\ref{secExtension}.
Finally in Sec.~\ref{secConclusions},
we present our summary and conclusions.

\section{Generalities}
\label{secGen}

We introduce now the formalism
associated with the 
$\gamma^\ast N \to R$ transition,
where $N$ is the nucleon ($J^P=\frac{1}{2}^+$)
and $R$ is a $J^P= \frac{3}{2}^\pm$ resonance.
The case  $J^P= \frac{3}{2}^+$ corresponds to the 
$\Delta(1232)$ resonance; the case  $J^P= \frac{3}{2}^-$ corresponds 
to the $N(1520)$ resonance.
The variable $M_R$ represents the 
mass of the resonance under discussion 
[$\Delta(1232)$ or $N(1520)$].

We start with the discussion 
of the relation between the 
electromagnetic transition current and 
the helicity amplitudes.
Next, we look for the properties of 
the amplitude $S_{1/2}$.
Before discussing in detail the transitions
$\gamma^\ast N \to \Delta(1232)$ 
and $\gamma^\ast N \to N(1520)$, 
we present some useful notation.

\subsection{Electromagnetic current and helicity amplitudes}

In general, the  $\gamma^\ast N \to R$ transitions can be characterized 
in terms of transition form factors, to be defined later, or 
by the helicity amplitudes defined at the resonance rest frame.
At the $R$ rest frame, 
the initial ($P_-$) and final ($P_+$) 
momenta can be represented, 
choosing the photon momentum, $q=P_+-P_-$, 
along the $z$-axis, as
\ba
& &
P_- = (E_N,0,0,-|{\bf q}|), \hspace{.5cm}
P_+= (M_R,0,0,0) \nonumber \\
& &
q=(\omega,0,0, |{\bf q}|). 
\ea
In the previous equations, 
$|{\bf q}|$ is the magnitude of the photon
(and nucleon) three-momentum, given by 
\ba
|{\bf q}|= \frac{\sqrt{Q_+^2 Q_-^2}}{2 M_R},
\ea
with $Q_\pm^2= (M_R\pm M)^2 + Q^2$.
The nucleon energy $E_N= \sqrt{M^2 + |{\bf q}|^2}$
and the photon energy $\omega = M_R - E_N$ can be 
expressed covariantly as
$E_N= \frac{M_R^2 + M^2 + Q^2}{2M_R}$ and 
$\omega =   \frac{M_R^2 - M^2 - Q^2}{2M_R}$
respectively.  

The transverse ($A_{1/2},A_{3/2}$) and the 
longitudinal ($S_{1/2}$) amplitudes,
are defined at the $R$ rest frame~\cite{Aznauryan12,N1520},
as 
\ba
& &
\hspace{-.6cm}
A_{1/2} = \sqrt{\frac{2\pi \alpha}{K}} 
\left< R, S_z^\prime = + \frac{1}{2} \right|
\varepsilon_+ \cdot J \left| 
N , S_z = - \frac{1}{2} \right>, 
\label{eqA12} \\
& &
\hspace{-.6cm}
A_{3/2} = \sqrt{\frac{2\pi \alpha}{K}} 
\left< R, S_z^\prime = + \frac{3}{2} \right|
\varepsilon_+ \cdot J \left| 
N , S_z = + \frac{1}{2} \right>, 
\label{eqA32}  \\
& &
\hspace{-.6cm}
S_{1/2} = \sqrt{\frac{2\pi \alpha}{K}} 
\left< R, S_z^\prime = + \frac{1}{2} \right|
\varepsilon_0 \cdot J \left| 
N , S_z = + \frac{1}{2} \right> \frac{|{\bf q}|}{Q}, \nonumber \\ 
\label{eqS12} 
\ea
where $S_z^\prime$ ($S_z$) is the final (initial) 
spin projection, $Q= \sqrt{Q^2}$, 
$\varepsilon_\lambda^\mu$ ($\lambda = 0,\pm 1$)
are the photon polarization vectors, 
and $J^\mu$ is the electromagnetic transition current operator 
in units of the elementary charge $e$.
In addition, $\alpha = \frac{e^2}{4\pi} \simeq 1/137$
is the fine-structure constant and 
$K= \frac{M_R^2-M^2}{2 M_R}$.

The properties associated with the
structure of the resonance $R$ are then encoded in 
the  electromagnetic transition current operator $J^\mu$.
In the case of a transition between a spin $\frac{1}{2}$ state ($N$)
and a spin $\frac{3}{2}$ state ($R$)
we can project the current into the asymptotic states 
using  
\ba
J_{NR}^\mu &\equiv & \left<R \right| J^\mu \left| N \right> 
\nonumber \\
& =& \bar u_\alpha (P_+) \Gamma^{\alpha \mu}(P,q) u(P_-), 
\label{eqJgen}
\ea
where $u_\alpha$ and $u$ are respectively 
the Rarita-Schwinger and the Dirac spinors,
$P = \sfrac{1}{2}(P_+ + P_-)$, and 
$\Gamma^{\alpha \mu}$ is an operator dependent 
of the parity, to be defined in the following sections
for the case of the $\Delta(1232)$ (positive parity)
and the $N(1520)$ (negative parity).

\subsection{Scalar amplitude}
\label{secS12}

In the case of the current conservation, 
one can replace 
$(\varepsilon_0 \cdot J ) \sfrac{|{\bf q}|}{Q} \to J^0$,
in the definition of the scalar amplitude (\ref{eqS12}),
and write 
\ba
S_{1/2} = \sqrt{\frac{2\pi \alpha}{K}} \left< J ^0\right>, 
\label{eqJ0gen}
\ea 
where the brackets represent the projection 
into the spin states with $S_z^\prime = S_z= + \sfrac{1}{2}$,
defined at the resonance rest frame.
If the current is not conserved, 
or the current operator is truncated, we cannot 
use Eq.~(\ref{eqJ0gen}), as 
discussed in Refs.~\cite{Drechsel84,Buchmann98}.

Using Eq.~(\ref{eqJ0gen}), we can conclude that 
the scalar amplitude, near the pseudo-threshold, 
can be expressed for $J^P=\frac{3}{2}^\pm$ cases, as 
\ba
S_{1/2} \propto G_5^P(\bar u_3 {\Unit}_P\, u) |{\bf q}|,
\ea
where $G_5^\pm$ is a form factor
(dependent on the parity $\pm$), to be defined later, 
and  ${\Unit}_P$ is a parity-dependent operator, given by 
${\Unit}_+ = \gamma_5$ and 
${\Unit}_- = \Unit$.

Using the properties of Dirac 
and Rarita-Schwinger spinors, 
we can conclude that 
$(\bar u_3 \gamma_5 u) = 
\sqrt{\sfrac{2}{3}} \sfrac{|{\bf q}|}{2M} 
\propto |{\bf q}|$ 
and  $(\bar u_3 u) = -\sqrt{\sfrac{2}{3}} = {\cal O}(1)$,
near the pseudo-threshold 
\cite{NDelta,NDeltaD}. 
Applying those results,  one obtains  
$S_{1/2} = {\cal O} (|{\bf q}|^2)$ 
for $\Delta (1232)$ resonance and  
$S_{1/2} = {\cal O} (|{\bf q}|)$ for $N(1520)$ 
resonance.

Note that, the result $\left< J^0\right> \propto G_5^\pm |{\bf q}|^n$,
with $n=2$ for positive parity, and $n=1$, 
for negative parity, leads to $\left< J^0\right>  \to 0$, 
in the pseudo-threshold limit,
if the form factor $G_5^\pm$ has no singularities 
in this limit, as expected~\cite{Devenish76}.
The result  $\left< J^0\right>  = 0$ 
is equivalent to the orthogonality 
between the nucleon and the resonance states.
The same property can be observed 
in the $\gamma^\ast N \to N(1535)$ transition,
where $N(1535)$ is a $J^P = \frac{1}{2}^-$ 
state~\cite{newPaper}.

It is interesting to note that, the 
dependence of a function $F(Q^2)$, near the 
pseudo-threshold,  can be inferred directly from 
the graph of the function in terms of $Q^2$.
Since the derivative in $Q^2$, can be 
determined by the derivative in $|{\bf q}|$, 
given by
\ba
\frac{d F}{d Q^2}= \frac{M_R^2 + M^2 + Q^2}{4M_R^2 |{\bf q}|}
\frac{d F}{d |{\bf q}|},
\ea
we conclude that, at the pseudo-threshold 
(limit $|{\bf q}| \to 0$), the derivative $\frac{d F}{d Q^2}$ 
will be infinite (vertical line), 
when $F = {\cal O}(|{\bf q}|)$, 
and finite, only when  $F = {\cal O}(|{\bf q}|^n)$ with $n\ge 2$
(we are interested only in the  natural powers $n$).
To summarize:
the graphs with an infinite derivative 
at the pseudo-threshold are the representation 
of functions ${\cal O}(|{\bf q}|)$;
the graphs with finite derivative at 
the  pseudo-threshold represent
functions  ${\cal O}(|{\bf q}|^n)$ with $n\ge 2$.

\subsection{Notation}

In the following sections, we 
will study separately the resonances $\Delta(1232)$ and $N(1520)$.
To convert helicity amplitudes 
into form factors, we use the factor~\cite{Aznauryan12,N1520,Delta1600}
\ba
F_\pm = \frac{1}{e} \frac{2 M}{M_R \pm M} 
\sqrt{\frac{M M_R K}{Q_\mp^2}}.
\label{eqFpm}
\ea
The factor $F_+$ will be used 
for the case $\frac{3}{2}^+$, and the factor $F_-$ 
will be used for the case $\frac{3}{2}^-$.
For convenience we also define $\tau = \frac{Q^2}{(M_R + M)^2}$.

In the next two sections, 
we will also define the magnetic, electric and scalar amplitudes:
$M_{l \pm}, E_{l \pm}$ and $S_{l \pm}$, where $l= J \mp  1/2$ 
is an integer and $P=\pm$ is the parity, 
for  the case $J=3/2$.
For a more detailed discussion about the  
multipole amplitude notation 
see Refs.~\cite{Devenish76,Drechsel2007}.

\section{$\gamma^\ast N \to \Delta(1232)$ transition}
\label{secDelta}

The $\gamma^\ast N \to \Delta(1232)$ transition current 
can be determined using Eq.~(\ref{eqJgen}),
with the operator~\cite{Aznauryan12,Devenish76,Jones73,NDelta}
\ba
& &
\Gamma^{\alpha \mu}(P,q)= \nonumber \\
& &
G_1 q^\alpha \gamma^\mu \gamma_5  + G_2 q^\alpha P^\mu \gamma_5+ 
G_3 q^\alpha q^\mu\gamma_5  - G_4 g^{\alpha \mu} \gamma_5, \nonumber \\
\ea 
where $G_i$ ($i=1,..,4$) are structure form factors 
dependent on $Q^2$.
The four form factors are not all independent,
only three of them are independent.
Using the current conservation condition, $q \cdot J=0$,
we can conclude that~\cite{NDelta,NDeltaD}
\ba
G_4= (M_R + M) G_1 + \frac{1}{2} (M_R^2- M^2) G_2 
- Q^2 G_3.
\label{eqG4D}
\ea

Instead of the {\it elementary} form factors $G_i$,
alternatively we can use the multipole form factors:
magnetic dipole ($G_M$), electric quadrupole ($G_E$) 
and Coulomb quadrupole ($G_C$), 
defined as~\cite{Jones73,NDelta,NDeltaD}
\ba
G_M &= & Z_R  \left[(M_R-M) G_5 + 4 M G_1 \frac{}{}  
\right.  \nonumber \\
& & \left. +
\frac{4 M_R^2 |{\bf q}|^2}{Q_+^2} 
\left(\frac{G_1}{2M_R} - G_3 
\right)\right], 
\label{eqGM}
\\
G_E &=&  Z_R  
\left[ (M_R-M) G_5 \frac{}{}  
\right. \nonumber \\
& & \left. 
- \frac{4 M_R^2 |{\bf q}|^2}{Q_+^2}
\left( \frac{G_1}{2 M_R} + G_3 \right)
 \right], 
\label{eqGE}\\
G_C & =& Z_R  
\left[ 2 M_R G_5 \frac{}{} \right. \nonumber \\
& &\left. 
+ \frac{4 M_R^2 |{\bf q}|^2}{Q_+^2} 
\left( \frac{1}{2} G_2 - G_3 \right)
 \right], 
\label{eqGC}
\ea
where $Z_R= \frac{2M}{3(M_R + M)}$ and
\ba
G_5= G_1 + \frac{1}{2}(M_R + M) G_2 + (M_R - M) G_3,
\ea
is a new auxiliary form factor.

For convenience Eqs.~(\ref{eqGM})-(\ref{eqGC}) 
are expanded in powers of $|{\bf q}|$.
For the sake of the discussion, we consider 
$G_1,G_2$ and $G_3$, as our base for the form factors,
following Jones and Scadron~\cite{Jones73},
but we use also $G_4$ and $G_5$, when necessary.
For the multipole form factors we choose 
the Jones and Scadron representation.
To convert to the alternative Ash representation,
the functions $G_M,G_E$ and $G_C$ should be divided
by the factor $\sqrt{1 + \tau}$~\cite{Drechsel2007}.

The helicity amplitudes (\ref{eqA12})-(\ref{eqS12})
can be obtained from the form factors~\cite{Aznauryan12,Drechsel2007}, 
using
\ba
& &
A_{1/2} =  \frac{1}{4F_+} (G_M- 3 G_E), \\
& &
A_{3/2} = - \frac{\sqrt{3}}{4F_+} (G_M + G_E), \\
& &
S_{1/2} = \frac{1}{\sqrt{2} F_+}  \frac{|{\bf q}|}{2 M_R} G_C, 
\ea
where $F_+$ is defined by Eq.~(\ref{eqFpm}).

The multipole amplitudes $M_{1+}, E_{1+}, S_{1+}$,
can be defined directly in terms of 
the multipole form factors, or 
as a combination of the amplitudes~\cite{Devenish76,Drechsel2007}, 
\ba
G_M &=& F_+ M_{1+} \nonumber \\
&\equiv & - F_+ (A_{1/2} +  \sqrt{3} A_{3/2} ), 
\label{eqMtil}\\
G_E & =&   F_+ E_{1+} \nonumber \\
&\equiv &  - F_+ (\sfrac{1}{\sqrt{3}}A_{3/2} - A_{1/2}), 
\label{eqEtil}  \\
\frac{|{\bf q}|}{2 M_R} G_C 
& =& F_+  S_{1+}
 \equiv 
\sqrt{2} F_+  S_{1/2}.
\label{eqStil}
\ea
The multipole amplitudes have the same dimensions as 
the helicity amplitudes.
In this work we define 
the multipole amplitudes with the sign of the form factors.
Other authors use different conventions 
of sign for the multipole amplitudes~\cite{Devenish76,Drechsel2007}.

\subsection{Pseudo-threshold limit}

Now we consider the pseudo-threshold limit.
Since the form factors $G_i$ ($i=1,2,3$), are defined 
with no kinematic singularity,
we can conclude from Eqs.~(\ref{eqGE})-(\ref{eqGC}) 
that~\cite{Jones73}
\ba
G_E= (M_R - M) Z_R G_5, 
\hspace{.5cm}
G_C=  2 M_R Z_R G_5,
\ea
when $|{\bf q}| \to 0$.
A simple consequence of this result~\cite{Jones73}, is
\ba
G_E = \frac{M_R-M}{2 M_R} G_C.
\label{eqJones}
\ea 
To express the relation (\ref{eqJones})
in terms of helicity amplitudes,
we use the relations (\ref{eqEtil})-(\ref{eqStil}) and 
$F_+ =   \frac{1}{e} \frac{M}{M_R + M} 
\sqrt{\frac{M K Q_+^2}{M_R}} \frac{1}{|{\bf q}|}$,
and obtain
\ba
\frac{E_{1+}}{|{\bf q}|} = 
\lambda_R
\frac{S_{1/2}}{|{\bf q}|^2}.
\label{eqSiegert-Amp}
\ea
In the previous relation we recall that $\lambda_R = \sqrt{2}(M_R -M)$
and $E_{1+} = A_{1/2} -\sfrac{1}{\sqrt{3}} A_{3/2}$.
Note in Eq.~(\ref{eqSiegert-Amp}), that the common factor, 
$1/|{\bf q}|$, cannot be eliminated, 
unless we can prove that $E_{1+} \propto |{\bf q}|^n$ 
and $S_{1/2} \propto |{\bf q}|^{(n+ 1)}$, with $n \ge 2$,
near the pseudo-threshold.

The relation (\ref{eqSiegert-Amp}) is consistent with  
\ba
E_{1+} = {\cal O}(|{\bf q}|),
\hspace{1cm}
S_{1/2} = {\cal O}(|{\bf q}|^2),
\ea
near the pseudo-threshold.
The previous forms were adopted by the MAID2007 
parametrization~\cite{MAID2009}.
As for the amplitude $M_{1+}$, the MAID2007 
parametrization gives $M_{1+} = {\cal O}(|{\bf q}|)$,
near the pseudo-threshold 
[which is equivalent to $G_M = {\cal O}(1)$].
The behavior of the multipole amplitudes 
near the pseudo-threshold is
consistent with the results expected when 
the form factors $G_i$ are free of kinematic 
singularities at the 
pseudo-threshold~\cite{Bjorken66,Devenish76}.

To satisfy the condition (\ref{eqSiegert-Amp}),
it is necessary that both sides of the equation 
give the same numerical value.
It is at that point that the MAID2007 
parametrization fails, as we will show next.

To summarize:
we conclude that the correlation 
between the form factors at the  pseudo-threshold
given by Eq.~(\ref{eqJones}), usually 
refereed as the Siegert's theorem,
is not equivalent to the condition 
$E_{1+} = \lambda_R \frac{S_{1/2}}{|{\bf q}|}$.
The equivalent condition is the one expressed by 
Eq.~(\ref{eqSiegert-Amp}).

The results of the MAID2007 parametrization for 
the form factors $G_E$ and $\kappa G_C$, where 
$\kappa= \frac{M_R-M}{2M_R}$, are presented in the 
top panel of Fig.~\ref{figDelta}, 
in comparison with the data from 
Ref.~\cite{MokeevDatabase}.
The database from Ref.~\cite{MokeevDatabase}
includes data for finite $Q^2$
from Refs.~\cite{Stave08,Data,Aznauryan09}, 
and the world data average of $G_E$ at $Q^2=0$,
extracted from the particle data group (PDG)
result for $G_E/G_M$ at $Q^2=0$~\cite{PDG}.

From the top panel of 
Fig.~\ref{figDelta}, we can conclude, 
that, although the MAID2007 describes 
well the data for $G_E$ and $G_C$,
it fails to describe the relation (\ref{eqJones}).
In the graph it is clear that 
$ \kappa G_C (Q_{PS}^2) > G_E (Q_{PS}^2)$.
We discuss now alternative parametrizations 
of the form factors $G_E$ and $G_C$, that are consistent with 
the Siegert's theorem expressed in the form (\ref{eqJones}).

\subsection{Improved parametrizations of $G_E$ and $G_C$}
\label{secDeltaR}

A parametrization consistent with Eq.~(\ref{eqJones}),
inspired in the MAID2007 form is
\ba
& &
\hspace{-0.9cm}
G_E = \frac{C_0}{K} b_0 (1 + b_1 Q^2 + b_2 Q^4 + b_3 Q^6) e^{-b_4 Q^2} G_D 
\label{eqGEsg} \\
& & 
\hspace{-0.9cm}
G_C= \frac{C_0}{K}\frac{2 M_R}{K}
 c_0 (1 + c_1 Q^2 + c_2 Q^4 + c_3Q^6) e^{-c_4 Q^2} G_D, \nonumber \\
\label{eqGCsg}
\ea
where $G_D = (1 + Q^2/0.71)^{-2}$ is a dipole 
form factor, and 
$C_0= \frac{1}{e} \left( \frac{M^3 K}{M_R}\right)^{1/2}$.
The parameters $(b_0, b_1,b_2,b_3,b_4)$ and  $(c_1,c_3,c_3,c_4)$ 
are adjustable.
There are two main differences between 
the MAID2007 expressions and our expressions,
apart from the constraint 
at pseudo-threshold discussed previously.
The first difference is that we omitted 
the factor $\sqrt{1 + \tau}$,
in Eqs.~(\ref{eqGEsg})-(\ref{eqGCsg}).
This factor appears in the MAID2007 parametrization 
because the form factors were defined 
originally in the Ash representation, 
and not in the  Jones and Scadron representation
(conversion factor)~\cite{Jones73}.
The second difference is the suppression 
of a factor $1/(1 + d \, Q^2/(4M^2))$,
where $d$ is a new parameter, 
used in the MAID2007 parametrization
of the function $G_C$.
This factor was added to the   MAID2007 parametrization
in an attempt to improve the quality of 
the fit near $Q^2=0$~\cite{Tiator16,Drechsel2007}.
We choose not to include that factor 
to avoid possible singularities 
in the timelike region, and also because 
the inclusion of higher powers in $Q^2$, 
as the terms associated with the coefficients $c_2$ and $c_3$,
may be sufficient to simulate 
the effect of an extra monopole factor in $Q^2$
(in MAID2007: $c_2=c_3=0$).

Apart from the two differences discussed above,
the relevant 
difference between the present forms and 
the MAID2007 parametrization is that 
the coefficient $c_0$ is fixed by Eq.~(\ref{eqJones}),
once defined the remaining coefficients. 
We label the improved parametrization given by 
Eqs.~(\ref{eqGEsg})-(\ref{eqGCsg}), 
as the MAID-SG parametrization,
since the new parametrization 
is consistent with the Siegert's theorem 
(SG holds for Siegert).
\begin{figure}[t]
\vspace{.6cm}
\centerline{\mbox{
\includegraphics[width=2.5in]{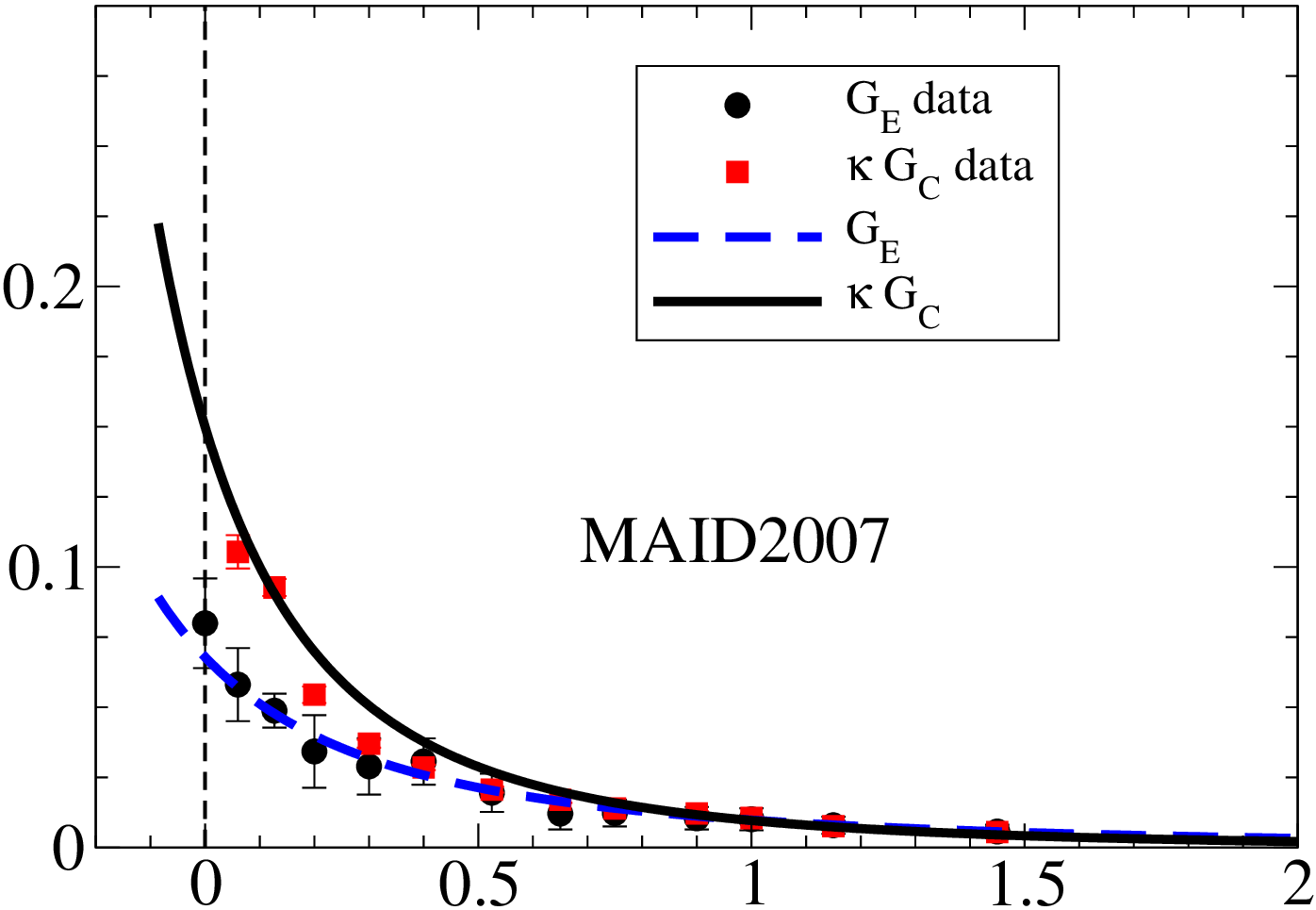}}}
\vspace{.9cm}
\centerline{\mbox{
\includegraphics[width=2.5in]{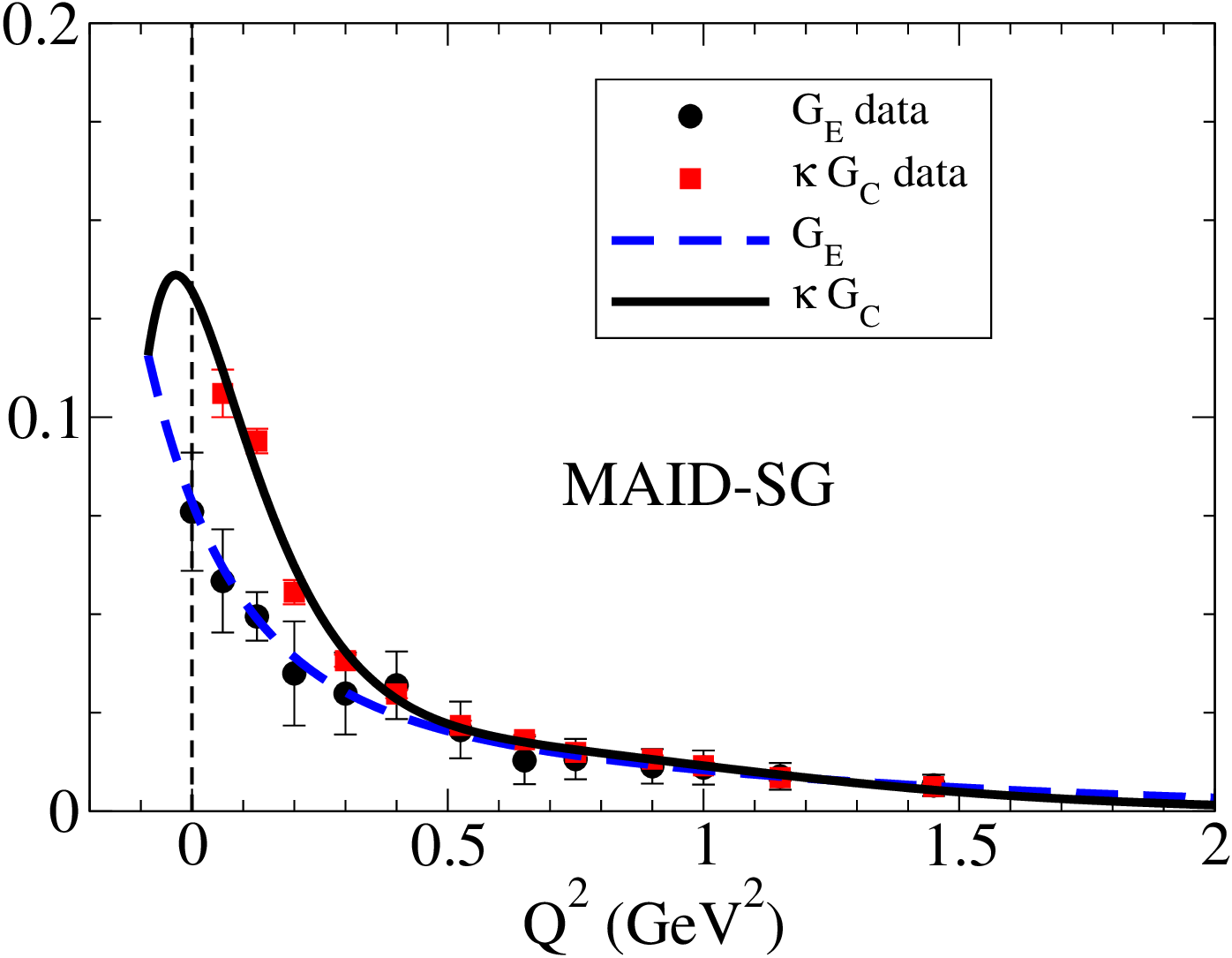}}}
\caption{\footnotesize
Electric and Coulomb quadrupole form factors 
for the $\gamma^\ast N \to \Delta(1232)$ transition.
$G_C$ is multiplied by $\kappa =\frac{M_R-M}{2M_R}$.
At the top: MAID2007 parametrization.
At the bottom: improved parametrization MAID-SG, 
consistent with the Siegert's theorem.
Data from Ref.~\cite{MokeevDatabase} 
(see description in the test).}
\label{figDelta}
\end{figure}

The coefficients defined by the best fit of the functions 
(\ref{eqGEsg})-(\ref{eqGCsg}) to the 
$Q^2 \le 2$ GeV$^2$ data,
 are presented 
in the Table~\ref{tabModel-Delta1}.
The coefficients associated with the 
MAID2007 parametrization are also included in the table.
Note, however, that only the coefficients $r_0$ 
can be directly compared, 
since different combinations of polynomials  
and exponentials may lead to similar functions.

The results for the MAID-SG parametrization
for the form factors $G_E$ and $G_C$
are presented in the lower panel of Fig.~\ref{figDelta}.
At this point we restrict the calculations 
to the region $Q^2 \le 2$ GeV$^2$, since 
the main goal at the moment is the study 
of parametrizations consistent with the Siegert's theorem, 
near $Q^2= Q^2_{PS} \simeq - 0.09$ GeV$^2$.
For larger $Q^2$ there are discrepancies 
between the data from different groups~\cite{Aznauryan09,NDeltaD},
which are not relevant for the discussion near the pseudo-threshold.
In  Fig.~\ref{figDelta}, one can see that 
the MAID-SG parametrization is consistent with 
the data from Ref.~\cite{MokeevDatabase} and with the Siegert's theorem.
Note that, compared to the MAID2007 parametrization, 
the  MAID-SG parametrization, gives smaller values for 
$G_C$ near the pseudo-threshold.

\begin{table}[t]
\begin{center}
\begin{tabular}{l  r r r r r}
\hline
\hline 
MAID-SG   & $r_0$ & $r_1$ & $r_2$ & $r_3$ & $r_4$  \\
\hline
$G_E$  & $14.72$ & $-0.0566$ & $1.91$  & $0.0164$ & \sp\sp$1.31$ \\
$G_C$  & \boldmath{$21.82$} & $4.38$ & $-13.54$ & $22.54$ & $3.33$ \\
\hline
MAID2007   & $r_0$ & $r_1$ & $r_2$ & $r_3$ & $r_4$  \\
\hline
$G_E$  & $12.74\, Z_1$ & $-0.021$ & --  & -- & \sp\sp $0.16$  \\
$G_C$  & $24.80\, Z_2$ & $0.12$ & --  & --  & $0.23$ \\
\hline
\hline 
\end{tabular}
\end{center}
\caption{
$\gamma^\ast N \to \Delta(1232)$ transition.
At the top: parameters used in the calculation of the 
form factors in the MAID-SG parametrization.
At the bottom: 
parameters used in the MAID2007 parametrization.
The labels $r_l$ ($l=0,1,2,3,4$) hold for $r_l=b_l,c_l$.
$r_0$ is in units $10^{-3}$ GeV$^{-1/2}$
($C_0/K= 5.32$ GeV$^{1/2}$).
$r_1$ and $r_4$ are in units GeV$^{-2}$,
$r_2$ is in units GeV$^{-4}$
and $r_3$ is in units  GeV$^{-6}$.
In the MAID-SG parametrization
the value of $c_0$ (at bold) is determined by Eq.~(\ref{eqJones}).
In the MAID2007 parametrization, the coefficients 
$r_0$ are corrected by the  functions $Z_1= \sqrt{1 + \tau}$
and $Z_2 =1/(1+ 4.9 Q^2/(4M^2))  Z_1$.}
\label{tabModel-Delta1}
\end{table}

\begin{figure}[t]
\vspace{.6cm}
\centerline{\mbox{
\includegraphics[width=2.6in]{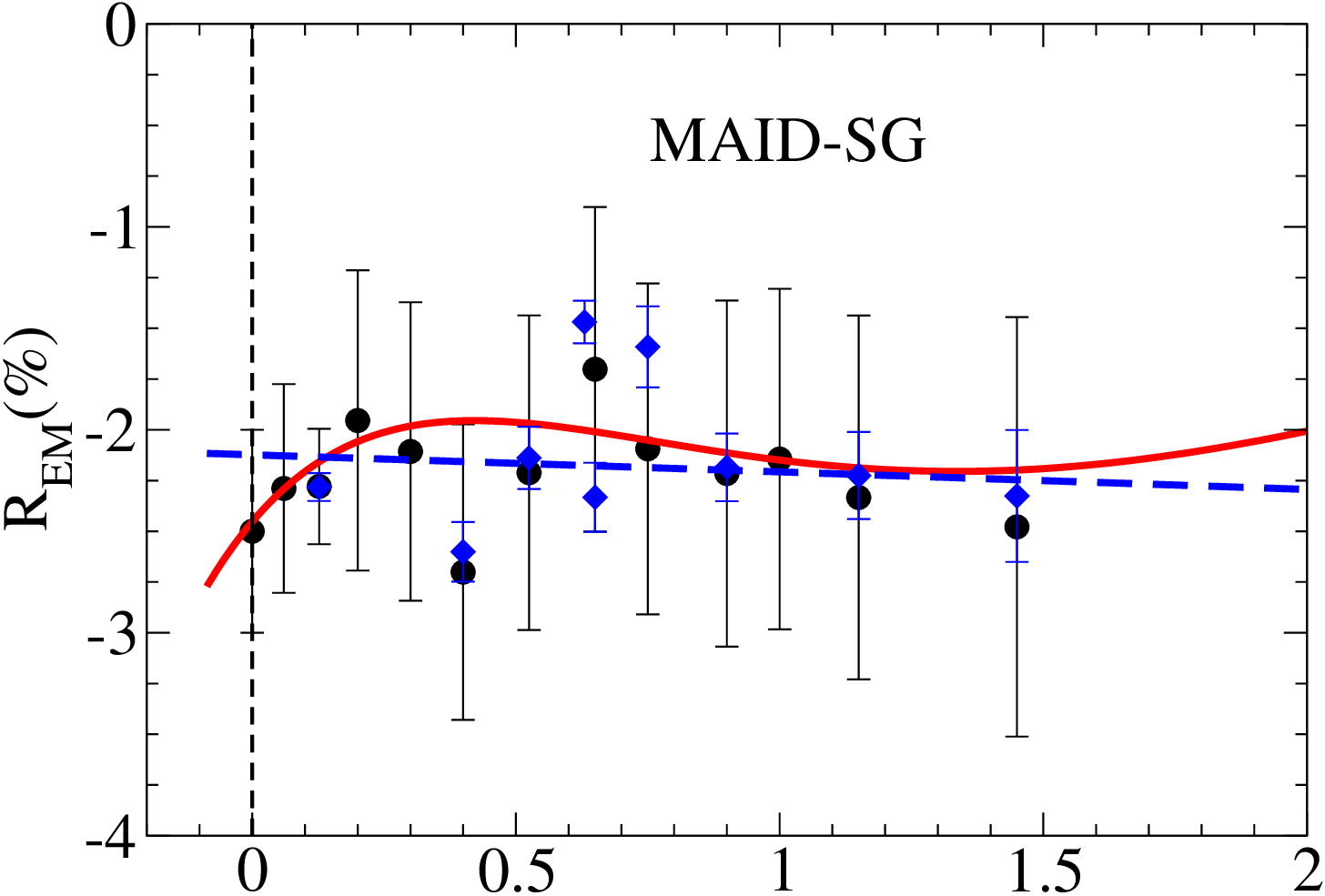}}}
\vspace{.8cm}
\centerline{\mbox{
\includegraphics[width=2.6in]{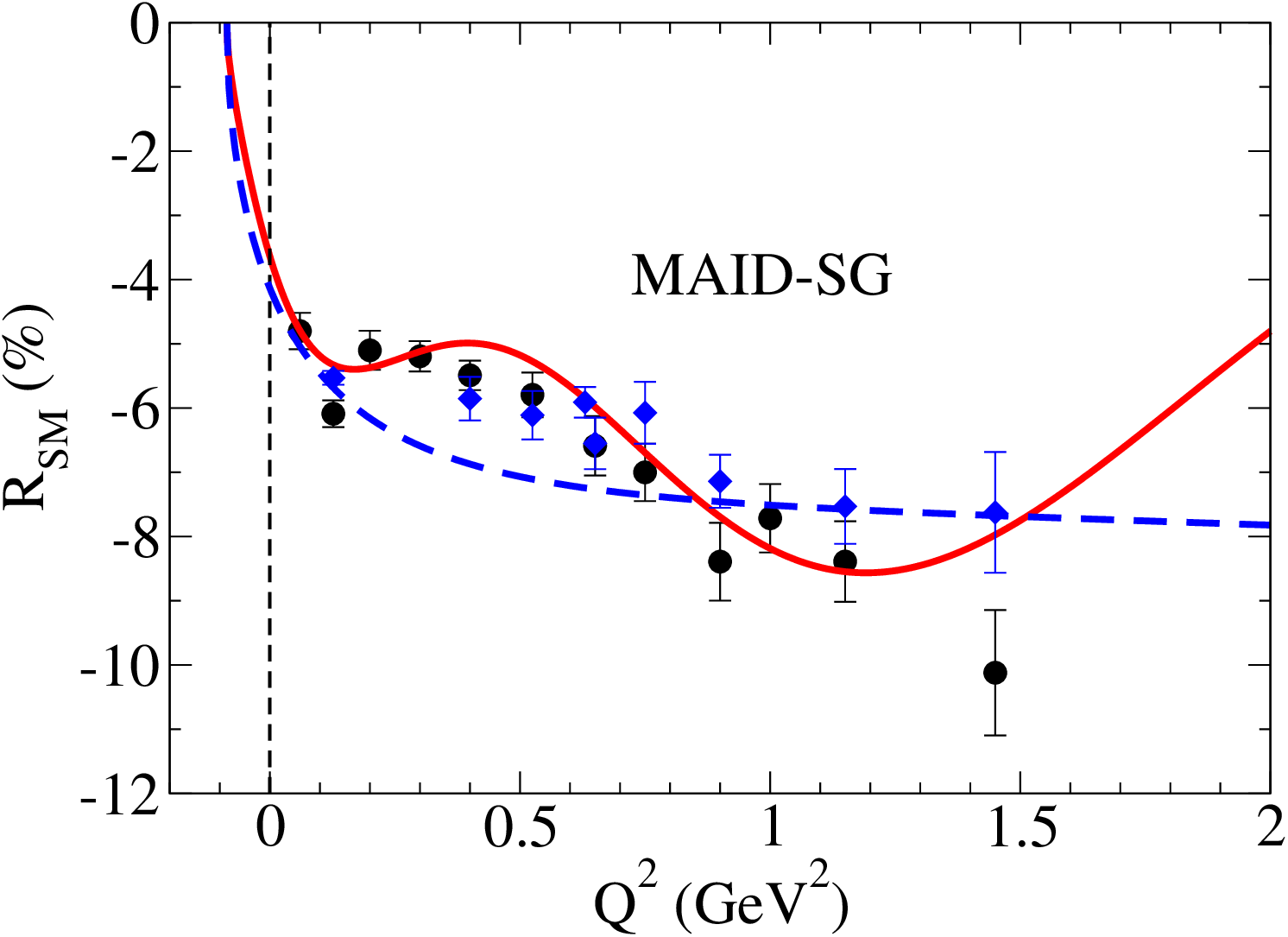}}}
\caption{\footnotesize
$\gamma^\ast N \to \Delta(1232)$ transition.
Ratios $R_{EM}$ and $R_{SM}$ compared 
with the MAID-SG (solid line) 
and MAID2007 (dashed line) parametrizations. 
The data is from the  MAID analysis (diamonds)~\cite{Drechsel2007}
and from the Ref.~\cite{MokeevDatabase} (circles).}
\label{figDeltaR}
\end{figure}

In order to check in more detail 
the implications of Eq.~(\ref{eqJones}),
instead of looking for the form factors, 
we compare the MAID-SG parametrization with the 
data for $R_{EM} = - \frac{G_E}{G_M}$ 
and $R_{SM} = - \frac{|{\bf q}|}{2 M_R} \frac{G_C}{G_M}$.
To calculate $G_M$ we use the MAID2007 parametrization,
since it gives a very  good description 
of the data and it is unconstrained at the pseudo-threshold.
The results for the ratios $R_{EM}$ and $R_{SM}$ 
are presented in Fig.~\ref{figDeltaR}.
In addition to the previous data, we present 
also the MAID data~\cite{Drechsel2007}.
In the figure, one note the different behavior 
between the MAID2007 parametrization 
and the MAID-SG parametrization.
Part of this difference is a consequence 
of the discrepancy between the MAID data 
and the data used in our fit~\cite{MokeevDatabase}.
We note, however, that, even the MAID2007 
parametrization has problems in describing 
the MAID data for $R_{SM}$ below 1 GeV$^2$.

From Figs.~\ref{figDelta} and \ref{figDeltaR},
we can conclude, that, the  MAID-SG parametrization
gives a very good description of the 
low $Q^2$ data ($Q^2 < 1$ GeV$^2$) for $G_E$ and $G_C$. 
Both functions 
are smooth near the pseudo-threshold. 
Looking in more detail in Fig.~\ref{figDeltaR},
we see that the MAID-SG parametrization  starts to fail 
when $Q^2 > 1$ GeV$^2$ for $R_{SM}$.
As for $R_{EM}$, one can see that the function  
starts to decrease in magnitude for $Q^2 > 2$ GeV$^2$.
One can also note, that the data for $R_{EM}$
is well approximated by a constant (similar to  MAID2007).
The failure of the parametrization for larger values of $Q^2$, 
is in part a consequence of the inclusion 
of exponential functions in the parametrization 
of the form factors.

The disadvantage of the use of parametrizations 
based on exponential factors is that 
those parametrizations are not valid, in general, 
for large interval in $Q^2$, or fail when $Q^2$ increases.
Latter on, in Sec.~\ref{secExtension},
we discuss the  possibility of extending 
the parametrization of the data for large $Q^2$.

We checked, however, if we can improve the description 
of the large $Q^2$ region ($Q^2 > 2$ GeV$^2$), 
by enlarging the range of the data used in the 
fit to $Q^2=3$ GeV$^2$, or  $Q^2=4.1$ GeV$^2$.
Overall, we can improve the description 
of the data for $Q^2> 1$ GeV$^2$, but the 
description for the low $Q^2$ region losses quality. 
In particular the result for $G_E(0)$ is overestimated 
($R_{EM}$ is underestimated), compared
to the PDG result
[$G_E(0)= 0.076 \pm 0.015$ and $R_{EM}(0)= -(2.5 \pm 0.5) \%$]~\cite{PDG}.
Note that, when we extend the range of the fit 
for larger values of $Q^2$, up to 3 GeV$^2$ or 4.1 GeV$^2$, 
we reduce the impact of the low $Q^2$ region in our fit,
which leads to a poor estimate 
of the form factors near the pseudo-threshold.
We may then conclude, that, with the parametrizations 
 (\ref{eqGEsg})-(\ref{eqGCsg}), 
we can not describe well the low and the large $Q^2$ regions
simultaneously.
For this reason we restrict, for now, 
our analysis to the low $Q^2$ region.

It is worth to mention, that, the fit 
based on Eqs.~(\ref{eqGEsg})-(\ref{eqGCsg}), 
is very sensitive to the low $Q^2$ data, 
in particular to the result at the photon point 
from PDG~\cite{PDG}.
If the datapoint from PDG is replaced
by another datapoint, or the errorbar is reduced in the fit, 
the results for the form factors may change significantly.
We note in particular that, in some 
experiments like in Ref.~\cite{LEGS01}, 
the value of $R_{EM}$ is larger in 
absolute value, 
$R_{EM}(0)= -(3.07 \pm 0.36)\%$.

The Siegert's theorem was already investigated 
in the context of the $\gamma^\ast N \to \Delta(1232)$ transition,  
within the quark model 
framework~\cite{Buchmann98,Drechsel84,Weyrauch86,Bourdeau87,Capstick90}.
It was found that the Siegert's theorem can be violated
when the operators associated with the current density 
or the charge density are  truncated, 
or expanded in different orders, 
inducing a violation of the current 
conservation condition~\cite{Buchmann98,Drechsel84,Weyrauch86}.
From those studies, one can conclude that 
a consistent calculation, where the current is conserved,
requires the inclusion of processes 
beyond the impulse approximation at the quark level 
(one-body currents), 
and that, the inclusion of higher order processes 
involving two-body currents, 
such as  processes with quark-antiquark states 
and/or meson cloud contributions, is necessary to ensure 
the conservation of the transition current 
and the Siegert's theorem~\cite{Buchmann98}.
Since the Siegert's theorem is defined 
at the pseudo-threshold, when $Q^2 \simeq -0.09$ GeV$^2$,
one may then conclude, that, 
processes beyond the impulse approximation are 
fundamental to describe the helicity 
amplitudes and the transition form factors at low $Q^2$.

The last conclusion is particularly important 
for the $\gamma^\ast N \to \Delta(1232)$ transition,
since there are strong evidences of importance 
of the meson cloud effects for all 
form factors at small 
$Q^2$~\cite{NSTAR,Aznauryan12,NDelta,NDeltaD,LatticeD,Delta1600}.
For the magnetic dipole form factor, $G_M$, 
it is known that the meson cloud effects are small for 
$Q^2> 2$ GeV$^2$~\cite{NSTAR,Aznauryan12,NDelta,NDeltaD}.
As for the quadrupole form factors  
$G_E$ and $G_C$, there are indications 
that the meson cloud contributions 
may be important up to 
4 GeV$^2$~\cite{NDeltaD,Buchmann04}.
The meson cloud contributions for the quadrupole form factors  
 will be discussed in detail in Sec.~\ref{secDiscussion2}.

It is important to mention that, the distinction 
between the valence quark degrees of freedom 
and the non-valence quark degrees of freedom
is only well defined in a given framework.
Therefore, valence quark contributions in a given 
model may appear as non-valence quark contributions in another model.
One can nevertheless conclude that, 
independent of the model, the meson cloud contributions 
help in general to approach the 
estimates from quark models to the experimental data.

\section{$\gamma^\ast N \to N(1520)$ transition}
\label{secN1520}

The current associated with the $\gamma^\ast N \to N(1520)$ 
transition is determined using Eq.~(\ref{eqJgen}),
with the operator~\cite{Aznauryan12,Devenish76,N1520}
\ba
\Gamma^{\alpha \mu}(P,q)= 
G_1 q^\alpha \gamma^\mu  + G_2 q^\alpha P^\mu + 
G_3 q^\alpha q^\mu  - G_4 g^{\alpha \mu}, \nonumber \\
\ea 
where $G_i$ ($i=1,..,4$) are structure form factors 
dependent on $Q^2$.
As in the case of the $\gamma^\ast N \to \Delta(1232)$,
the four form factors are not all independent.
In this case  the current conservation implies that~\cite{N1520}
\ba
G_4= (M_R - M) G_1 + \frac{1}{2} (M_R^2- M^2) G_2 
- Q^2 G_3.
\label{eqG4N}
\ea

The multipole form  factors 
can be represented~\cite{Aznauryan12,N1520}, as
\ba
G_M & = & - Z_R \frac{4 M_R |{\bf q}|^2}{Q_+^2} G_1, 
\label{eqGMn} \\
G_E & = & -  Z_R \left[ 4 (M_R -M)G_5  \frac{}{}
\right.  \nonumber \\
& & \left.
- \frac{4 M_R^2 |{\bf q}|^2}{Q_+^2} 
\left( \frac{G_1}{M_R} + 4 G_3 \right)
\right], 
\label{eqGEn} \\
G_C  &= & -Z_R \left[  
4 M_R G_5 \frac{}{}\right. \nonumber \\
& & \left.
+ \frac{4 M_R^2 |{\bf q}|^2}{Q_+^2} 
\left(G_2 - 2 G_3 \right)
\right],
\label{eqGCn}
\ea
where $Z_R = \frac{1}{\sqrt{6}} \frac{M}{M_R -M}$
and
\ba
G_5 = G_1 + \frac{1}{2}(M_R -M) G_2 + (M_R+ M)G_3.
\ea
Once again, the form factors 
are decomposed  in powers of $|{\bf q}|$.

The helicity amplitudes 
can be determined~\cite{Aznauryan12,N1520} by
\ba
& &
A_{1/2} = \frac{1}{4 F_-} (3 G_M - G_E), \\
& &
A_{3/2} =-  \frac{\sqrt{3}}{4 F_-} (G_M + G_E), 
\label{eqA32b} \\
& &
S_{1/2} = \frac{1}{\sqrt{2} F_-} 
\frac{|{\bf q}|}{2 M_R} G_C.
\ea
Similar  to the $\Delta(1232)$ case 
we also can define the amplitudes
$M_{2-},E_{2-},S_{2-}$~\cite{Devenish76} 
\ba
G_M &= &   F_- M_{2-} \nonumber \\
& \equiv & 
- F_- \left(\frac{1}{\sqrt{3}} A_{3/2} - A_{1/2}    \right), 
\label{eqGMn2} \\
G_E &= & F_- E_{2-} \nonumber \\
& \equiv &
- F_- ( A_{1/2} + \sqrt{3} A_{3/2}),
\label{eqGEn2}  \\
\frac{|{\bf q}|}{2 M_R} G_C &=& 
 F_- S_{2-} \equiv  \sqrt{2} F_- S_{1/2}.
\label{eqGCn2}
\ea

\subsection{Pseudo-threshold limit}

From Eqs.~(\ref{eqGMn})-(\ref{eqGCn}), 
we can conclude that, at the pseudo-threshold~\cite{Devenish76,Note1} 
\ba
& &
G_M=0, \label{eqM2m}\\
& &
G_E = \frac{M_R - M}{M_R} G_C.
\label{eqDeven0}
\ea 
In particular, Eq.~(\ref{eqDeven0}) 
is a consequence of the result
\ba
& &
G_E= - 4(M_R-M) Z_R G_5, 
\nonumber \\
& &
G_C= - 4M_R Z_R G_5, 
\ea
when $|{\bf q}| \to 0$, 
since the form factors $G_i$ ($i=1,2,3$) 
have no kinematic singularities at 
the  pseudo-threshold~\cite{Devenish76}.

\begin{figure}[t]
\vspace{.2cm}
\centerline{\mbox{
\includegraphics[width=2.8in]{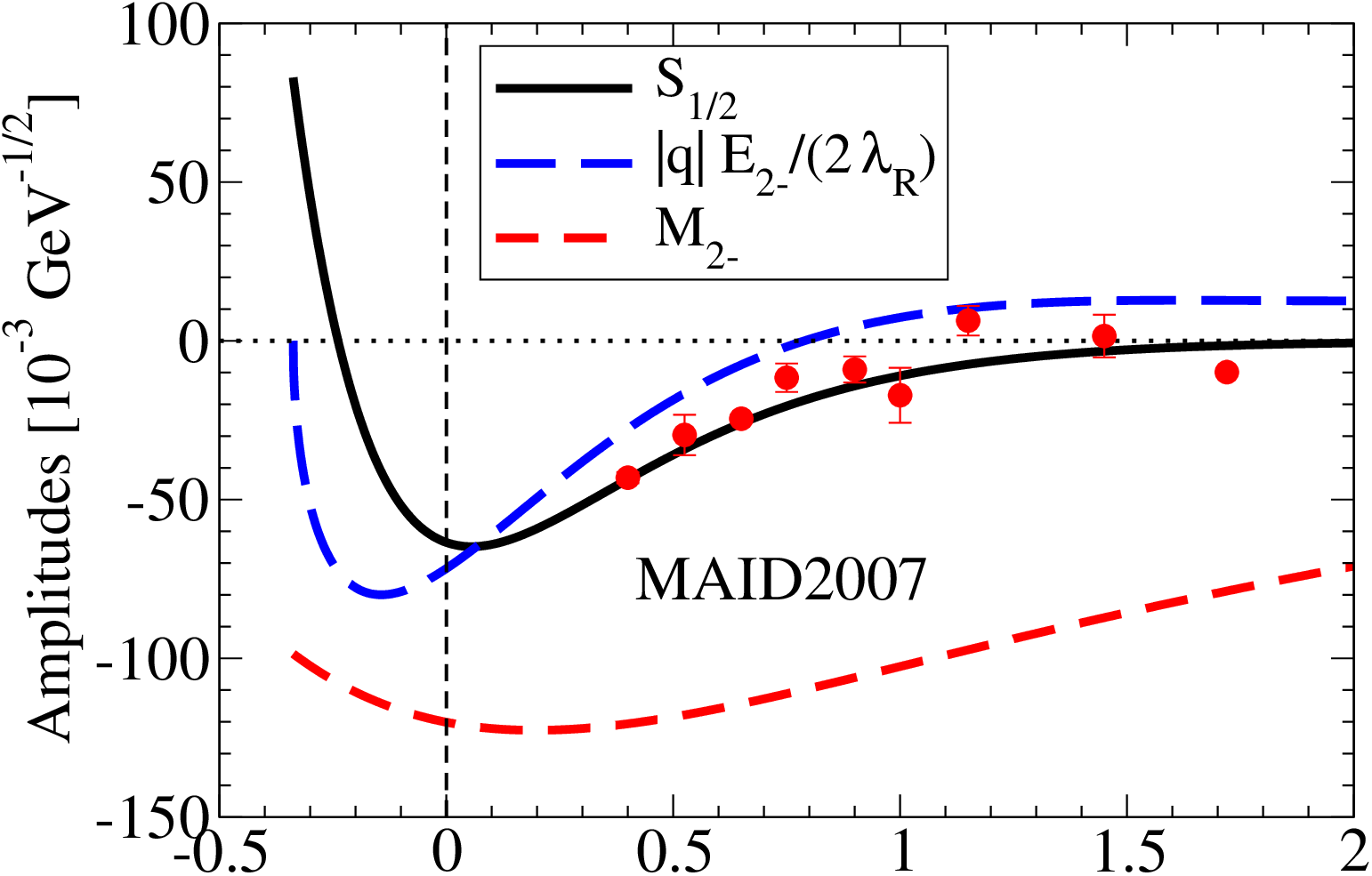}}}
\vspace{.8cm}
\centerline{\mbox{
\includegraphics[width=2.8in]{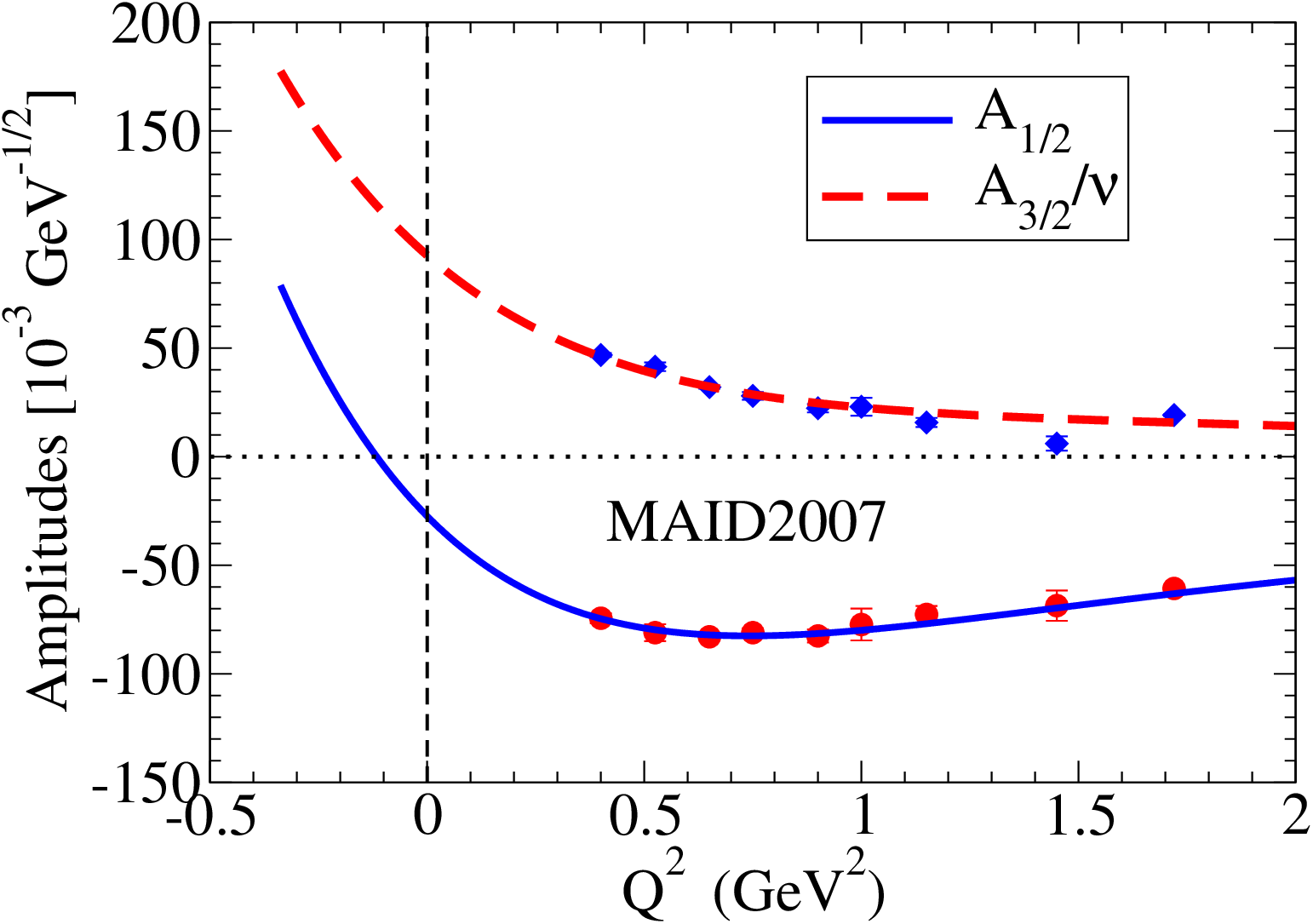}}}
\caption{\footnotesize
$\gamma^\ast N \to N(1520)$ transition.
Results of the MAID2007 parametrization 
for the amplitudes $A_{1/2},A_{3/2}$ and $S_{1/2}$
(the data represented is for $S_{1/2}$).
At the top: $S_{1/2}$ is compared with 
$|{\bf q}| E_{2-}/(2 \lambda_R)$ and $M_{2-}$.
At the bottom:
comparison between $A_{1/2}$ (circles) and $A_{3/2}/\nu$ (diamonds),
where $\nu = \sqrt{3}$. 
Data from the MAID analysis~\cite{Drechsel2007,MAID2009}.}
\label{figS11-MAID}
\end{figure}

A consequence of Eq.~(\ref{eqDeven0}) is, that, 
at the pseudo-threshold
\ba
\frac{1}{2} E_{2-} = \lambda_R \frac{S_{1/2}}{|{\bf q}|},
\label{eqDeven1}
\ea
where, as before $\lambda_R = \sqrt{2}(M_R-M)$.
As for the result from Eq.~(\ref{eqM2m}), it implies that   
\ba
A_{1/2} = \frac{1}{\sqrt{3}} A_{3/2},
\label{eqDeven2}
\ea
at the pseudo-threshold [see Eq~(\ref{eqGMn2})].

The relations (\ref{eqDeven1}) and (\ref{eqDeven2})
are consistent with the   
following behavior of the functions
near the pseudo-threshold~\cite{Devenish76}
\ba
& &
M_{2-} = {\cal O}(|{\bf q}|^2), \hspace{1cm}
E_{2-} = {\cal O}(1), \nonumber \\
& &
S_{1/2} = {\cal O}(|{\bf q}|).
\label{eqMult-S11}
\ea
The relation for $M_{2-}$ 
can be derived directly from Eq.~(\ref{eqGMn})
when $G_1 = {\cal O}(1)$.
The dependences of $M_{2-}$ and 
$E_{2-}$ are consistent with 
the expected result for 
the transverse amplitudes 
$A_{1/2},A_{3/2} = {\cal O}(1)$~\cite{Bjorken66}.

Since the available data for the $\gamma^\ast N \to N(1520)$
transition at finite $Q^2$ is restricted to the 
reactions with proton target we will restrict 
our analysis to that case.

The results of the MAID2007 parametrization 
for the amplitudes $A_{1/2},A_{3/2}$ and 
$S_{1/2}$  are presented in Fig.~\ref{figS11-MAID}.
At the top, we test the relation (\ref{eqDeven1}),
multiplied by the factor $|{\bf q}|$.
If the relation is satisfied, the solid line ($S_{1/2}$) 
and the long-dashed line ($|{\bf q}| E_{2-}/(2 \lambda_R)$)
should converge to zero, both, at the pseudo-threshold, 
when $Q^2 \simeq -0.34$ GeV$^2$.
In the same graph we also represent  $M_{2-}$, 
that should also vanish at 
the pseudo-threshold, according to Eq.~(\ref{eqM2m}).
At the bottom, we compare 
$A_{1/2}$ with $A_{3/2}/\sqrt{3}$, 
in order to test how  broken is the relation (\ref{eqDeven2}), 
as a consequence of the violation 
of the condition  $M_{2-}=0$, at the pseudo-threshold.

The results from Fig.~\ref{figS11-MAID} show 
that the relations (\ref{eqDeven1}) and (\ref{eqDeven2}),
which are consequence of the pseudo-threshold limit 
(Siegert's theorem), 
are broken by the MAID2007 parametrization.
The failure of the  MAID2007 parametrization 
is then the consequence of the dependences of
$M_{2-}, E_{2-}, S_{1/2}= {\cal O}(1)$,
near the  pseudo-threshold.
Those dependences are in conflict with the 
expected dependence  of the multipole amplitudes, 
expressed in Eqs.~(\ref{eqMult-S11}).
Now we consider alternative parametrizations 
that are consistent with Eqs.~(\ref{eqM2m}) and (\ref{eqDeven0})
[or alternatively by Eqs.~(\ref{eqDeven1}) and (\ref{eqDeven2})].

\subsection{Improved parametrizations of $G_M,G_E$ and $G_C$}

To obtain parametrizations 
of the  $G_M,G_E$ and $G_C$ data, 
that include the constraints at the  pseudo-threshold,
we consider the following representations
of the helicity amplitudes
\ba
& &
\hspace{-1.2cm}
A_{1/2}= D a_0 (1+ a_1 Q^2 + a_2 Q^4 + a_3 Q^6) e^{-a_4 Q^2}, 
\label{eqA12new}\\
& &
\hspace{-1.2cm}
A_{3/2}= D b_0 (1+ b_1 Q^2 + b_2 Q^4  + b_3 Q^6) e^{-b_4Q^2}, 
\label{eqA32new}\\
& &
\hspace{-1.2cm}
S_{1/2}= \frac{|{\bf q}|}{K} c_0 (1+ c_1 Q^2 + c_2 Q^4 
+ c_3 Q^6) e^{-c_4 Q^2}, 
\label{eqS12new}
\ea 
where $D= K/\sqrt{Q_+^2}$, 
which can also be written in the form $(M_R + M) D = K/\sqrt{1 + \tau}$.
In the expressions (\ref{eqA12new})-(\ref{eqS12new}),
the coefficients
$(a_0,a_1,a_2,a_3,a_4)$, $(b_1,b_2,b_3,b_4)$ and 
$(c_1,c_2,c_3,c_4)$
are adjustable parameters
and $b_0,c_0$ are determined from 
the constraints at the pseudo-threshold,
once  the values of the remaining coefficients are fixed.
Since, at the pseudo-threshold, 
we can write $E_{2-} = - 4 A_{1/2}$, 
using Eq.~(\ref{eqDeven2}), we can conclude, 
that both $b_0$ and $c_0$ can be expressed 
in terms of $a_0$ 
[because  $A_{3/2} \propto A_{1/2}$
and $S_{1/2}/|{\bf q}| \propto E_{2-} \propto A_{1/2}$].

\begin{table}[t]
\begin{center}
\begin{tabular}{l  r r r r r}
\hline
\hline 
Jlab-SG   & $r_0$ & $r_1$ & $r_2$ & $r_3$ & $r_4$  \\
\hline
$A_{1/2}$  & $-97.40$ & $14.62$ & $-9.49$  & $4.40$ & \sp\sp$1.22$ \\
$A_{3/2}$  & \boldmath{$731.50$} & $0.346$ & $0.0399$  & $1.62$ & $2.32$ \\
$S_{1/2}$  & \boldmath{$-65.17$} & $-0.148$ & $1.01$  & -- & $2.46$ \\
\hline 
MAID-SG   & $r_0$ & $r_1$ & $r_2$ & $r_3$ & $r_4$  \\
\hline
$A_{1/2}$  & $-90.43$ & $18.16$  & $-8.61$  & $3.51$ & $1.07$  \\
$A_{3/2}$  & \boldmath{$740.15$} & $-0.443$  & $0.677$ & --    & $1.29$  \\
$S_{1/2}$  & \boldmath{$-73.82$} & $-1.26$ & $0.839$  & $-0.185$ &  $1.41$ \\
\hline
MAID2007   & $r_0$ & $r_1$ & $r_2$ & $r_3$ & $r_4$  \\
\hline
$A_{1/2}$  & $-143.37 \, Z$ 
& $8.580$  & $-0.252$  & $0.357$  & $1.20$  \\
$A_{3/2}$  & $840.31 \, Z$ & 
$-0.820$  &  $0.541$ & $-0.016$  & $1.06$  \\
$S_{1/2}$  & $-63.5 \, X$ & $4.19$ & --  & -- &  $3.40$ \\
\hline
\hline
\end{tabular}
\end{center}
\caption{
$\gamma^\ast N \to N(1520)$ transition.
Parameters used in the calculation of the 
amplitudes $A_{1/2}$, $A_{3/2}$ and $S_{1/2}$, 
for the  Jlab-SG, MAID-SG 
and MAID2007 parametrizations.
The labels $r_l$ ($l=0,1,2,3,4$) hold for $r_l=a_l,b_l,c_l$.
$r_0$ is in units $10^{-3}$ GeV$^{-1/2}$, $r_1,r_4$ are in units GeV$^{-2}$,
$r_2$ is in units GeV$^{-4}$
and $r_3$ is in units  GeV$^{-6}$.
The values in bold are determined by Eqs.~(\ref{eqDeven1}) 
and (\ref{eqDeven2}).
In the MAID2007 parametrization, $a_0,b_0$ are corrected 
by the function $Z=\sqrt{1 + \tau}$, and $c_0$ 
by the function $X=K/|{\bf q}|$.}
\label{tabModels}
\end{table}

In Eqs.~(\ref{eqA12new})-(\ref{eqA32new}),
the factor $1/\sqrt{Q_+^2}$, 
is included in $D$ for convenience, in 
order to generate simpler analytic 
expressions for the form factors $G_M,G_E$ and $G_C$,
after the multiplication by the factor $F_-$,
defined by Eq.~(\ref{eqFpm}).
The conversion into the form factors can be done 
using Eqs.~(\ref{eqGMn2})-(\ref{eqGCn2}).
Alternatively we can use parametrizations 
for the form factors.
However, in the present case, the inclusion 
of the constraints at the pseudo-threshold 
is simplified when we use the 
helicity amplitude representation, combined 
with Eqs.~(\ref{eqDeven1}) and (\ref{eqDeven2}).

\begin{figure}[t]
\vspace{.6cm}
\centerline{\mbox{
\includegraphics[width=2.8in]{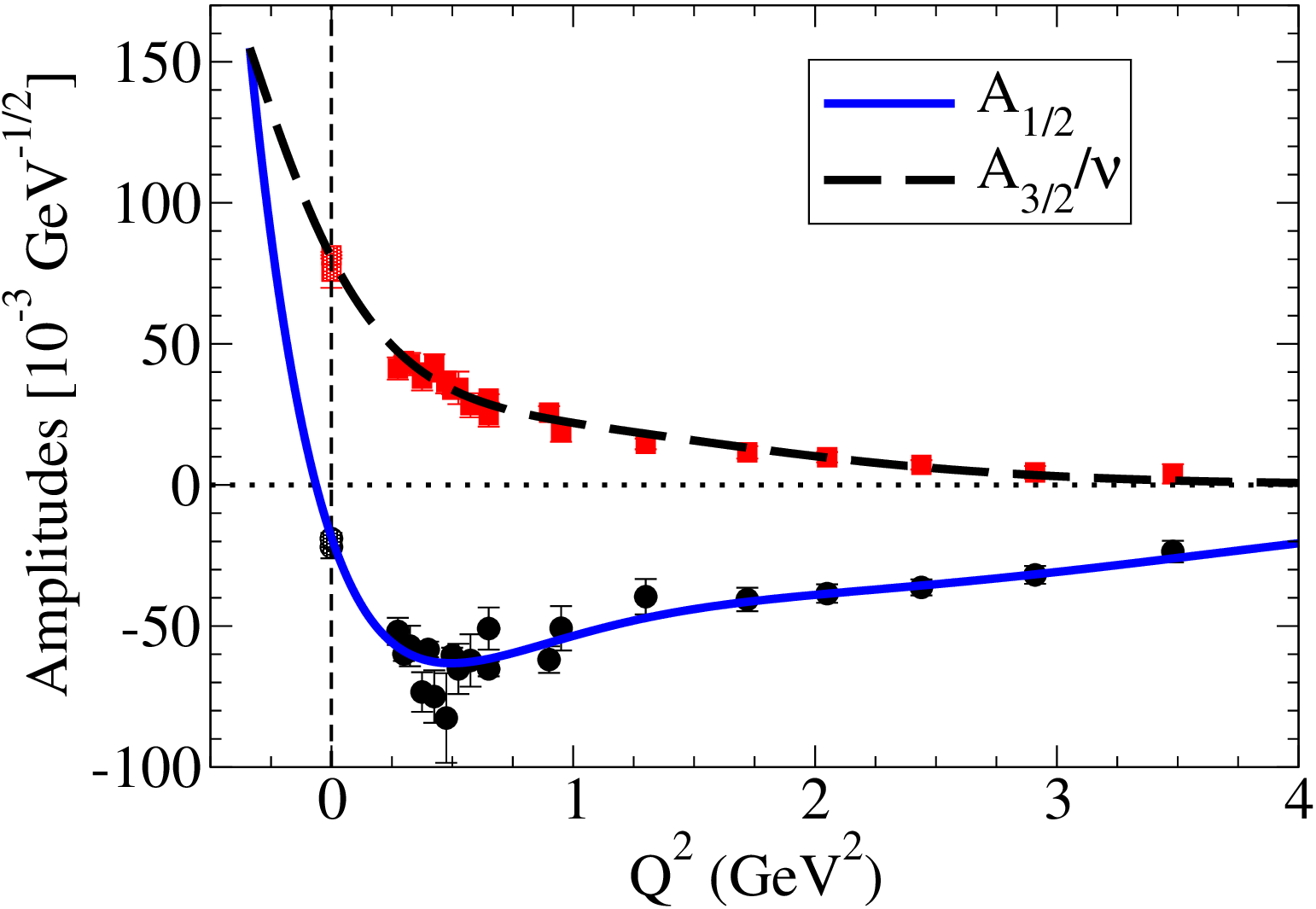}}}
\caption{\footnotesize
$\gamma^\ast N \to N(1520)$ transition.
Results of the Jlab-SG  parametrization for
the amplitudes $A_{1/2}$ (circles) and $A_{3/2}/\nu$ (squares),
where $\nu =\sqrt{3}$.
Data from Jlab~\cite{Aznauryan09,Mokeev12,Mokeev15} 
and  Refs.~\cite{Anisovich12,Workman12} ($Q^2=0$).}
\label{figA12A32}
\end{figure}
\begin{figure}[t]
\vspace{.6cm}
\centerline{\mbox{
\includegraphics[width=2.6in]{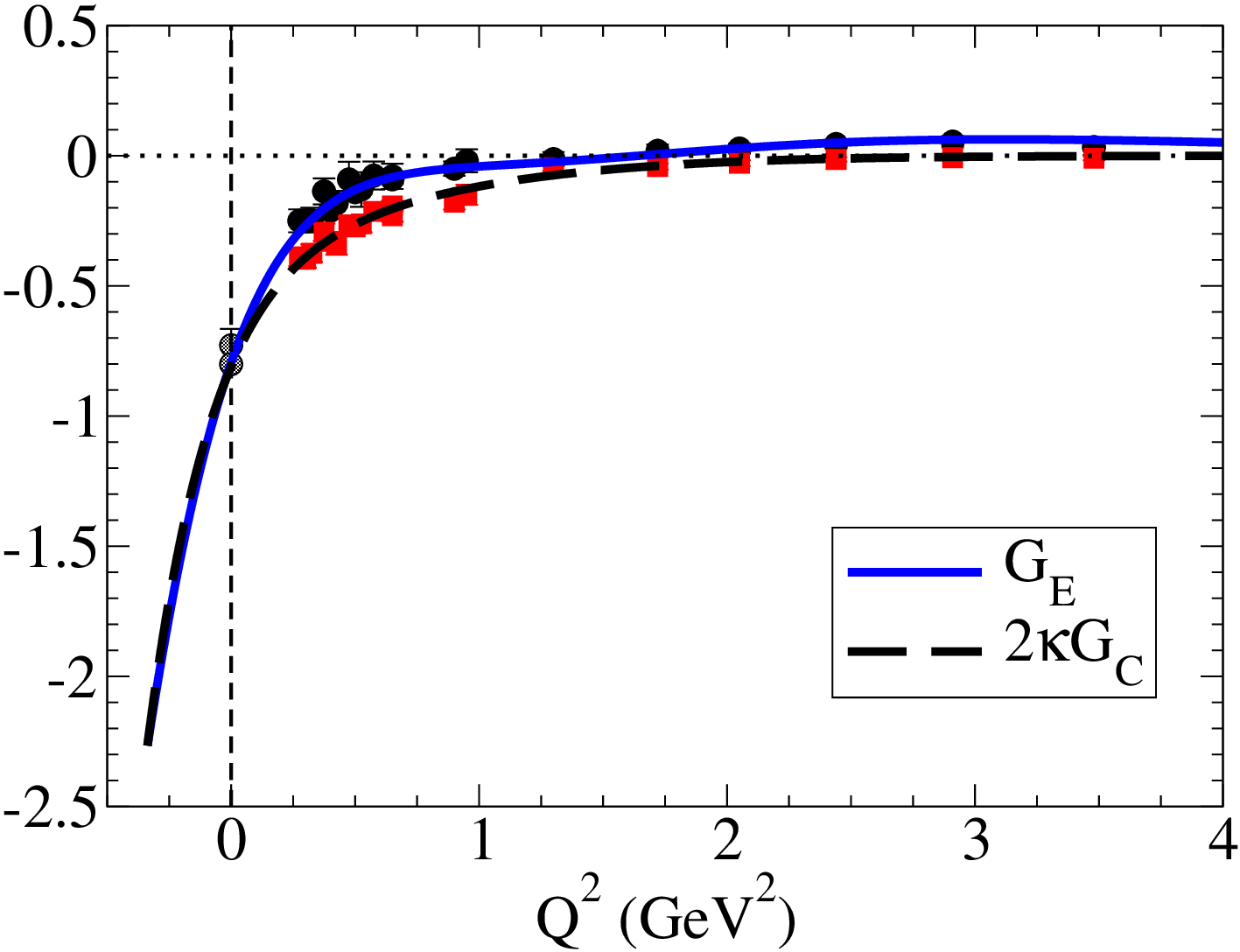}}}
\caption{\footnotesize
$\gamma^\ast N \to N(1520)$ transition.
Results of the  Jlab-SG  parametrization
for the form factors $G_E$ (circles) and $G_C$ (squares)
[multiplied by $2\kappa= (M_R-M)/M_R$].
Data from Jlab~\cite{Aznauryan09,Mokeev12,Mokeev15}
and  Refs.~\cite{Anisovich12,Workman12}
($Q^2=0$).}
\label{figGCGE}
\end{figure}

The parametrization (\ref{eqA12new})-(\ref{eqS12new}),
where $A_{1/2},A_{3/2} = {\cal O}(1)$ and $S_{1/2} = {\cal O}(|{\bf q}|)$,
is compatible with the pseudo-threshold limit, 
when we impose (\ref{eqDeven2}), leading to 
$M_{2-} = {\cal O} ( |{\bf q}|^2 )$~\cite{Note2}.
As for the electric amplitude $E_{2-}$,
the result $E_{2-}= {\cal O}(1)$, is the direct consequence of 
the results of the amplitudes $A_{1/2}$ and $A_{3/2}$.

In order to test if the parametrizations 
(\ref{eqA12new})-(\ref{eqS12new}) are compatible with 
the data, we fitted the coefficients 
from the expressions (\ref{eqA12new})-(\ref{eqS12new}) 
to the available data.

Apart from the data for $Q^2=0$, there are 
two main datasets available:
the data measured at CLAS/Jlab~\cite{Aznauryan09,Mokeev12,Mokeev15},
and the data from the MAID analysis~\cite{Drechsel2007,MAID2009}.
Since there are discrepancies between 
the results from Jlab and MAID~\cite{N1520,MAID2009},
we study the two datasets separately.

There are no data for the region 
$Q^2 < 0.28 $ GeV$^2$,
except for the measurements at the photon point ($Q^2=0$).
As a consequence, the fits to the data, 
and the extrapolation to the pseudo-threshold 
depend crucially of the data used at $Q^2=0$.

Since the average presented by the 
PDG is affected by a large errorbar,
which will leave the region near  $Q^2=0$ 
weakly constrained,
we take into account only 
the most recent measurements selected by PDG,
presented in  Refs.~\cite{Anisovich12,Workman12}.

In the following, we look for possible parametrizations 
of the  data from Jlab and MAID.
We analyze first the data from Jlab,  
defining the Jlab-SG  parametrization.
Next, we consider the MAID data, 
deriving the  MAID-SG parametrization.
At the end we compare the results 
from the two parametrizations near 
the pseudo-threshold.

\subsubsection{Jlab data}

We started the fitting process using the Jlab 
data, combined with the $Q^2=0$ data from 
Refs.~\cite{Anisovich12,Workman12}, as discussed above.
The parameters associated with the 
Eqs.~(\ref{eqA12new})-(\ref{eqS12new}) 
are presented in the Table~\ref{tabModels}, 
under the label Jlab-SG.
In the table, $r_l$ ($l=0,1,2,3,4$) 
holds for $a_l,b_l$ and $c_l$.
The values in bold are not the result of the fit. 
As mentioned, those values are 
determined by Eqs.~(\ref{eqDeven1}) and (\ref{eqDeven2}).
The coefficients of the MAID2007 parametrization
are  also presented for comparison.
Some $Q^2$ dependent factors 
used in MAID2007 are included in the caption.

\begin{figure}[t]
\vspace{.6cm}
\centerline{\mbox{
\includegraphics[width=2.8in]{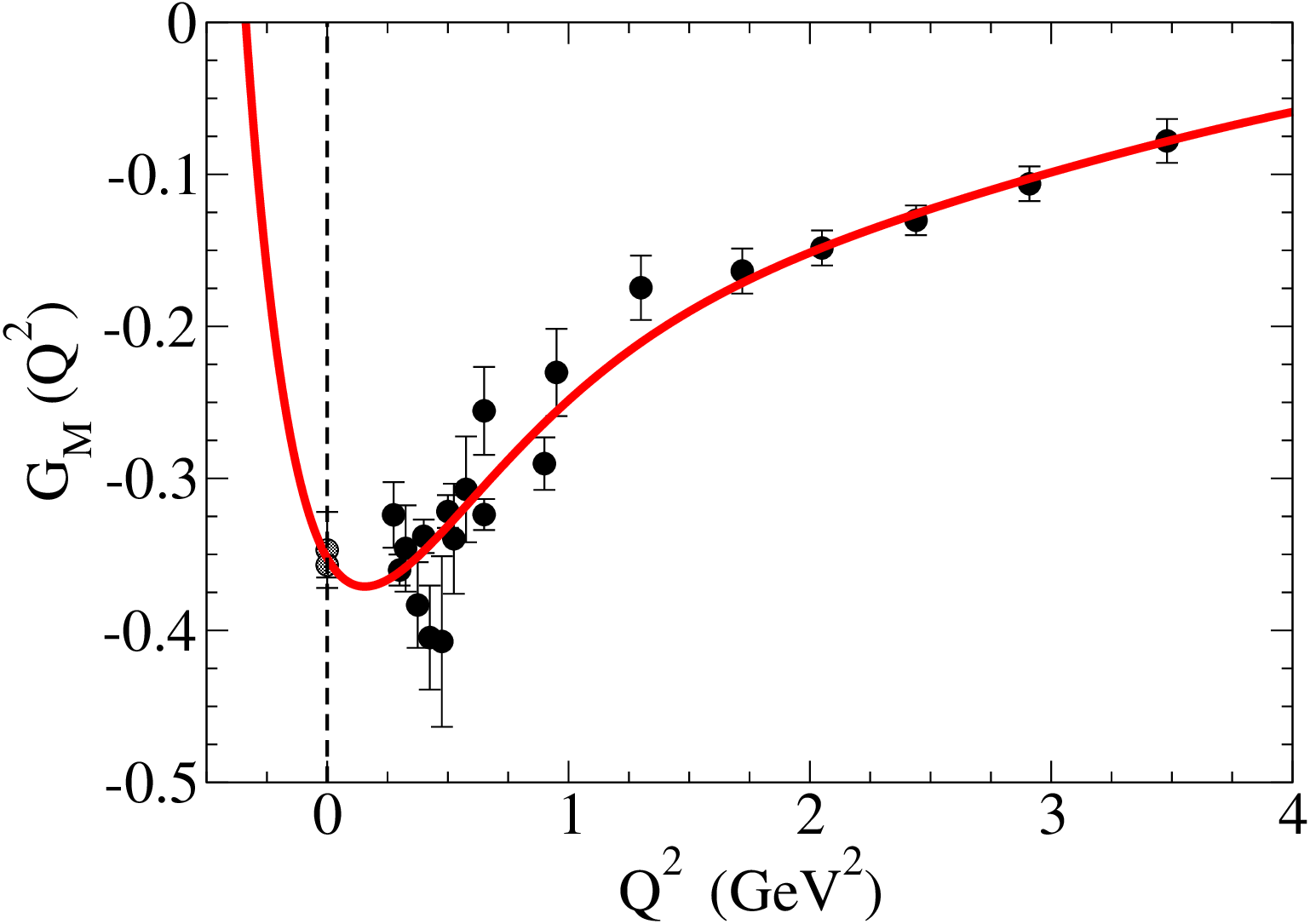}}}
\caption{\footnotesize
$\gamma^\ast N \to N(1520)$ transition.
Results of the  Jlab-SG  parametrization
for the form factor $G_M$.
Data from Jlab~\cite{Aznauryan09,Mokeev12,Mokeev15}
and Refs.~\cite{Anisovich12,Workman12}
($Q^2=0$).}  
\label{figGM}
\end{figure}

\begin{figure}[t]
\vspace{.6cm}
\centerline{\mbox{
\includegraphics[width=2.8in]{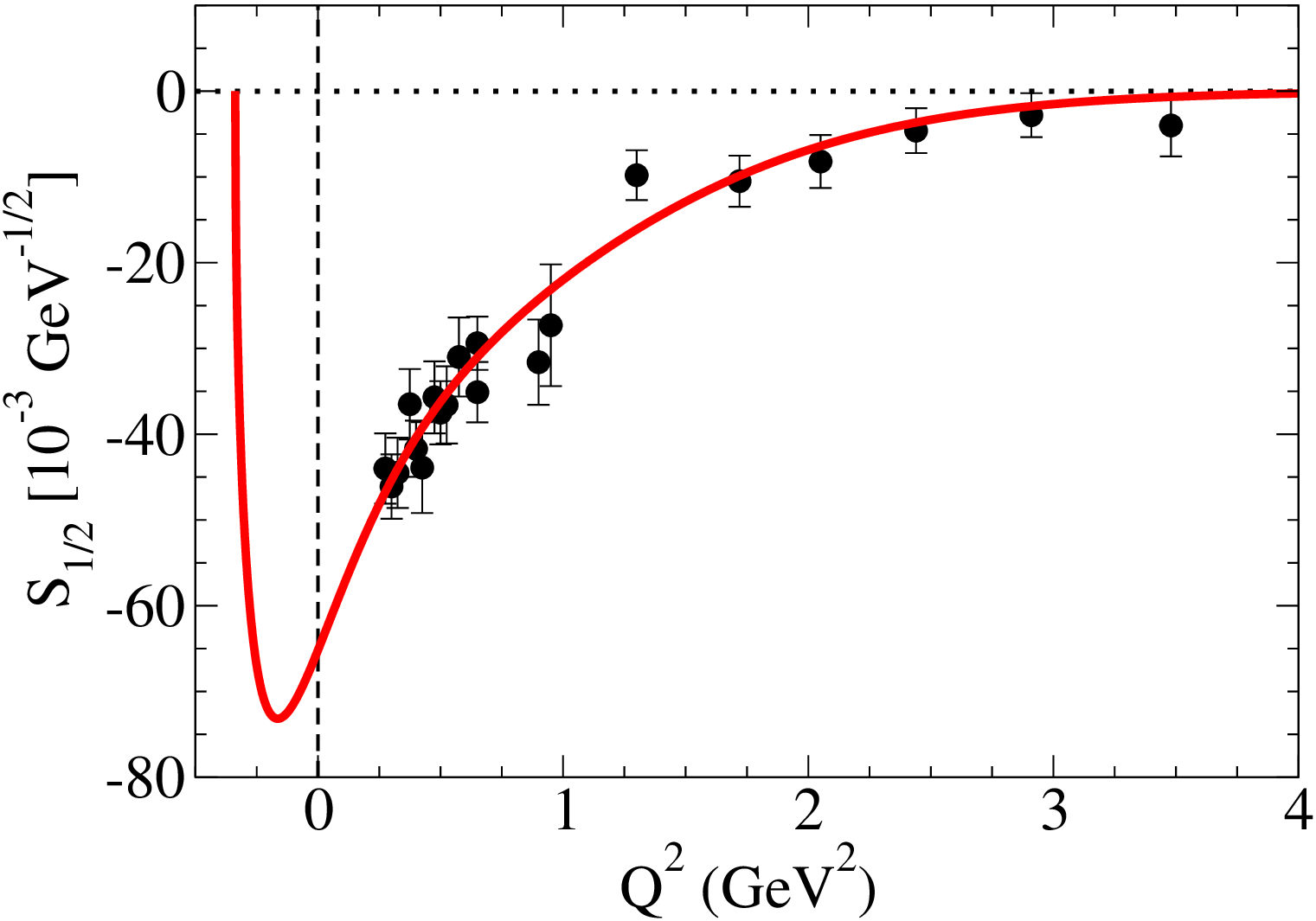}}}
\caption{\footnotesize
$\gamma^\ast N \to N(1520)$ transition.
Results of the parametrization Jlab-SG 
for the amplitude $S_{1/2}$.
Data from Jlab~\cite{Aznauryan09,Mokeev12,Mokeev15}.}
\label{figS12}
\end{figure}

The results for the  Jlab-SG parametrization 
are presented  in Figs.~\ref{figA12A32},
\ref{figGCGE}, \ref{figGM} and \ref{figS12}.
In Fig.~\ref{figA12A32}, we compare 
the amplitudes $A_{1/2}$ and $A_{3/2}$.
In the figure one can see that $A_{1/2}$ 
and $A_{3/2}/\sqrt{3}$ have the same value 
at the pseudo-threshold, 
as expected from Eq.~(\ref{eqDeven2}).
In Fig.~\ref{figGCGE}, we compare the results
for $G_E$ and $G_C$ 
[corrected by the factor $2 \kappa = (M_R-M)/M_R$].
It is clear from the figure, that, 
those form factors have a strong variation 
near $Q^2=0$ (see inflection at low $Q^2$), and that 
$G_E = 2 \kappa G_C$, at the pseudo-threshold,
consistent with the relation (\ref{eqDeven0}).

In Figs.~\ref{figA12A32} and \ref{figGCGE}, 
we tested directly the constraints of our fit. 
We can now observe the consequences of the constraints 
for the remaining functions.
In Fig.~\ref{figGM}, we show the result of the 
parametrization for the magnetic form factor $G_M$.
In the graph we can confirm that $G_M=0$, 
at the  pseudo-threshold, which 
is the consequence of the relation Eq.~(\ref{eqDeven2}),
tested in Fig.~\ref{figA12A32}.

Finally, in Fig.~\ref{figS12}, we show 
the results for the amplitude $S_{1/2}$.
Although the information included in  $S_{1/2}$
is almost the same as the one included in the form 
factor $G_C$, since $G_C \propto (F_- /|{\bf q}|) S_{1/2} $,
it is interesting to see the behavior
near $Q^2=0$, and the convergence to zero 
at the pseudo-threshold. 

Overall, the  Jlab-SG parametrization 
gives a  very good description 
of the Jlab data, both at low and large $Q^2$.
In addition, the Jlab-SG parametrization 
also provides regular extensions 
for the region $Q^2 < 0$, including the pseudo-threshold limit.


\subsubsection{MAID data}

We now look  for the results from 
the MAID analysis~\cite{Drechsel2007,MAID2009}.
We checked if the data can be described by  
Eqs.~(\ref{eqA12new})-(\ref{eqS12new}).
The results obtained for the coefficients 
of the fit are presented in the Table~\ref{tabModels},
under the label MAID-SG.
Since we concluded already  that 
the relations (\ref{eqA12new})-(\ref{eqS12new})
are compatible with the Siegert's theorem,
there is no need to test 
those relations for the new parametrization.

\begin{figure}[t]
\vspace{.6cm}
\centerline{\mbox{
\includegraphics[width=2.8in]{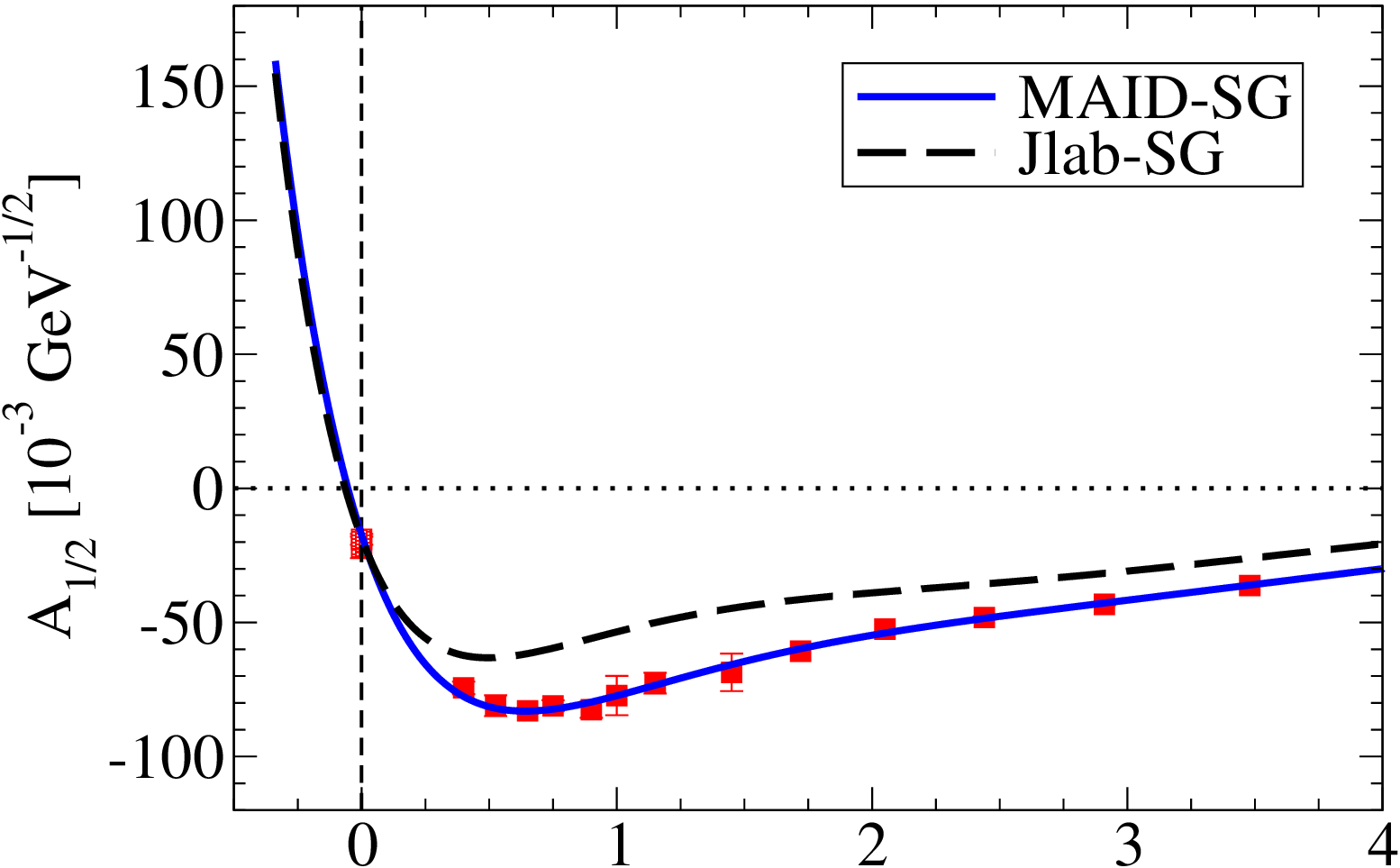}}}
\vspace{.8cm}
\centerline{\mbox{
\includegraphics[width=2.8in]{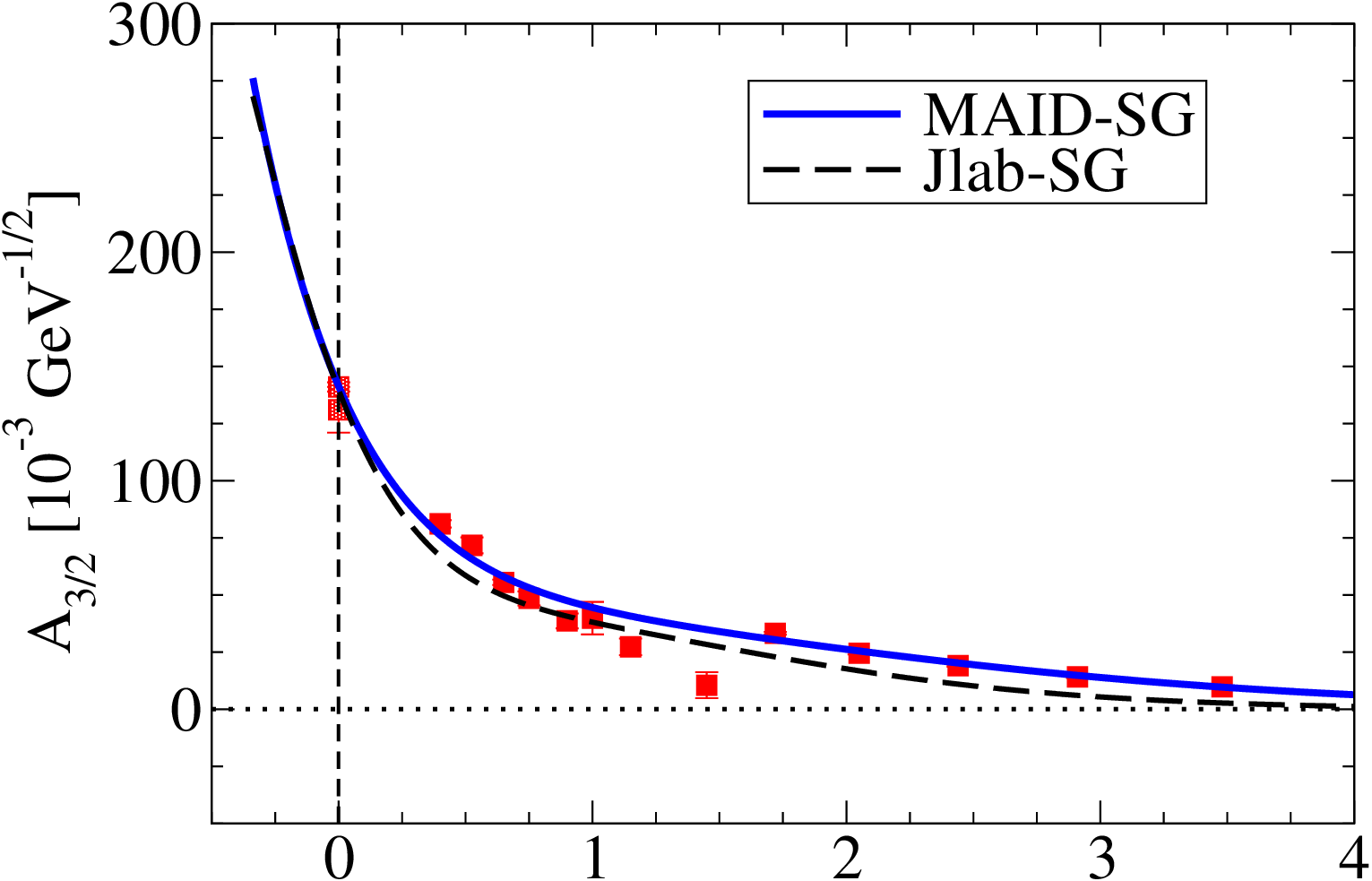}}}
\vspace{.9cm}
\centerline{\mbox{
\includegraphics[width=2.8in]{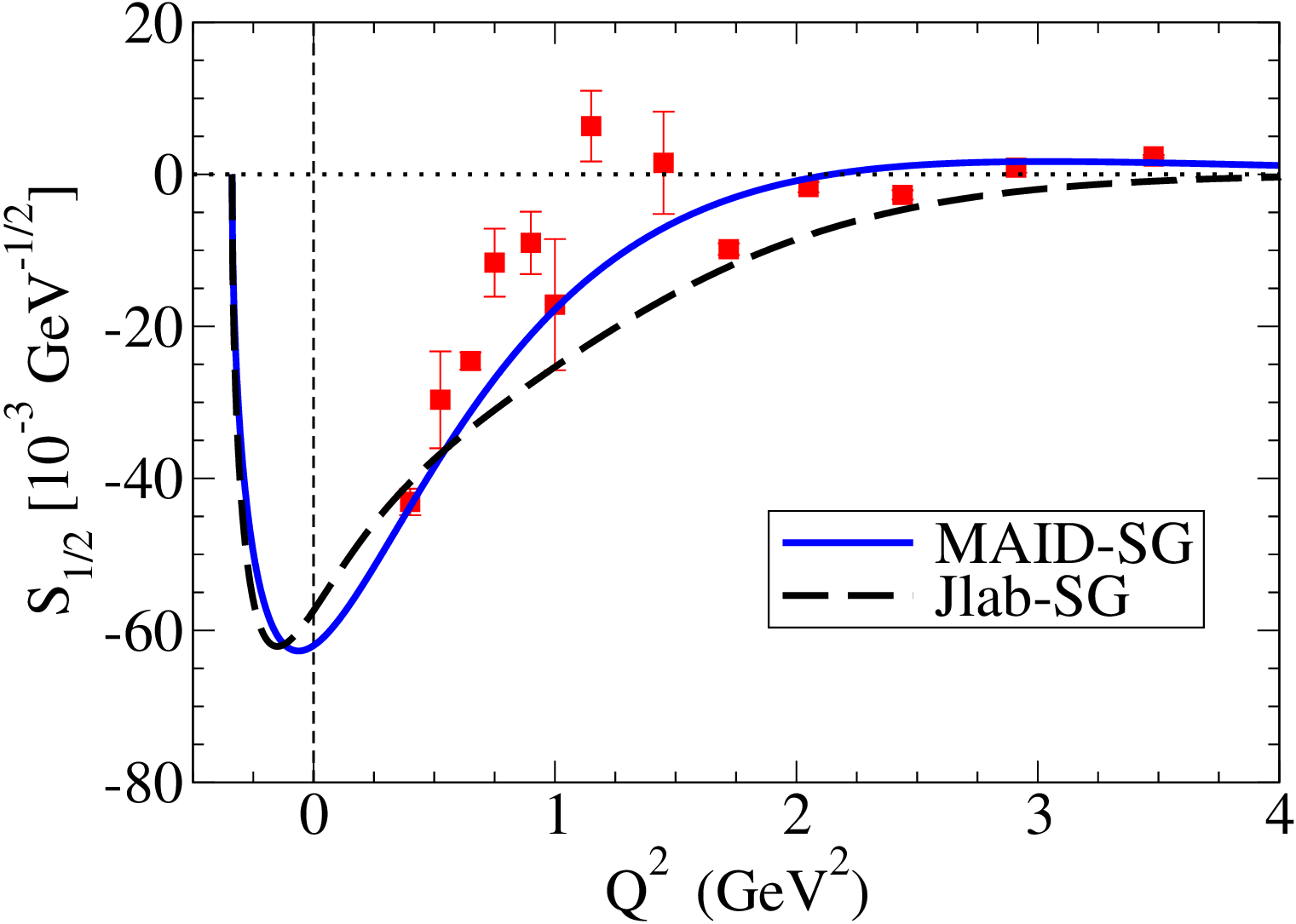}}}
\caption{\footnotesize
$\gamma^\ast N \to N(1520)$ transition.
Results for the MAID-SG amplitudes compared with the 
MAID data~\cite{Drechsel2007,MAID2009}.
The Jlab-SG  parametrization is also presented.}
\label{figN1520amps}
\end{figure}
\begin{figure}[h]
\vspace{.6cm}
\centerline{\mbox{
\includegraphics[width=2.8in]{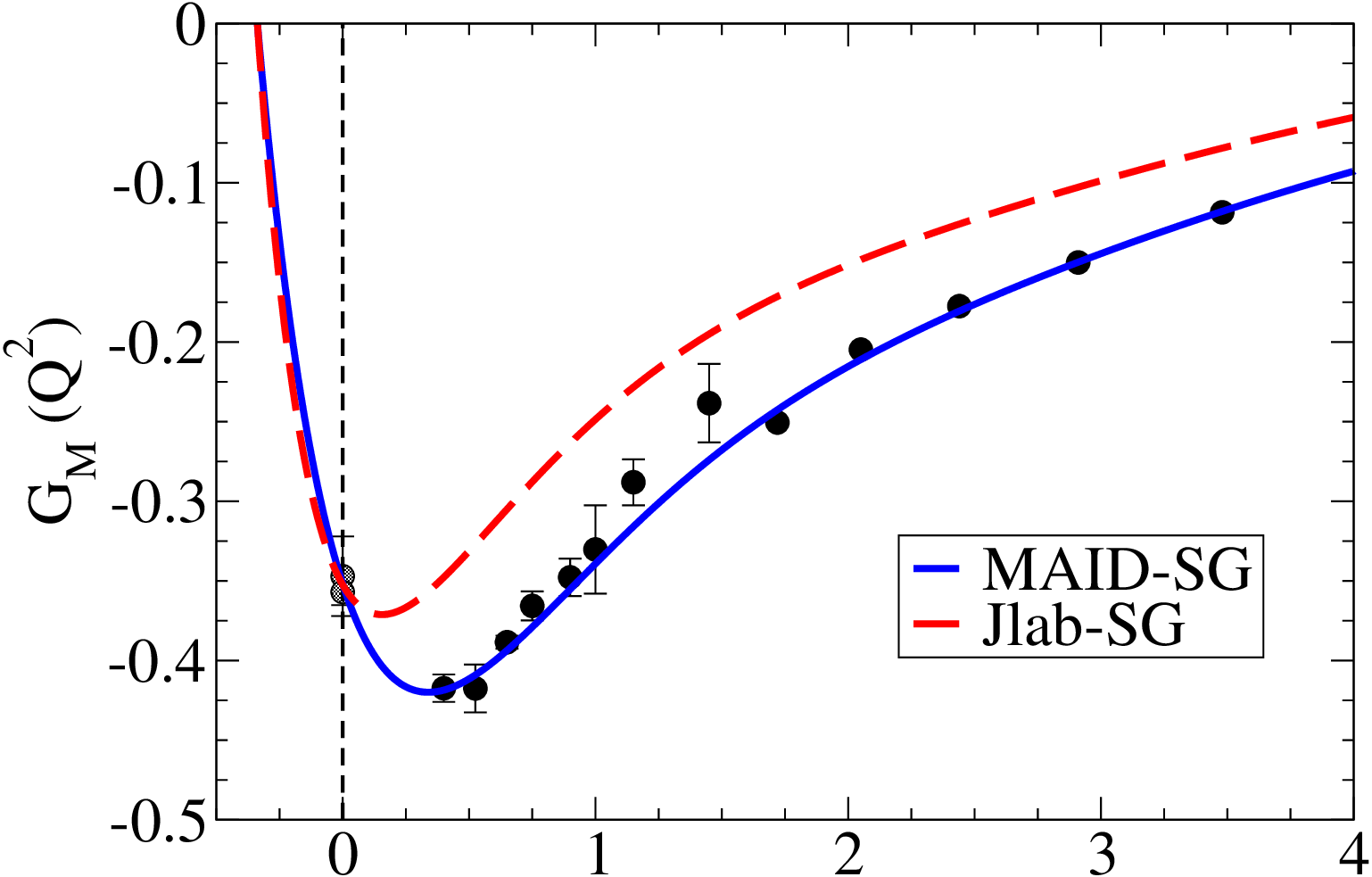}}}
\vspace{.8cm}
\centerline{\mbox{
\includegraphics[width=2.8in]{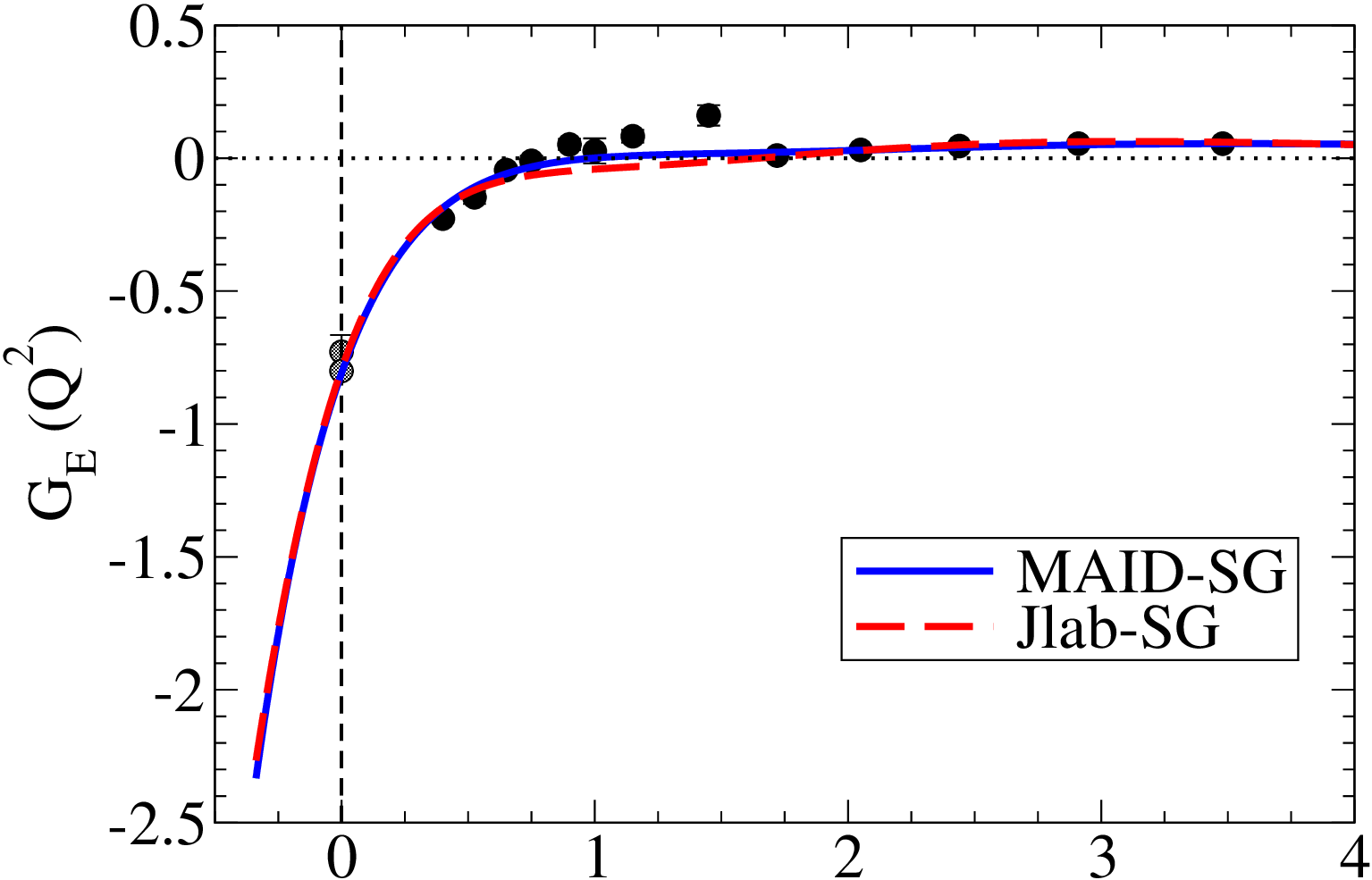}}}
\vspace{.9cm}
\centerline{\mbox{
\includegraphics[width=2.8in]{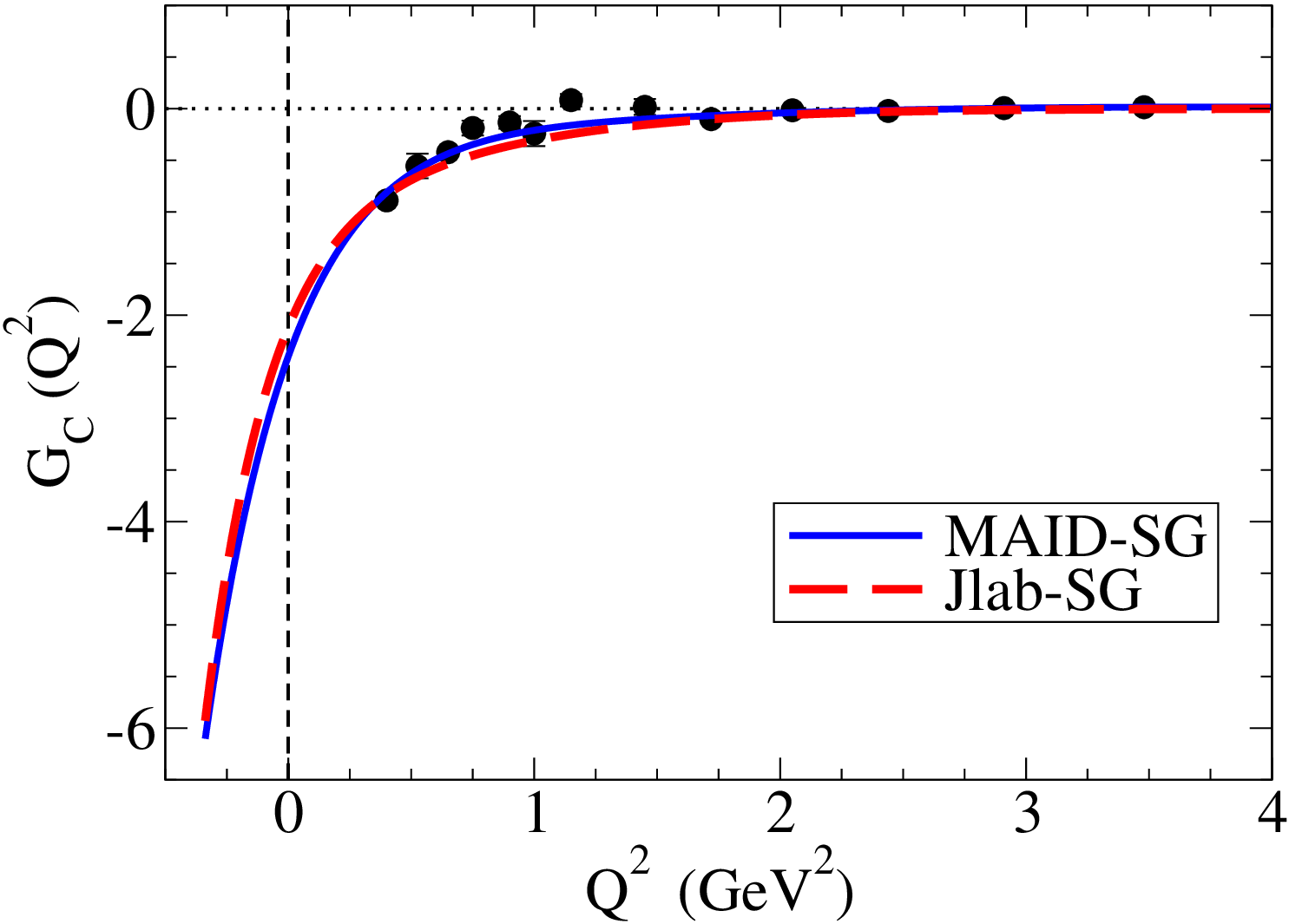}}}
\caption{\footnotesize
$\gamma^\ast N \to N(1520)$ transition.
Results for the MAID-SG form factors compared with the 
MAID data~\cite{Drechsel2007,MAID2009}.
The  Jlab-SG parametrization is also presented.}
\label{figN1520FF}
\end{figure}

The results of the MAID-SG parametrization for the 
helicity amplitudes are presented in Fig.~\ref{figN1520amps}
and the corresponding results for 
the form factors are presented in Fig.~\ref{figN1520FF}.

Contrary to to the case of the Jlab data, it is not possible 
in this case to obtain a high quality fit,
based on the form (\ref{eqA12new})-(\ref{eqS12new}).
There are two main reasons for that.
One of the reasons is that the data 
in the interval $Q^2=1.4$--1.8 GeV$^2$,
in the transition between the 
two MAID datasets~\cite{Drechsel2007,MAID2009},
shows fluctuations 
that are not compatible 
with the simple third order polynomial form, 
assumed in the present parametrization.
The second reason is that the MAID data 
for $Q^2 > 1.5$ GeV$^2$~\cite{MAID2009},
have very small errorbars, which impose a strong 
constrain in the large $Q^2$ region
(dominance of the large $Q^2$ data in the fit).

We conclude however, that, a high quality fit 
can be obtained in the region $Q^2 < 1.5$ GeV$^2$~\cite{Drechsel2007}.
In this case, the extrapolation 
for higher $Q^2$ generates very large contributions 
for the form factors, that are also incompatible 
with the MAID data for the region $Q^2 > 1.5$ GeV$^2$~\cite{MAID2009}.
As in the case of $\gamma^\ast N \to \Delta$ 
quadrupole form factors, 
it is not possible to derive a 
parametrization appropriated 
simultaneously for small and large $Q^2$,
based on functions with exponential factors.
Alternative parametrizations 
will be discussed  in Sec.~\ref{secExtension}.

In Figs.~\ref{figN1520amps} and \ref{figN1520FF}, 
we can observe that the  MAID-SG  parametrization
describes very well the $A_{1/2}$ and the $G_M$ data.
As for the amplitudes $A_{3/2}$ and $S_{1/2}$, 
one can observe some fluctuations 
of the data in the region  $Q^2=1.4$--1.8 GeV$^2$.
As already discussed, 
those fluctuations are the 
reason why 
it is not possible to obtain a very accurate description of the MAID data.
The result obtained for the amplitude 
$S_{1/2}$ differs significantly from the 
 MAID2007 parametrization.
The poor description of the MAID data given by the 
MAID-SG parametrization is a 
consequence of trying to describe 
the data and be consistent with the Siegert's theorem, at the same time.
The fluctuations of the data in region  $Q^2=1.4$--1.8 GeV$^2$,
and the deviation from the fit 
can also be seen in the graphs for the 
form factors $G_E$ and $G_C$.
In any case, if we look for the helicity amplitudes, 
or for the form factors, the MAID-SG parametrization  
describes well the large $Q^2$ region.

In the Figs.~\ref{figN1520amps} and \ref{figN1520FF},  
we also include the results 
of the parametrization Jlab-SG.
We avoid to include the Jlab data for a question 
of legibility, but, we recall that the 
Jlab-SG parametrization follows closely the Jlab data.
From the comparison between the 
MAID-SG and Jlab-SG parametrizations, we can conclude, 
that, the two parametrizations,
although very different for $Q^2 > 1$ GeV$^2$, 
have a very similar behavior near the 
pseudo-threshold.

\subsubsection{Discussion}

From the analysis of the Jlab and MAID data,
we already noticed, 
that both parametrizations lead to very similar 
extensions for the $Q^2 < 0$ region. 
We may be tempted to conclude, that, 
the closeness between the two parametrizations 
for small $Q^2$ is essentially the consequence 
of the data considered at the photon point.
This conclusion is however incorrect.
The inclusion of the more recent CLAS data, 
with three datapoints in the region $Q^2=0.65$--1.30 GeV$^2$~\cite{Mokeev15}, 
has a significant impact in the Jlab-SG fit.
When we include the corresponding data 
in the fit, the values of the amplitudes $A_{1/2}$ 
and $A_{3/2}$ increase by 15\%, at the pseudo-threshold,
and the amplitude $S_{1/2}$ decreases  at $Q^2=0$ by 13\%.
In both cases, the amplitudes are modified 
below the region $Q^2 < 0.3$ GeV$^2$ 
(the threshold of the data for finite $Q^2$).

We then conclude that, the behavior near 
the pseudo-threshold, is not only 
the consequence of the low $Q^2$ data, 
but that the fit is also constrained by data up to 1 GeV$^2$. 
For this reason, it is remarkable 
that the Jlab-SG and the MAID-SG parametrizations 
are so close near the pseudo-threshold.

As discussed in Sec.~\ref{secS12}, some 
of the properties  near the pseudo-threshold 
of the amplitudes and form factors can be observed directly in the graphs.
In Fig.~\ref{figN1520amps}, 
one can see in the graph for the 
amplitude $S_{1/2}$, 
the convergence at the pseudo-threshold, $S_{1/2} \to 0$, 
with an infinite derivative  (consequence of $S_{1/2} \propto |{\bf q}|$).
Also in Fig.~\ref{figN1520FF},
we can note, in the graph for $G_M$,
that, the derivative of $G_M$
near the pseudo-threshold is finite,
as we could anticipate from $G_M = {\cal O} (|{\bf q}|^2)$.

To finalize the discussion of the $\gamma^\ast N \to N(1520)$  transition, 
it is interesting to discuss the results for the amplitude $A_{3/2}$
in the context of a quark model framework.
In Fig.~\ref{figN1520amps}, one can see 
that the amplitude $A_{3/2}$ is very large near $Q^2=0$,  
compared to the amplitude $A_{1/2}$.
It is however known that estimates of the meson cloud effects
predict in general large contributions for the amplitude $A_{3/2}$ near $Q^2=0$,
and that the valence quark contributions 
are only about 1/3 of the 
total~\cite{N1520,Diaz08,Santopinto12,Ronniger13,Golli13,Mokeev16}.
Since $A_{3/2}(0)$ and the extrapolation 
of the function $A_{3/2}$ for the pseudo-threshold limit, 
differs significantly, 
we may conclude that, as for the case 
of the $\gamma^\ast N \to \Delta(1232)$ transition,
discussed in Sec.~\ref{secDeltaR},
the amplitude $A_{3/2}$ is also dominated by 
processes beyond the impulse approximation.
The stronger candidate for those contributions 
are the meson cloud contributions, as 
discussed in Ref.~\cite{N1520}.

In contrast with the previous discussion 
are results of the constituent quark model
with two-body exchange currents from Ref.~\cite{Meyer01}.
In this model the valence quark contribution 
is dominant near $Q^2=0$
and the non valence quark degrees of freedom 
contribute to less than 20\%.

\section{Extension of the MAID-type parametrizations for large $Q^2$}
\label{secExtension}

We now discuss alternative 
parametrizations to the usual MAID parametrizations,
which are based on the combination 
of polynomial and exponential functions of $Q^2$ 
(MAID-type parametrizations).
 
As already discussed, 
an important disadvantage of the parametrizations 
based on exponential factors, $e^{-\beta Q^2}$,
where $\beta$ represents the coefficient 
associated with any form factor, 
it is that the exponential factor cuts the 
form factors above a certain value of $Q^2$.
Depending on the value of $\beta$, the 
parametrizations of the form factors can fall 
slower or faster, 
but decrease after a certain value of $Q^2$.
The polynomial factor can change the 
threshold of the falloff with $Q^2$, 
but at some point the exponential factor dominates.
A simple example of the exponential 
falloff effect is shown in Fig.~\ref{figDeltaR},
for the function $R_{SM}$ when $Q^2 > 1.5$ GeV$^2$.
Since in that case, the value of $\beta$ for $G_C$ 
is larger than the value for $G_E$ 
($\beta= 3.33$ GeV$^{-2}$ to be compared with $\beta=1.31$ GeV$^{-2}$),
it is expected that $R_{SM}$ falls to zero 
faster than $R_{EM}$, as we can see 
in the $Q^2>$ 1.5 GeV$^2$ region.
[We also need to take into account the falloff of $G_M$.
The magnetic form factor $G_M$, is however 
regulated by a very small value, $\beta \simeq 0.2$ GeV$^{-2}$,
according with the MAID2007 parametrization].

Another disadvantage of the use the MAID-type parametrizations 
is the extension of the  large $Q^2$ region,
where the MAID parametrizations cannot 
be compared with the expected 
leading order power laws of pQCD 
(apart from logarithmic corrections)~\cite{Carlson}.

We then propose new parametrizations 
for the amplitudes and  form factors, 
that differs from the MAID form, and corrects 
the high $Q^2$ behavior of the MAID parametrizations.
As example, we will use 
the $\gamma^\ast N \to \Delta (1232)$ transition, 
since it is the transition with more 
accurate data at large $Q^2$, and can therefore 
be tested with more precision.
The methods proposed can, however, 
be generalized for other transitions.

We divide our analysis into the 
magnetic form factor $G_M$ (or amplitude $M_{1+}$) 
and into the quadrupole form factors $G_E$ and $G_C$.

\subsection{Magnetic form factor ($G_M$)}

In the $\gamma^\ast N \to \Delta (1232)$ transition,
we can represent the magnetic form factor $G_M$ in terms 
of the magnetic amplitude $M_{1+}$, 
using $G_M= F_+ \, M_{1+}$, 
according with Eq.~(\ref{eqMtil}),
where $M_{1+} =- (A_{1/2}  +   \sqrt{3} A_{3/2}) $.

In the original MAID2007 parametrization, 
$M_{1+}$ is represented by
\ba
M_{1+} = \frac{|{\bf q}|}{K} G_D \tilde M_0,
\label{eqM1p-MAID}
\ea
where $\tilde M_0 = a_0 (1 + a_1 Q^2) e^{- a_4 Q^2}$.
We now propose the form
\ba
M_{1+}^\prime & = & \frac{1}{\sqrt{1 + \tau}}  M_{1+} 
\nonumber \\
&=& \frac{M_R + M}{2 M_R} \frac{\sqrt{Q_-^2}}{K} G_D \tilde M_0.
\label{eqM1p-new}
\ea 
The parametrization (\ref{eqM1p-new}) 
has two main advantages compared
to the original form.
The first advantage is the inclusion 
of the factor $\sqrt{Q_-^2}$, 
which it will be canceled by the factor 
$F_+ \propto 1/\sqrt{Q_-^2}$, 
in the conversion from amplitudes to form factors.
Since the factor $\sqrt{Q_-^2}$ is eliminated 
in the conversion to the form factors, 
we obtain simpler expressions for the form factors.
The second advantage is the parametrization 
of the very large $Q^2$ region in a form more compatible  
with the power laws of pQCD~\cite{Carlson}, 
given by $M_{1+}^\prime \propto 1/Q^3$, as we explain next.

From Eq.~(\ref{eqM1p-new}), we can immediately
conclude that,  $\sqrt{Q_-^2} G_D \propto 1/Q^3$,
for large $Q^2$.
Thus, in order to obtain the expected $1/Q^3$ falloff 
for $M_{1+}^\prime$, it is necessary that $\tilde M_0$ behaves as a constant, 
meaning that $\tilde M_0 = {\cal O}(1)$, for very large $Q^2$.
Since the function  $\tilde M_0$ 
includes an exponential factor, 
it is obvious that we cannot have 
 $\tilde M_0 = {\cal O}(1)$.
We can correct this limitation 
replacing $\tilde M_0$ by an analytic expression 
that is close to the MAID form in the range 
of validity of the parametrization, $Q^2=0$--10 GeV$^2$,
but behaves like a constant for much higher $Q^2$.
This goal can be achieved with the replacement
\ba
\tilde M_0 &=& a_0 \frac{1 + a_1 Q^2}{e^{a_4 Q^2}} 
\nonumber \\
& =&  a_0 \frac{1 + a_1 Q^2}{ 1 + a_4 Q^2 + 
\frac{1}{2!} a_4^2 Q^4 + \frac{1}{3!} a_4^3 Q^6 + ...} 
\nonumber \\
 & \to &
a_0
\frac{ \sum_{k=0}^n \frac{(a_1 Q^2)^k}{k!} }{
\sum_{k=0}^n \frac{(a_4 Q^2)^k}{k!}},
\label{eqM0LQ}
\ea 
where $n \ge 1$ is an integer.
In the last line, we {\it expanded}
the  factor $(1 + a_1 Q^2)$, using the exponential series,
and truncated the exponential factor in the denominator, 
in the same order.
Using the replacement (\ref{eqM0LQ}), we obtain
a simple rational function that converges
to a constant for very large $Q^2$.
The dependence $M_{1+}^\prime \propto 1/Q^3$ 
for very large $Q^2$ is then naturally generated.

\begin{figure}[t]
\vspace{.6cm}
\centerline{\mbox{
\includegraphics[width=2.8in]{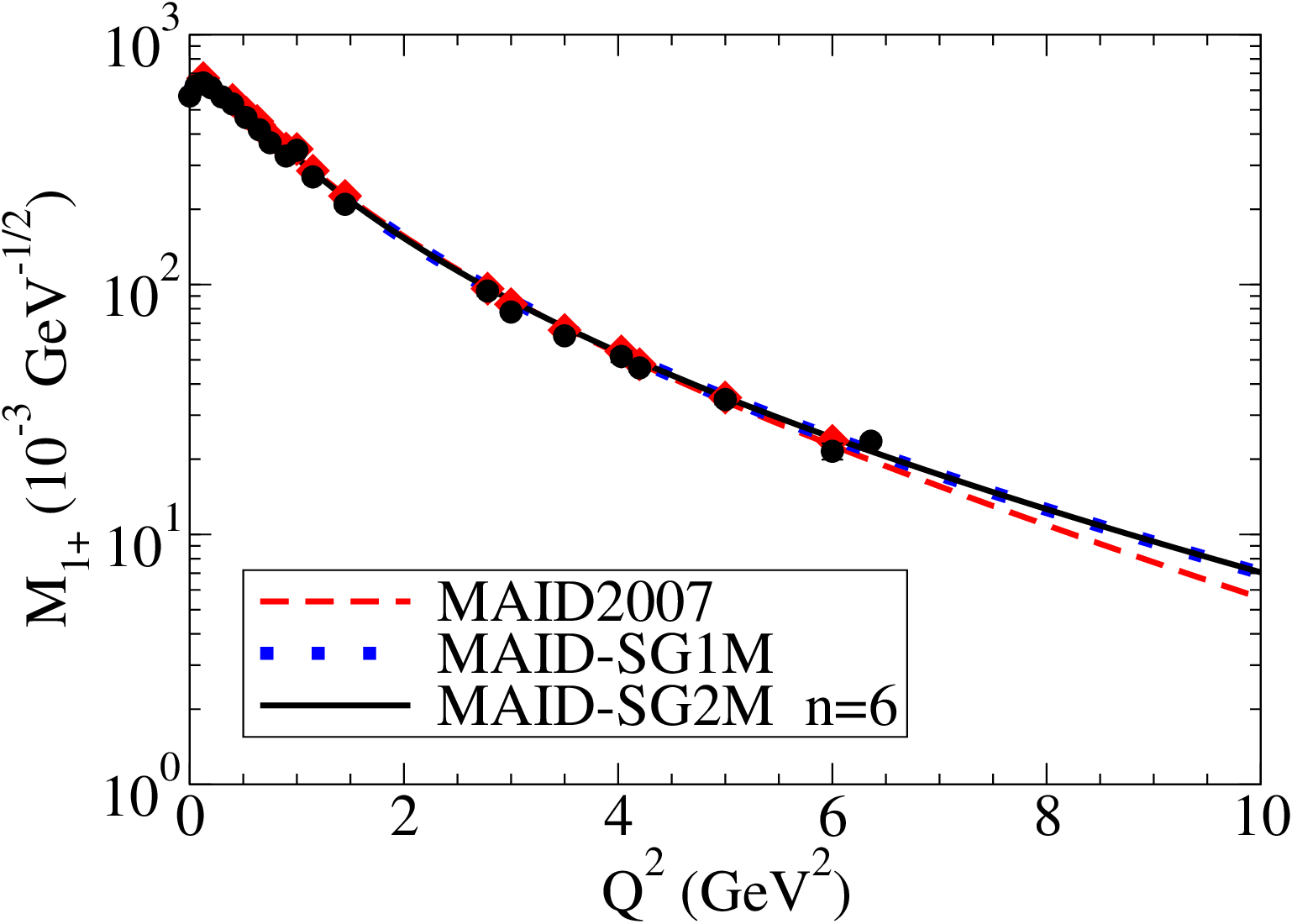}}}
\vspace{.7cm}
\centerline{\mbox{
\includegraphics[width=2.8in]{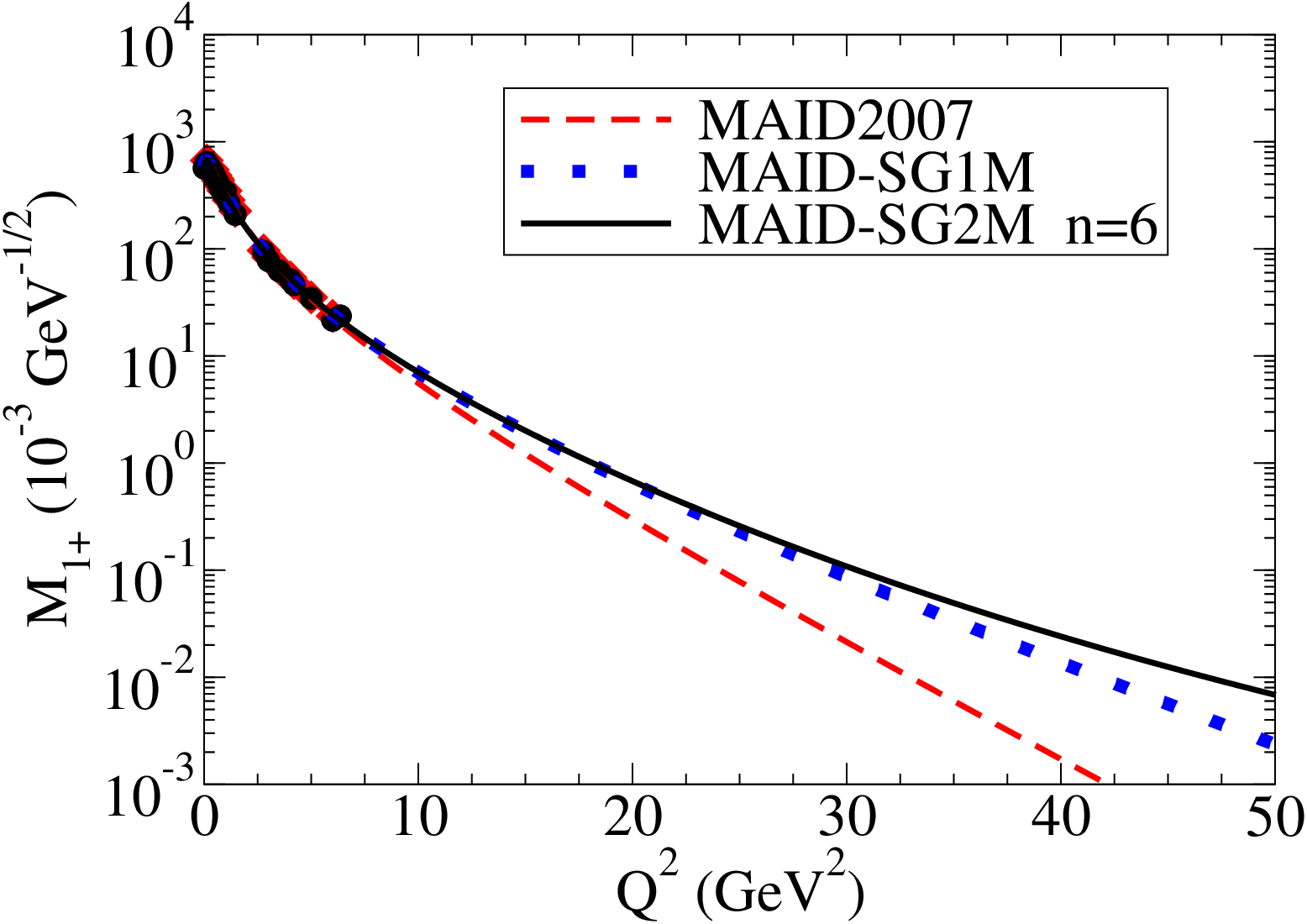}}}
\caption{\footnotesize
$\gamma^\ast N \to \Delta(1232)$ transition.
Magnetic amplitude $M_{1+}$ compared 
with the MAID2007 parametrization 
and two alternative parametrizations
(MAID-SG1M and MAID-SG2M). 
See discussion in the text.
Data from MAID analysis (diamonds)~\cite{Drechsel2007,MAID2009} and 
Ref.~\cite{MokeevDatabase} (circles). }
\label{figM1p}
\end{figure}

Note that, the truncation of the exponential series 
in the denominator is particularly convenient, 
as far as it does not induce any singularity in the region  $Q^2< 0$.
This can be assured if we restrict 
the expansion to even powers of $n$,
as discussed  in Appendix in detail.

The results of the MAID2007 parametrization 
for $M_{1+}$ are presented in Fig.~\ref{figM1p}.
In addition to the MAID2007 parametrization 
defined by Eq.~(\ref{eqM1p-MAID}), where 
$a_4=0.23$ GeV$^{-2}$, we consider a 
modification of the MAID2007 parametrization 
given by Eq.~(\ref{eqM1p-new}),
with the replacement $a_4 \to 0.15$ GeV$^{-2}$.
We label this new parametrization 
as MAID-SG1M (M stands for magnetic amplitude).
In the upper panel of Fig.~\ref{figM1p},
one can see, that, the new parametrization 
also gives a very good description of the data.
We can then conclude that, 
Eqs.~(\ref{eqM1p-MAID}) and (\ref{eqM1p-new})
provide equivalent parametrizations of the data, 
in the region $Q^2=0$--10 GeV$^2$.

Finally, in order to check if a consistent 
description of the data is possible replacing the 
factor $\tilde M_0$ by a rational expression, 
as suggested by Eq.~(\ref{eqM0LQ}),
we consider the modification of 
the  MAID-SG1M parametrization
by the expansion with $n=6$.
We label the new parametrization as MAID-SG2M.
In the upper and lower panels of Fig.~\ref{figM1p},
one can see the comparison between 
MAID-SG1M and MAID-SG2M.
In the upper panel, in the range $Q^2 < 10$ GeV$^2$, 
the two models are almost not distinguishable.
In the lower panel, when the scale of $Q^2$ 
is extended up to 50 GeV$^2$, one can see 
that MAID-SG1M falls off faster than MAID-SG2M,
as expected from the comparison 
between a rational and an exponential function. 
In the same graph, one can observe the linear behavior of the 
MAID2007 and MAID-SG1M parametrizations for $Q^2 > 20$ GeV$^2$, 
due to the logarithmic scale used for $M_{1+}$.
The graph also shows that, the difference between 
a parametrization based on powers of $Q^2$
(rational functions) and exponential functions, 
can in some cases, be disentangled only for very large $Q^2$.

The parametrizations based on rational functions 
are efficient for moderate values of the argument $a_4Q^2$,
because the denominator in Eq.~(\ref{eqM0LQ})
converges smoothly, 
and faster than the expansion based on 
exponential series with negative arguments ($e^{-a_4 Q^2}$). 

In the overall, we conclude, that 
the magnetic amplitude $M_{1+}$
can be conveniently parametrized by the form (\ref{eqM1p-new}),
which behaves for large $Q^2$ 
as $M_{1+} \propto \tilde M_0/Q^3$.
Combining this result with 
the rational parametrization given by  Eq.~(\ref{eqM0LQ}),
one obtains at the end 
a parametrization consistent with 
$M_{1+} \propto 1/Q^3$ and $G_M \propto 1/Q^4$.

The procedure used for the parametrization $M_{1+}$ ($M_{1+}^\prime$)
can be used for the amplitudes 
$E_{1+}$ (electric) and $S_{1+}$ (scalar).
The behavior of the amplitudes 
$A_{1/2}$, $A_{3/2}$ and $S_{1/2}$ for large $Q^2$
can also be estimated and/or extracted from the data. 
A note of caution about the results 
for very large $Q^2$ is  in order.
Since it is expected from pQCD arguments 
that $M_{1+} \simeq - E_{1+}$ for very large $Q^2$~\cite{Jones73,Carlson},
a consistent parametrization of the data 
requires the correlation between 
the coefficients of both parametrizations.
Since there are no experimental evidences of 
the scaling $M_{1+} \simeq - E_{1+}$, 
we cannot, at the moment, 
do more than estimate the expected shape of $E_{1+}$
for very large $Q^2$ based on the 
knowledge of $M_{1+}$ in the range $Q^2 \approx 10 $ GeV$^2$.
Nevertheless, we can test if the quality of 
the fit for the form factors $G_E$ and $G_C$,
at low $Q^2$, namely when $Q^2 < 1$ GeV$^2$, 
can be extended  for larger values of $Q^2$.

\subsection{Electric and Coulomb quadrupole form factors ($G_E$ and $G_C$)}
\label{secLargeQ2}

\begin{table}[t]
\begin{center}
\begin{tabular}{l  r r r r r}
\hline
\hline 
MAID-SG0   & $r_0$ & $r_1$ & $r_2$ &  $r_3$ &$r_4$  \\
\hline
$G_E$  & $21.64$ & $-2.91$ & $-5.72$  & -- &  \sp\sp$1.79$ \\
$G_C$  & \boldmath{$23.75$} & $-0.911$ & $1.36$  & -- & $1.23$ \\
\hline
MAID-SG1   & $r_0$ & $r_1$ & $r_2$ &  $r_3$ &$r_4$  \\
\hline
$G_E$  & $13.59$ & $15.43$ & $146.9$  & $400.1$ &  \sp\sp$15.78$ \\
$G_C$  & \boldmath{$29.16$} & $-1.20$ & $3.30$  & $0.690$ & $2.47$ \\
\hline
MAID-SG2   & $r_0$ & $r_1$ & $r_2$ &  $r_3$ &$r_4$  \\
\hline
$G_E$  & $13.84$ & $2.220$ & $7.44$  & $14.79$ &  \sp\sp$6.13$ \\
$G_C$  & \boldmath{$22.33$} & $4.66$ & $-15.64$  & $33.65$ & $6.22$ \\
\hline
\hline
\end{tabular}
\end{center}
\caption{
$\gamma^\ast N \to \Delta(1232)$ transition.
Coefficients used in the calculation of 
$G_E$ and $G_C$ 
in the MAID-SG0 parametrization,
defined by Eqs.~(\ref{eqGEsg})-(\ref{eqGCsg});
in the MAID-SG1 parametrization,
defined by Eqs.~(\ref{eqMAID-SG1a})-(\ref{eqMAID-SG1b});
and  MAID-SG2 parametrization, 
defined by Eqs.~(\ref{eqMAID-SG2a})-(\ref{eqMAID-SG2b}).
The labels $r_l$ ($l=0,1,2,3,4$) holds for $r_l=b_l,c_l$.
$r_0$ is in units $10^{-3}$ GeV$^{-1/2}$
($C_0/K= 5.32$ GeV$^{1/2}$).
$r_1$ and $r_4$ are in units GeV$^{-2}$,
$r_2$ in units GeV$^{-4}$,
and $r_3$ in units  GeV$^{-6}$.
The values in bold are determined by Eqs.~(\ref{eqJones}).}
\label{tabModel-Delta2}
\end{table}

Inspired by  Eq.~(\ref{eqM0LQ}), 
we check if similar extensions can be used 
for the form factors $G_E$ and $G_C$,
defined by Eqs.~(\ref{eqGEsg})-(\ref{eqGCsg}).
For simplicity we restrict 
the discussion to the polynomial 
$(1 + r_1 Q^2 + r_2 Q^4 + r_3 Q^6)$
and the factors $G_D$, $e^{-r_4 Q^2}$,
since the reaming factors are constant.
As before $r_l=b_l,c_l$ ($l=1,2,3,4$).

The simplest extension, inspired by Eq.~(\ref{eqM0LQ}),
is given by the replacement
\ba
& &
(1 + b_1 Q^2 + b_2 Q^4 + b_3 Q^6) \, e^{-b_4 Q^2} G_D 
\nonumber \\
& &
\to
\frac{1 + b_1 Q^2 + b_2 Q^4 + b_3 Q^6}{
1 + b_4 Q^2 + \frac{1}{2!}b_4^2 Q^4 + \frac{1}{3!}b_4^3 Q^6}G_D,
\label{eqMAID-SG1a}
\ea
for $G_E$, and 
\ba
& &
(1 + c_1 Q^2 + c_2 Q^4 + c_3 Q^6) \, e^{-c_4 Q^2} G_D 
\nonumber \\
& &
\to
\frac{1 + c_1 Q^2 + c_2 Q^4 + c_3 Q^6}{
1 + c_4 Q^2 + \frac{1}{2!}c_4^2 Q^4 + \frac{1}{3!}c_4^3 Q^6 +
\frac{1}{4!} c_4^4 Q^8  
}G_D, 
\nonumber \\
\label{eqMAID-SG1b}
\ea
for $G_C$.
Using Eqs.~(\ref{eqMAID-SG1a})-(\ref{eqMAID-SG1b}) 
we obtain the expected pQCD 
falloff: $G_E \propto 1/Q^4$ and $G_C \propto 1/Q^6$
for very large $Q^2$~\cite{Carlson}.
We label this new parametrization 
for the quadrupole form factors as the 
MAID-SG1 parametrization.

Note that in the MAID-SG1 parametrization, 
we still include the dipole form factor 
$G_D$, where the cutoff $\Lambda^2=0.71$ GeV$^2$, 
is extracted from the studies of 
the nucleon form factors.
Since there is no particular reason to 
expect that the scale of the 
form factors $G_E$ and $G_C$ is  
related to the scale of the nucleon form factors, 
we eliminate the factor $G_D$ 
from the parametrizations for the quadrupole form factors, 
and transpose the $1/Q^4$  falloff 
to the remaining rational factors.
A simple extension of the MAID-SG1 
parametrization that excludes $G_D$ is 
\ba
& &
(1 + b_1 Q^2 + b_2 Q^4 + b_3 Q^6) \, e^{-b_4 Q^2} G_D
\nonumber \\
& &
\to
\frac{1 + b_1 Q^2 + b_2 Q^4 + b_3 Q^6}{
1 + b_4 Q^2 + \frac{1}{2!}b_4^2 Q^4 + \frac{1}{3!}b_4^3 Q^6 
+ \frac{1}{4!}b_4^4 Q^8  + \frac{1}{5!}b_4^5 Q^{10}}, \nonumber \\
\label{eqMAID-SG2a}
\ea
for $G_E$, and 
\ba
& &
\hspace{-.4cm}
(1 + c_1 Q^2 + c_2 Q^4 + c_3 Q^6) \, e^{-c_4 Q^2} G_D
\nonumber \\
& &
\hspace{-.4cm}
\to
\frac{1 + c_1 Q^2 + c_2 Q^4 + c_3 Q^6}{
1 + c_4 Q^2 + \frac{c_4^2}{2!}Q^4 + \frac{c_4^3}{3!} Q^6 +
\frac{c_4^4}{4!} Q^8    + \frac{c_4^5}{5!} Q^{10} +
\frac{c_4^6}{6!} Q^{12}}, 
\nonumber \\
\label{eqMAID-SG2b}
\ea
for $G_C$.
Equations~(\ref{eqMAID-SG2a})-(\ref{eqMAID-SG2b}) define the 
MAID-SG2 parametrization.
As in the MAID-SG1 case, one obtains $G_E \propto 1/Q^4$ 
and $G_C \propto 1/Q^6$, for very large $Q^2$.

\begin{figure}[t]
\vspace{.6cm}
\centerline{\mbox{
\includegraphics[width=2.6in]{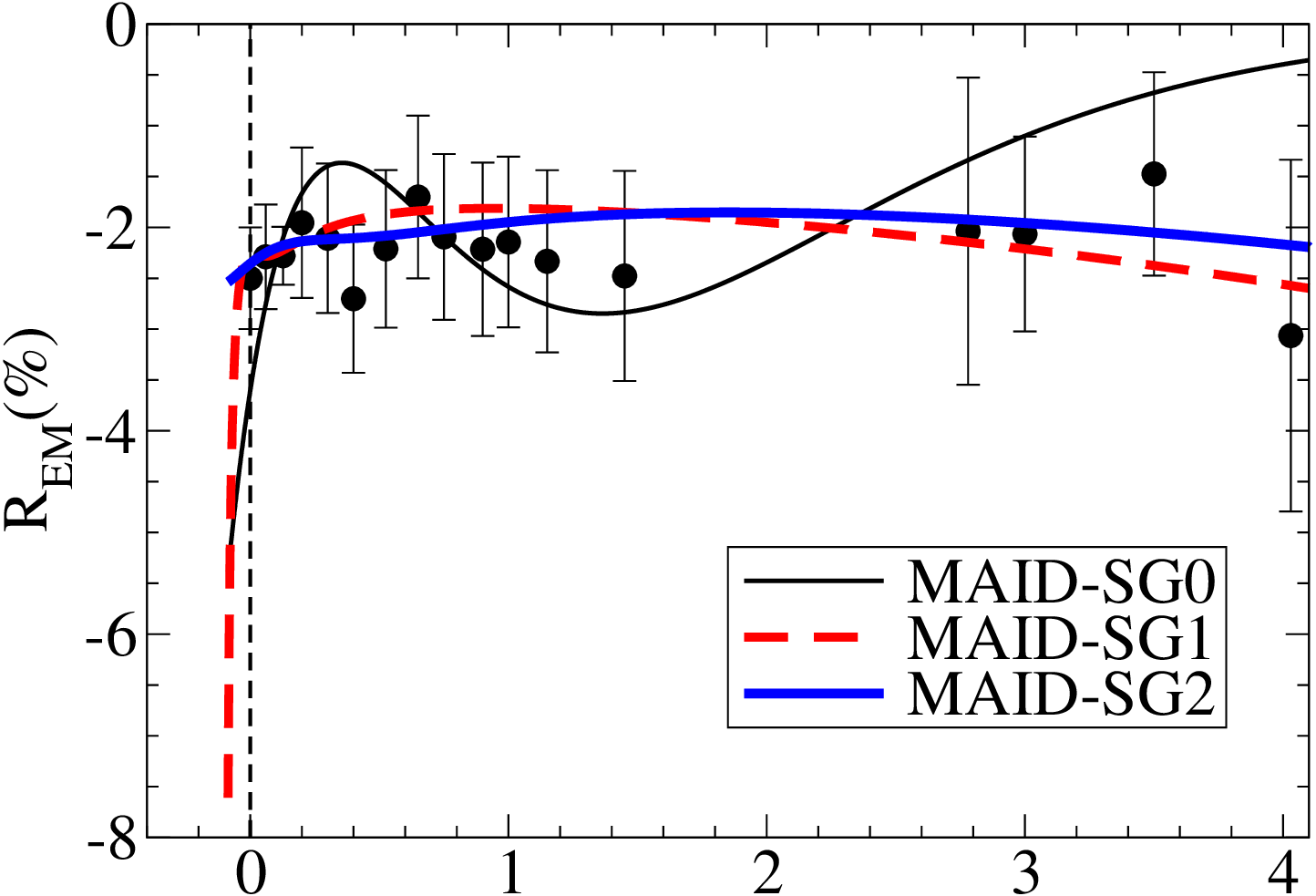}}}
\vspace{.8cm}
\centerline{\mbox{
\includegraphics[width=2.6in]{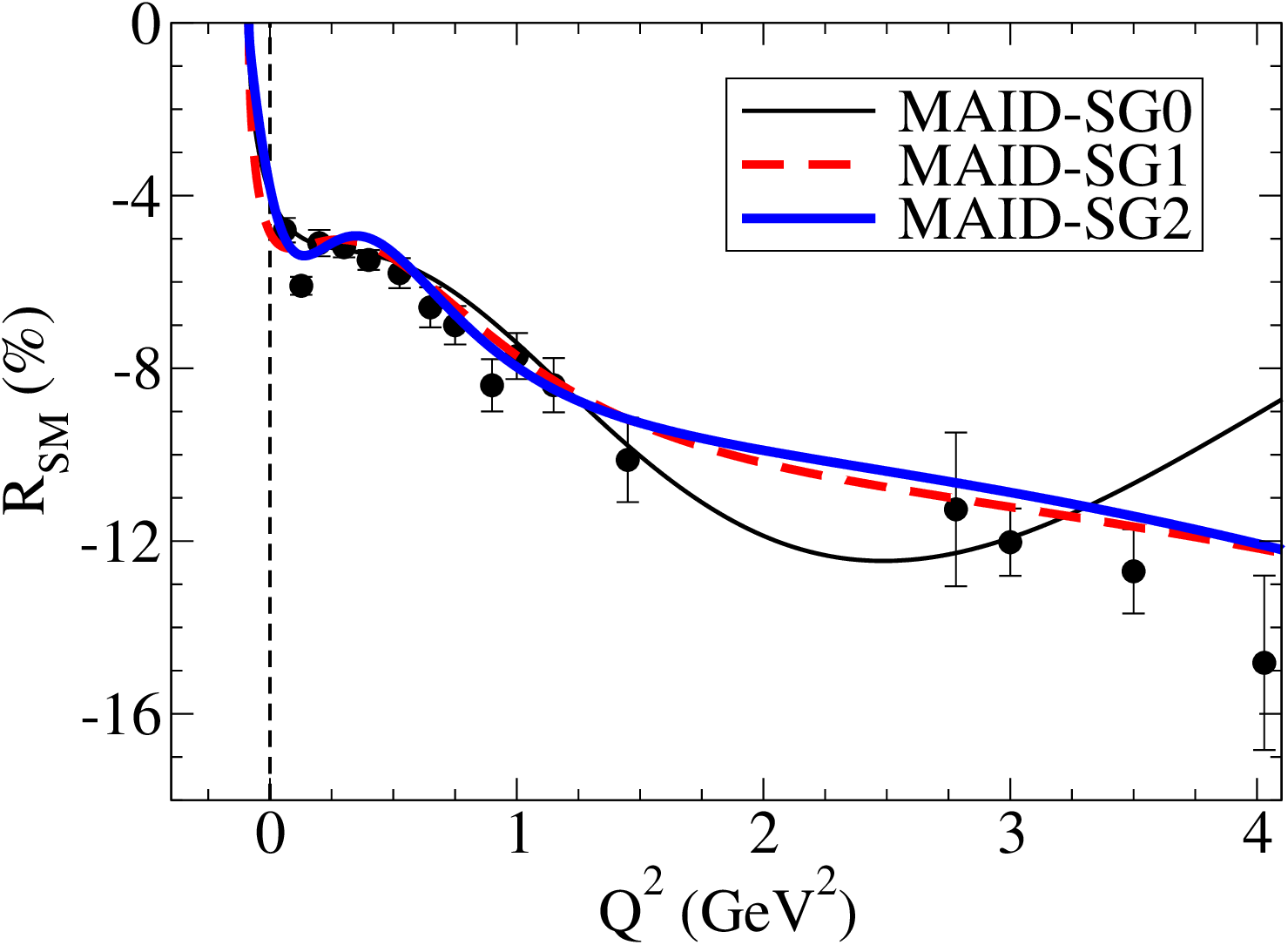}}}
\caption{\footnotesize
$\gamma^\ast N \to \Delta(1232)$ transition.
Ratios $R_{EM}$ and $R_{SM}$ given by the MAID-SG0, 
MAID-SG1 and MAID-SG2 parametrizations.
Data from Ref.~\cite{MokeevDatabase} (circles).}
\label{figDeltaR2}
\end{figure}

\begin{figure}[t]
\vspace{.6cm}
\centerline{\mbox{
\includegraphics[width=2.5in]{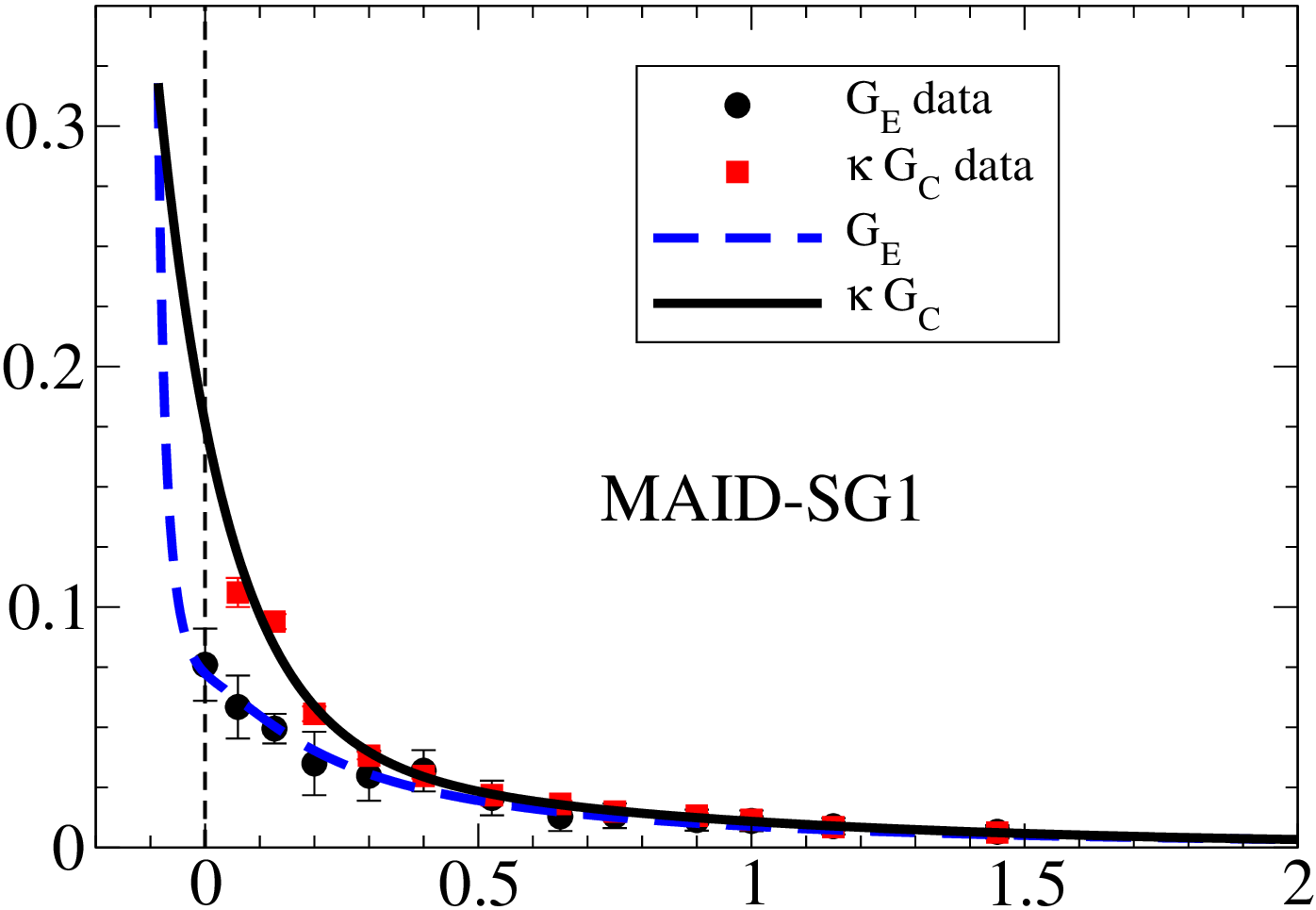}}}
\vspace{.9cm}
\centerline{\mbox{
\includegraphics[width=2.5in]{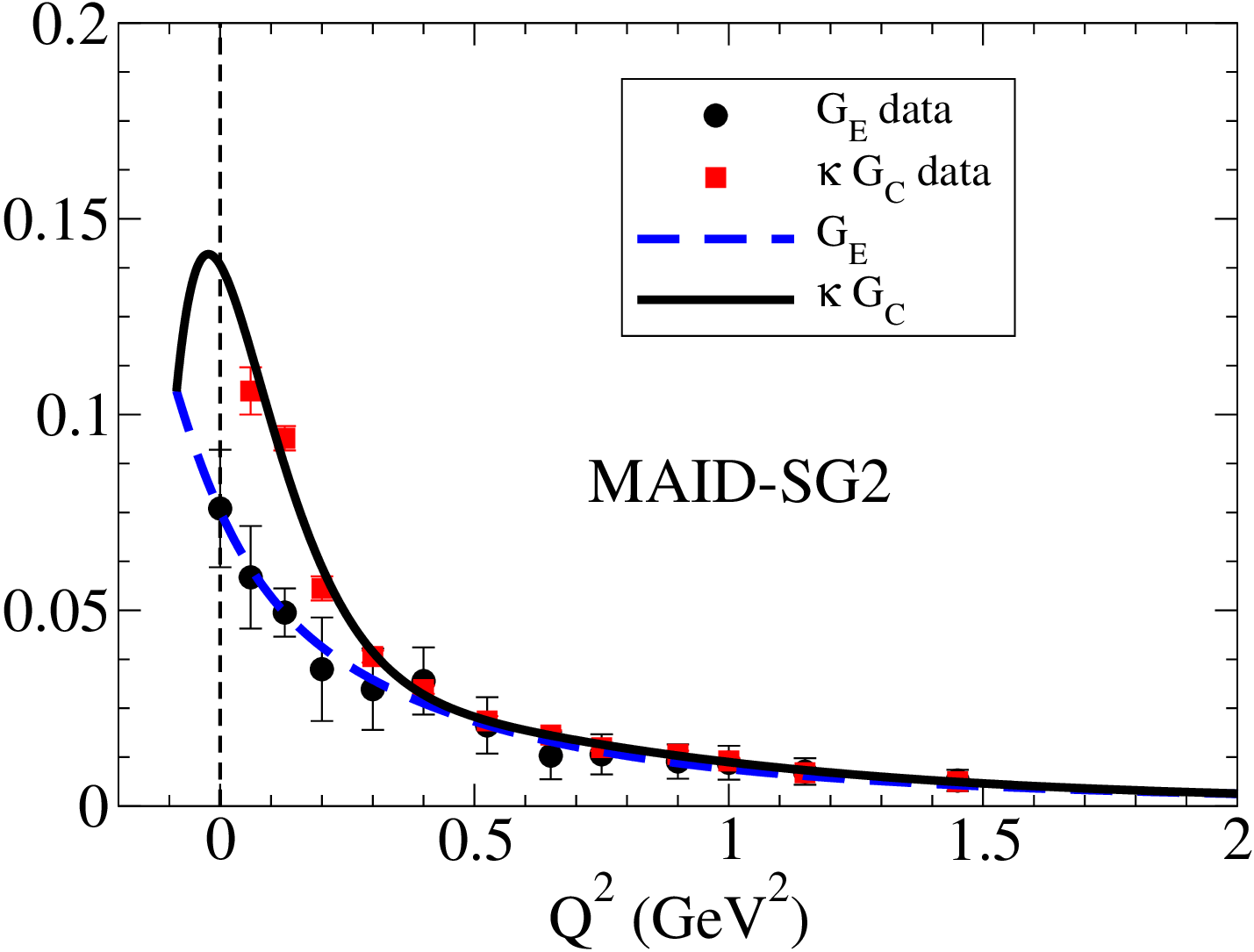}}}
\caption{\footnotesize
Electric and Coulomb quadrupole form factors 
for the $\gamma^\ast N \to \Delta(1232)$ transition.
$G_C$ is multiplied by $\kappa= \frac{M_R -M}{2 M_R}$.
At the top: MAID-SG1 parametrization.
At the bottom: MAID-SG2 parametrization.
Data from Ref.~\cite{MokeevDatabase}.}
\label{figDelta2}
\end{figure}

In the parametrizations~(\ref{eqMAID-SG1a})-(\ref{eqMAID-SG1b})
and~(\ref{eqMAID-SG2a})-(\ref{eqMAID-SG2b}),  
it is important to ensure that no singularities 
are introduced in the denominator 
in the extension for the region $Q^2 < 0$.
The analysis of the possible singularities is done 
in Appendix~\ref{appA}.
For the new parametrizations for $G_C$, 
it is possible to show that the functions 
are free from singularities, since the truncated 
expansion in powers $Q^{2n}$ generates positive functions,
(higher power $Q^{2n}$ with even $n$),
as shown in Appendix~\ref{appA1}.
As for the parametrizations for $G_E$ 
given by Eqs.~(\ref{eqMAID-SG1a}) and (\ref{eqMAID-SG2a}),
one needs to be more careful, since there is
the possibility of singularities for 
larger values of $b_4$, as discussed in Appendix~\ref{appA2}.
We can  however show that, the values of $b_4$ 
obtained in our fits, are consistent with 
parametrizations free of singularities 
(see the details of the discussion in Appendix~\ref{appA2}).
In alternative to Eqs.~(\ref{eqMAID-SG1a}) and  (\ref{eqMAID-SG2a}),
we can also define parametrizations 
of the functions $G_E$ automatically singularity-free, 
if we remove or add a parameter to the 
expressions (\ref{eqMAID-SG1a}) and (\ref{eqMAID-SG2a}).
This alternative is discussed in Appendix~\ref{appA3}.

As in Sec.~\ref{secDeltaR}, we tested if the 
MAID-SG1 and MAID-SG2 parametrizations, 
can give a good description of the data. 
Compared to Sec.~\ref{secDeltaR}, we 
increased the range of the data included 
in the fit from 2 GeV$^2$ to 4.1 GeV$^2$.
As  already discussed, there are discrepancies 
between the different analysis for large $Q^2$.
For this reason we avoid the use 
of $Q^2 > 4.1$ GeV$^2$ data.

In Sec.~\ref{secDeltaR}, we conclude, 
that, one should not expect a good global 
description of the data simultaneously 
for small and large $Q^2$,  
based on a MAID-type parametrization. 
To test the quality of the new parametrizations,
MAID-SG1 and MAID-SG2,
we extend first the previous MAID-SG 
parametrization to the range $Q^2=0$--4.1 GeV$^2$.
We label this new parametrization as MAID-SG0.
The coefficients obtained in the refit are presented in 
the Table~\ref{tabModel-Delta2}, under the label MAID-SG0.
It is interesting to note in the table 
that the best fit is obtained when $b_3, c_3 \simeq 0$.

The results for the ratios $R_{EM}$ and $R_{SM}$ 
for the three parametrizations are presented in Fig.~\ref{figDeltaR2}.
In the figure one can see that the description 
of the MAID-SG parametrization, that is good only up to 1.5 GeV$^2$,
as shown in Fig.~\ref{figDeltaR}, 
it is improved up to about 3 GeV$^2$, 
with the MAID-SG0 parametrization.
At the same time, one can confirm, from the graph for $R_{EM}$,
that, a MAID-type parametrization (MAID-SG0),
overestimates the value for $G_E(0)$
[underestimation of $R_{EM}(0)$]. 

We now look for the results obtained with 
the  MAID-SG1 and MAID-SG2  parametrizations.
The values of the coefficients determined by 
the best fit to the data, are also presented in 
the Table~\ref{tabModel-Delta2}.
In Fig.~\ref{figDeltaR2}, one can see, 
that, the MAID-SG1 and MAID-SG2 parametrizations 
give a very good description of the data,
including the region $Q^2 > 2$ GeV$^2$.
Qualitatively, the  MAID-SG2 parametrization
gives a better description of the $G_E$, $G_C$ data
(smaller chi-squared for $G_E$ and $G_C$ subsets,  
and smaller total chi-squared).
The largest difference between 
the parametrizations  MAID-SG1 and MAID-SG2
occurs for $R_{EM}$ near the pseudo-threshold.
The values of $R_{EM}(0)$, for both parametrizations, 
are very similar, but the results at the pseudo-threshold 
are very different.
One has $R_{EM}(Q_{PS}^2) \simeq -5.34\%$ 
for MAID-SG0,  $R_{EM}(Q_{PS}^2) \simeq -7.60\%$
for MAID-SG1 and  $R_{EM}(Q_{PS}^2) \simeq -2.54\%$ for MAID-SG2.

The difference between the 
MAID-SG1 and MAID-SG2 parametrizations can be better observed 
in Fig.~\ref{figDelta2}, where we present 
the results for $G_E$ and $G_C$, where $G_C$ is 
corrected by $\kappa= \frac{M_R-M}{2 M_R}$.
In this representation 
the difference between the two parametrizations becomes clear. 
the difference between the two parametrizations.
[The graph for  MAID-SG0 is similar to the graph for  MAID-SG1].
One can now see that, 
the MAID-SG2 parametrization gives a smother description of the data, 
with smaller values for $G_E$ and $G_C$, 
near the pseudo-threshold.
In contrast the  MAID-SG1 parametrization
shows a stronger variation of the form factors 
near the pseudo-threshold,
that leads to a larger magnitude of $R_{EM}(Q_{PS}^2)$,
and a larger difference to $R_{EM}(0)$.

Note, that, the  MAID-SG2 parametrization
has a behavior close to the 
 MAID-SG  parametrization
(data up to 2 GeV$^2$), as shown in Fig.~\ref{figDelta},
but contrary to MAID-SG, it provides an accurate description 
for larger $Q^2$.

We can try to interpret the differences between the 
two global parametrizations: MAID-SG1 and MAID-SG2, 
recalling that MAID-SG2 gives the best fit.
The large values obtained in the 
MAID-SG1 parametrization
for $G_E$ and $G_C$, near the pseudo-threshold,
are mainly a consequence of the factor $G_D$,
included in the parametrizations 
(\ref{eqMAID-SG1a})-(\ref{eqMAID-SG1b}).
The function $G_D$ is enhanced
below $Q^2 = 0$ due to the pole $Q^2 = - 0.71$ GeV$^2$,
inducing the large values for $G_E$ 
and $G_C$ near $Q^2 \simeq -0.09$ GeV$^2$.
We however note that, the factor $G_D$
is not only responsible 
for large values of $G_E$ and $G_C$, 
since  the MAID-SG parametrization, 
restricted to the range $Q^2=0$--2 GeV$^2$,
leads also to smaller values for  $G_E$ and $G_C$, 
near the pseudo-threshold.
When we increase the range of the fit,  
the factor $G_D$ becomes more relevant.
In any case, as discussed previously,
since there is no reason to relate 
the function $G_D$ to the $\gamma^\ast N \to \Delta(1232)$
quadrupole form factor data, 
a parametrization that avoids a reference to $G_D$
and transfers the $Q^2$ dependence 
for the coefficients of the polynomial factors is preferable.
For all the above reasons, MAID-SG2 is preferable over MAID-SG1.

\subsection{Comparison with the literature}
\label{secDiscussion2}

We now compare  the parametrizations 
for $G_E$ and $G_C$ with alternative 
descriptions presented in the literature.

The ratio $R_{SM}$ is calculated 
using pQCD in Ref.~\cite{Idilbi04}.
We do not compare our results 
directly with pQCD, since the result 
depends on a normalization at large $Q^2$,
and in the present work our main focus 
is the low and intermediate $Q^2$ region.

Alternative descriptions of $G_E$ and $G_C$ 
comes from the large $N_c$ limit 
and from constituent quark models.
Using the  large $N_c$ limit 
it is possible to relate the quadrupole form factors $G_E$ and $G_C$
with the neutron electric form factor 
$G_{En}$~\cite{Pascalutsa07,Buchmann02}, by
\ba
& &
G_C (Q^2)= \sqrt{\frac{2M}{M_R}} M_R M 
\frac{G_{En}(Q^2)}{Q^2}, 
\label{eqGCbuch}\\
& &
G_E (Q^2)= \left( \frac{M}{M_R}\right)^{3/2}
\frac{M_R^2 - M^2}{2 \sqrt{2}} \frac{G_{En}(Q^2)}{Q^2}.
\label{eqGEbuch}
\ea 
In the large $N_c$ limit the 
form factors $G_E$ and $G_C$ 
appear as higher order corrections in $1/N_c$ 
to the leading order form factor $G_M$~\cite{Pascalutsa07,Jenkins02a}.
The relations (\ref{eqGCbuch})-(\ref{eqGEbuch}) are 
sometimes modified by the factor  $\left(\frac{M_R}{M}\right)^{3/2}$,
which corresponds to  $1/N_c^2$ 
correction~\cite{Buchmann04,Pascalutsa07b}.

The derivation of Eqs.~(\ref{eqGCbuch})-(\ref{eqGEbuch})
is guided by the observation that 
within a $SU(6)$ spin-flavor symmetry model, 
the neutron would have a symmetric spatial 
distribution of charge, 
leading to $G_{En}(Q^2)=0$.
In addition, the electric and Coulomb quadrupole 
moments would both vanish [$G_E(0) = G_C(0)=0$].
Non-zero results for $G_{En},G_E(0)$ and $G_C(0)$
are then a consequence of the $SU(6)$ 
symmetry breaking~\cite{Buchmann09a,Buchmann97a,Buchmann04,Grabmayr01}.

\begin{figure}[t]
\vspace{.6cm}
\centerline{\mbox{
\includegraphics[width=2.6in]{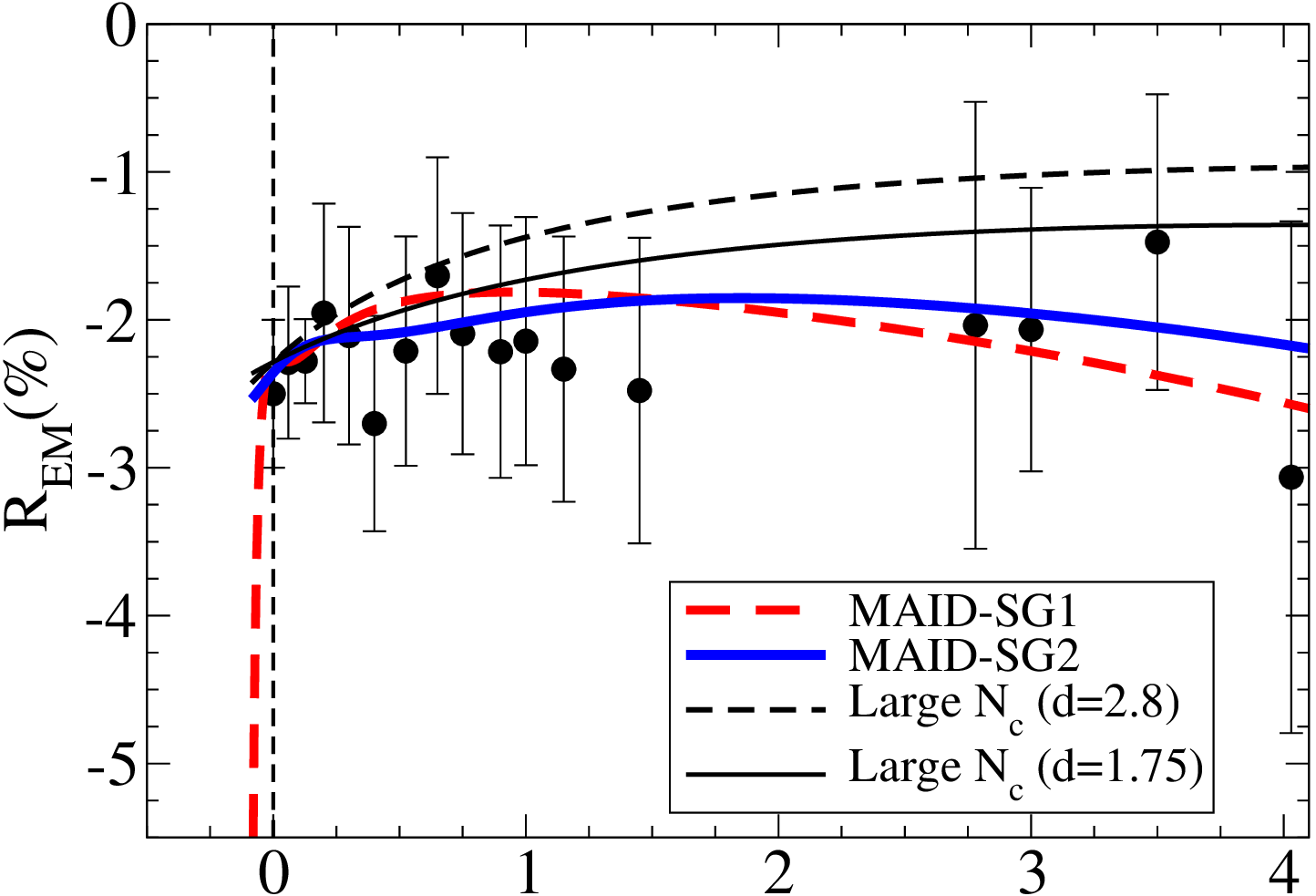}}}
\vspace{.8cm}
\centerline{\mbox{
\includegraphics[width=2.6in]{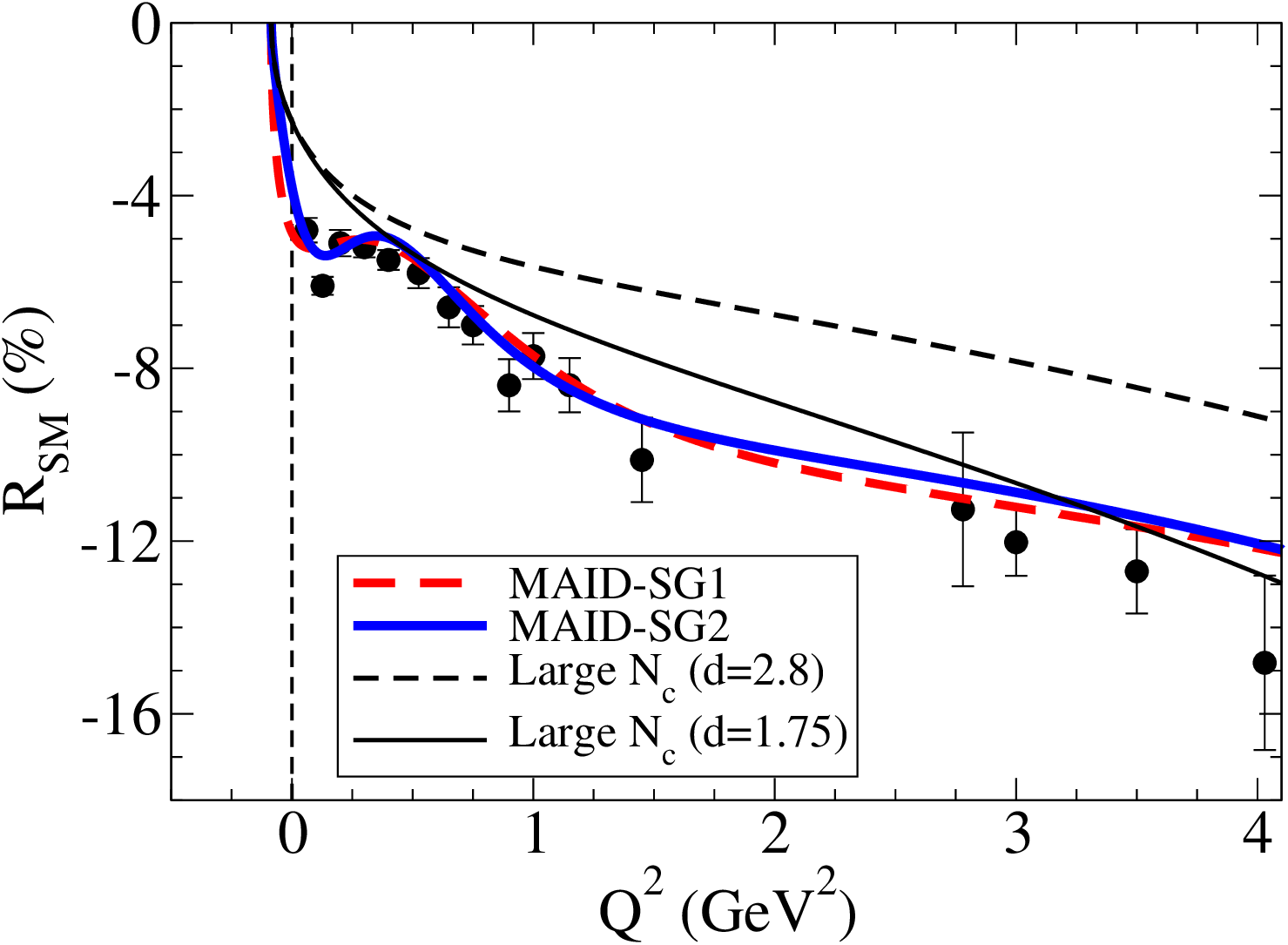}}}
\caption{\footnotesize
$\gamma^\ast N \to \Delta(1232)$ transition.
Ratios $R_{EM}$ and $R_{SM}$ given by
two large $N_c$ parametrizations 
characterized by the values of $d$.
The MAID-SG1 and MAID-SG2 parametrizations
are also presented for comparison.
Data from Ref.~\cite{MokeevDatabase} (circles).}
\label{figDeltaR3}
\end{figure}

The relation (\ref{eqGCbuch}) was derived 
for the first time in the context 
of a constituent quark model with two-body 
exchange currents~\cite{Grabmayr01,Buchmann97a,Buchmann04}. 
Since the two-body currents are connected 
with diagrams involving $q \bar q$ pairs,
those contributions can be regarded 
as the contributions of the cloud 
of  quark-antiquark pairs.
This mechanism is usually refereed to  
as meson cloud effects.
In the case of the nucleon and the first 
nucleon excitations, the dominant meson is the pion.
Combining the low $Q^2$ expansion of the electric form factor,
 $G_{En}(Q^2) = -r_n^2 Q^2/6 $,
where $r_n^2$ is the neutron squared radius,
one can show in the context of a constituent quark model 
with two-body exchange currents, that 
$G_E(0),G_C(0) \propto r_n^2$~\cite{Pascalutsa07,Grabmayr01,Buchmann09a}.
It then becomes clear that, the  
mechanisms responsible for the symmetry breaking,
which induce $r_n^2 \ne 0$,  
are also the mechanisms responsible for the non-zero quadrupole 
moments~\cite{Grabmayr01,Buchmann09a,Buchmann04,Buchmann97a}.
The relation  (\ref{eqGCbuch}) is still valid 
when the three-body exchange currents are included 
for both observables,
as can be shown using the expansion in 
$1/N_c$~\cite{Buchmann02,Buchmann02b}.

Although the relations (\ref{eqGCbuch})-(\ref{eqGEbuch})
may be derived from a constituent quark model
with two-body exchange currents,
for simplicity we will refer those relations as 
large $N_c$ (limit) parametrizations.

To represent the electric form factor of the neutron
in Eqs.~(\ref{eqGCbuch})-(\ref{eqGEbuch}),
one can consider the Galster parametrization~\cite{Galster71} 
\ba
G_{En}(Q^2) = 
- \mu_n \frac{a \tau_N}{1 + d \tau_N}G_D,
\label{eqGEnX}
\ea 
where $\mu_n = -1.913$  
is the neutron magnetic moment,
$\tau_N= \frac{Q^2}{4M^2}$ and $a,d$ are 
two parameters.
The parameters $a$ and $d$ 
are  related to the lowest radial moments 
of the neutron electric charge distribution.
In particular $a$ is determined by the 
second momentum $r_n^2$ 
(neutron squared radius)
through $a= \frac{2M^2}{3 \mu_n} r_n^2$.
As for $d$, it is determined by $r_n^2$ 
and the fourth momentum $r_n^4$~\cite{Grabmayr01,Buchmann09a}.
The MAID2007 
parametrization for $G_C$ 
is inspired by the  Galster form (\ref{eqGEnX})~\cite{Drechsel2007}.

For the numerical evaluation of form factors $G_E$ and $G_C$ 
we consider two parametrizations with $a=0.9$.
One then has 
$r_n^2 \simeq - 0.114$ fm$^2$, close to experimental 
result $r_n^2 \simeq - 0.116$ fm$^2$~\cite{PDG}.
The best description of the neutron data is achieved with $d=2.8$.
A good description of the data can also be obtained with 
$d=1.75$~\cite{Buchmann04}.

The results from the  
(\ref{eqGCbuch})-(\ref{eqGEbuch})
parametrizations 
combined with Eq.~(\ref{eqGEnX}) for $G_{En}$
are presented in Fig.~\ref{figDeltaR3}.
In the figure, one can notice that both 
parametrizations give a good description 
of the $R_{EM}$ data, although the results 
of the parametrization with $d=1.75$
are closer to the data.
As for $R_{SM}$, the parametrization with $d=1.75$
is also the one closer to the data, 
particularly for large $Q^2$.
For $R_{SM}$, however, both parametrizations,
underestimate the low $Q^2$ data, in absolute value.
We note that the ratio $R_{SM}$ is usually 
estimated using parametrizations for $G_M$ 
derived from $SU(6)$ or large $N_c$, 
as $G_M \to - \sqrt{2} G_{Mn}$, where $G_{Mn}$ is the 
neutron magnetic form factor~\cite{Buchmann04}, 
or $G_M \to (F_{2p} - F_{2n})/\sqrt{2}$, where $F_{2p}, F_{2n}$ are 
the Pauli form factor of the proton and neutron 
respectively~\cite{Pascalutsa07}.
In those estimates the magnitude of  $R_{SM}$ 
increases by about 15\%, improving 
the agreement with the data.
This observation also holds for $R_{EM}$\footnote{
Assuming an error of 10\% 
in a large $N_c$ expansion of a form factor,
for a  term of the order ${\cal O}(1/N_c^2)$,
one can expect an error of 20\% 
in a ratio between two form factors.}.

Going back to the discussion of $R_{SM}$,
although there is some underestimation of the data at low $Q^2$,
and for the case $d=2.8$ at large $Q^2$,
our results show that the parametrization (\ref{eqGCbuch}) 
provides a good first estimate 
of the function $G_C$.
Since Eq.~(\ref{eqGCbuch}) can be 
interpreted as the pion cloud contribution 
for $G_C$, we can conclude that 
the difference between the parametrization 
and the data can be the consequence of 
the valence quark contributions.
In Ref.~\cite{LatticeD} it was shown, that 
a very good description of the $G_E$ and $G_C$
data can be obtained, 
when we add an extrapolation of the valence 
quark contributions from the lattice QCD data.  

It is worth to mention that, the estimation 
of pion cloud contribution for $G_C$
presented here is based on the 
Galster parametrization of the $G_{En}$ data,
which considers only two parameters.
It is then possible that 
a more detailed fit to the $G_{En}$ data, 
as the ones proposed in Refs.~\cite{Friedrich03,Kelly04},
reveals also a more detailed structure of $G_C$ for finite $Q^2$.

Overall one can conclude that, 
the pion cloud contributions 
for the  $G_E$  and $G_C$ form factors,  
given by Eqs.~(\ref{eqGCbuch})-(\ref{eqGEbuch}),
takes into account the dominate contribution 
of  those form factors~\cite{Buchmann04,Pascalutsa07}.
Thus, contrary to the case of the magnetic 
dipole form factor $G_M$, 
the  quadrupole form factors (electric and Coulomb) 
are not dominated by valence quark effects
in the range $Q^2=1$--4 GeV$^2$, but are instead 
dominated by pion cloud effects~\cite{Buchmann04}.
The magnitude of the 
valence quark contributions for 
the quadrupole form factors 
can be estimated as about 
10-20\%~\cite{Buchmann97a,Buchmann00a,LatticeD,NDeltaD}.
Note that this magnitude  is comparable with the 
difference between large $N_c$ estimates and the data.
 
The results for $R_{EM}$, $R_{SM}$ at low $Q^2$
can be better understood if we look for the limit $Q^2=0$.
In that case we can conclude in the large $N_c$ limit, that 
$R_{EM}(0) = R_{SM}(0)= \frac{1}{12 \sqrt{2}}
\left(\frac{M}{M_R} \right)^{3/2} \frac{M_R^2 -M^2}{G_M(0)} r_n^2$,
is a consequence of 
$G_E(0) = \frac{M_R^2- M^2}{4M_R^2}G_C(0)
$~\cite{Pascalutsa07}.
The correlation between the ratios however
seems  in conflict with the experimental data,
since $R_{EM}(0) = -(2.5 \pm 0.5) \%$
and  $R_{SM}(0.06 \, \mbox{GeV}^2) = -(4.54 \pm 0.26) \%$.
One can then conclude that, the large $N_c$ limit 
underestimates $G_C$ near $Q^2=0$.
As mentioned, this can be a sign 
that the pure valence quark contributions
may be more important for this form factor,
as shown in particular in Ref.~\cite{LatticeD}.

The analysis of the functions $G_E$ and $G_C$    
at low $Q^2$ can be simplified if we look for the 
squared radius, that measures the slope of those functions
at $Q^2=0$.
The squared radius 
associated with the quadrupole 
form factors $G_E$ and $G_C$ is defined by
\ba
r_X^2= - \frac{14}{G_X(0)} 
\left.
\frac{d G_X}{d Q^2} \right|_{Q^2=0},
\label{eqRX2}
\ea
where $X=E,C$.
In the previous equation, the factor $14$ 
replaces the factor $6$ used in the leading order 
form factors (electric charge and magnetic dipole).
This correction is the result 
of the expansion of the quadrupole operators 
in powers of $Q^2$ with 
the proper normalization~\cite{Buchmann09a,Forest66}. 
A direct consequence of Eq.~(\ref{eqRX2}) 
is that in the case of the large $N_c$ parametrizations
we can write the Coulomb quadrupole squared 
radius as $r_C^2= \frac{7}{10} \frac{r_n^4}{r_n^2}$~\cite{Buchmann09a}.

Since the parametrizations MAID2007, MAID-SG1 and MAID-SG2 
are analytic, both radius can be calculated 
using the coefficients of the parametrizations.
The results are presented in Table~\ref{tabRE2RC2}.
In the case of the large $N_c$ parametrizations, 
the results for $r_E^2$ and $r_C^2$ are identical 
because both functions are correlated with $G_{En}$.

\begin{table}[t]
\begin{center}
\begin{tabular}{l  c c}
\hline
\hline 
 & $r_{E}^2$ (fm$^2$) &  $r_C^2$ (fm$^2$) \\
\hline 
\hline 
MAID2007 & 1.63 &  2.35 \\
MAID-SG1 & 1.73 &  3.54 \\
MAID-SG2 & 2.13 &  0.85 \\
Large $N_c$ ($d=2.8$) & 1.97 & 1.97 \\
Large $N_c$ ($d=1.75$) & 1.81 & 1.81 \\
\hline
\hline 
\end{tabular}
\end{center}
\caption{
$\gamma^\ast N \to \Delta(1232)$ transition.
Results for the electric quadrupole and 
Coulomb quadrupole squared radius defined according 
to Eq.~(\ref{eqRX2}).} 
\label{tabRE2RC2}
\end{table}

In Table~\ref{tabRE2RC2} we can note 
that the Coulomb squared radius, $r_C^2$, 
is large in general, 
compared to the proton squared radius 
($r_p^2= 0.76$ fm$^2$).
The exception is the MAID-SG2 parametrization.
The analysis of the values obtained for $r_C^2$ 
is interesting due to the suggestion that
a large $r_C^2/r_p^2$ ratio  is a manifestation 
of a large spatial extension of 
the charge distribution due to the $q \bar q$ pair distribution 
in the nucleon~\cite{Buchmann09a}.
Recall that the $q \bar q$ cloud is the reason 
why $G_{En}$, $G_{E}$ and $G_C$ are 
non vanishing functions of $Q^2$. 
Large values for $r_C^2$ reflect 
the increment of the size 
of the constituent quarks due 
to the $q \bar q$ pair/meson cloud 
dressing~\cite{Grabmayr01,Buchmann00a,Buchmann09a}.
In particular the results $r_C^2 \approx 2$ fm$^2$ 
can be interpreted as $r_C^2 \simeq r_\pi^2$, 
where $r_\pi = 1/m_\pi$ is the pion Compton wavelength,
that characterizes the pion cloud
distribution inside the nucleon~\cite{Buchmann09a}.

The large values obtained for $r_C^2$ may then reflect 
the connection between 
the neutron charge distribution and $r_C^2$.
There is therefore a strong motivation 
to determine this 
radius experimentally~\cite{Grabmayr01,Buchmann09a}.

The previous discussion about $r_C^2$
can be generalized to $r_E^2$ 
(electric quadrupole squared radius),
except that the range of the pion cloud effect 
is shorter in this case.

As mentioned, concerning the large extension 
of the pion cloud for $G_C$, the MAID-SG2 parametrization 
is the exception, since the value obtained for 
$r_C^2$ is closer to the proton squared radius ($r_p^2= 0.76$ fm$^2$).
This result suggests that the 
extension of the pion cloud effect is shorter  
in this parametrization for $G_C$.
As for $G_E$, we still expect a long 
distribution of the pion cloud.
 
To summarize the discussion about the 
quadrupole form factors $G_E$ and $G_C$,
we obtain the best description 
of the data, with a model compatible 
with the Siegert's theorem, 
when we use the MAID-SG2 parametrization.
The  MAID-SG2 parametrization 
gives a very good description 
of the form factors $G_E$ and $G_C$ at small $Q^2$, 
with smoother functions near the pseudo-threshold, 
similarly to MAID-SG parametrization, 
but provides at the same time a very good description 
of the large $Q^2$ data. 
 
The smoother behavior of the  MAID-SG2 parametrization, 
is however characterized by a small value 
for $r_C^2$, which suggests that the effect of the pion cloud 
is shorter for $G_C$. 
If the data is constrained by large values 
for $r_C^2$, then MAID-SG2 does not provide the 
the best description of the data 
and the MAID-SG1 parametrization is preferable.

\section{Summary and conclusions}
\label{secConclusions}

In the present article we study the properties 
of the helicity amplitudes in the 
$\gamma^\ast N \to \Delta(1232)$ 
and $\gamma^\ast N \to N(1520)$ transitions, 
at the pseudo-threshold.
One of the consequences of the pseudo-threshold limit 
 is that the correlation between 
the electric amplitude $E$ and the scalar 
amplitude $S_{1/2}$, in the long-wavelength limit ($|{\bf q}| \to 0$),
which is usually refereed to as the Siegert's theorem.

We conclude, that, the analytic properties 
of the electromagnetic transition form factors 
imply that 
$\frac{E}{|{\bf q}|} = \sqrt{2}(M_R -M) \frac{S_{1/2}}{|{\bf q}|^2}$ 
for the $\gamma^\ast N \to \Delta(1232)$ transition 
and 
$E= 2\sqrt{2}(M_R -M) \frac{S_{1/2}}{|{\bf q}|}$
for the   $\gamma^\ast N \to N(1520)$ transition.
In the case of the $\gamma^\ast N \to \Delta(1232)$ transition, 
there is a additional $1/|{\bf q}|$ 
in both sides of the equation, relative to 
the form usually discussed in the literature, 
$E= \sqrt{2}(M_R -M) \frac{S_{1/2}}{|{\bf q}|}$.
The result discussed in the present article 
is the consequence of 
the correlation between electric and Coulomb 
quadrupole form factors $G_E = \frac{M_R-M}{2M_R} G_C$.
As for the $\gamma^\ast N \to N(1520)$ transition, 
one has in addition, the correlation between the transverse amplitudes 
$A_{1/2} = A_{3/2} /\sqrt{3}$, at the pseudo-threshold, 
which is equivalent to $G_M=0$.

We tested the previous relations between amplitudes 
for the two transitions, 
and derived parametrizations of the available data, 
valid for small and large $Q^2$,
compatible with the constraints at the 
pseudo-threshold (Siegert's theorem). 

The analytic  form of our parametrizations 
can be used in future analysis of the 
$\gamma^\ast N \to N(1650)$ transition,
similarly to the  $\gamma^\ast N \to N(1520)$ transition
[both are $\frac{1}{2}^+ \to \frac{3}{2}^-$ transitions].

Our parametrizations are compared directly  
with the MAID2007 parametrizations.
The features of the  MAID2007 parametrizations
are discussed and the failure 
for the  $\gamma^\ast N \to \Delta(1232)$
and  $\gamma^\ast N \to N(1520)$ transitions,
are explained in detail.
In addition, we propose parametrizations 
similar to the usual MAID forms,
that avoid the use of exponential factors.
The new parametrizations are compatible with the low $Q^2$ data,
the large $Q^2$ data, and also  with the 
expected behavior of pQCD, for very large $Q^2$. 
We conclude, that, the parametrizations 
based on rational functions of $Q^2$ are more appropriate 
for the description of the data in a wide region of $Q^2$,
as far as there are no singularities in the $Q^2 < 0$ region.
This can be ensured, by making use of expansions 
of the exponential series with even powers of $Q^2$
in the denominator of the parametrizations 
for the helicity amplitudes or the transition form factors.

Our best parametrization for the 
$\gamma^\ast N \to \Delta(1232)$ 
amplitudes is consistent with smooth 
$G_E$ and $G_C$ form factors at low $Q^2$, and 
near the pseudo-threshold,
contrary to the MAID2007 parametrization.
Our best parametrization provides a very good description 
of the low and large $Q^2$ data 
for the $\gamma^\ast N \to \Delta(1232)$ transition.
The value obtained for Coulomb quadrupole square radius $r_C^2$
is close to the proton squared radius,
suggesting that the effect of the pion cloud 
for $G_C$ is shorter than in other parametrizations based 
on the $q \bar q$ effects.

As for the  $\gamma^\ast N \to N(1520)$ transition,
although there are some conflicts between different data analysis, 
it is possible to conclude that, 
different parametrizations of the different datasets, 
lead to the almost same extrapolation for the low $Q^2$ region,
and similar behavior near the pseudo-threshold. 
 
The comparison between the 
parametrizations based on the Siegert's theorem, 
and estimates from valence quark models, 
suggests that, near the pseudo-threshold,
processes beyond the impulse approximation 
are essential for the interpretation of the empirical data.

The methods proposed in the present article 
can be extended to the study of the 
helicity amplitudes and transition form factors 
associated with other nucleon excitations,
as shown already, in particular for the 
$\gamma^\ast N \to N(1535)$ transition~\cite{newPaper}.

\begin{acknowledgments}
The author thanks Lothar Tiator for the 
useful discussions
and Pulak Giri for useful suggestions.
This work is supported by the Brazilian Ministry of Science,
Technology and Innovation (MCTI-Brazil).
\end{acknowledgments}

\appendix

\section{Study of the truncated exponential series}
\label{appA}

In this appendix, we discuss the possible singularities 
associated with the truncation of the exponential series
used in the definition of the form factors $G_E$ and $G_C$,
for the MAID-SG1 and MAID-SG2 parametrizations,
presented in Sec.~\ref{secLargeQ2}.

We recall that those functions 
are based on the expansion
\ba
F(Q^2) = 
1 + r_4 Q^2 + \frac{r_4^2 }{2!}  Q^4 + ... +  \frac{r_4^n}{n!}  Q^{2n},
\label{eqFgen}
\ea
where the value of $n$ depends of the particular 
function/model under discussion.

We need to check if there are 
zeros in the function $F(Q^2)$. 
To simplify the analysis, we look for the worst case 
scenario,  the point $Q^2= Q_{PS}^2= -(M_R -M)^2$.
If there is no zeros, at the  point $Q^2= Q_{PS}^2$,
there are no zeros for the values of $Q^2$
larger than $Q^2_{PS}$.
We then look for the polynomial function
\ba
P_{n}(x)= 1 - x + \frac{ x^2}{2!} +  ... + (-1)^n \frac{x^{n}}{n!} ,
\ea
where $x=r_4(M_R -M)^2$.

Since the function $P_{n}(x)$ is in the denominator 
of the functions $G_E$ or $G_C$, it is 
important to know if there are singularities 
in those functions, given by 
the zeros of the function $P_{n}(x)$.

To study the possible zeros of  $P_{n}(x)$,
we divide the discussion into two cases:
the case $n=2,4,6,...$ ($n$ even), 
and the case $n=1,3,5,...$ ($n$ odd).

\subsection{Even powers $n$}
\label{appA1}

When $n=2,4,6,...,$ it is trivial to show 
that there are no solutions for $P_n(x)=0$.
We start by checking if there is a 
minimum for the function $P_n(x)$ when $x > 0$.
The values of possible minima are determined 
by the zeros of the derivative $P_n^\prime (x)$.
The zeros of the derivative $P_n^\prime (x)$, $x_0$, 
are represented in Table~\ref{tabPNx}, 
together with the value of the function $P_n$ at the same point.
The derivative is calculated from 
$P_n^\prime (x)= - P_{n-1}(x)$.
From the table, we can conclude that 
the minimum of $P_n(x)$ is positive,
therefore there are no singularities 
in the functions defined by $1/P_n(x)$,
when $n$ is even.

\begin{table}[t]
\begin{center}
\begin{tabular}{l  c c}
\hline
\hline 
$n$    &  $x_0$ & $P_n(x_0)$ \\
\hline
$2$   &  $1$  & $0.5$ \\
$4$   &  $1.596$ & $0.270$ \\
$6$   &  $2.180$  & $0.149$  \\
\hline
\hline
\end{tabular}
\end{center}
\caption{Point of minimum ($x_0$) and value 
for the function $P_n(x_0)$, for even values of $n$.}
\label{tabPNx}
\end{table}

\subsection{Odd powers $n$}
\label{appA2}

We now consider the case $n=3,5,...,$ when $n$ is odd.
It is easy to conclude that $P_n(x)=0$,
for some values of $x$, since we start 
with $P_n(0)=1$, and the last coefficient 
of the sum (term in $x^n$) has a negative coefficient.
Therefore for $x$ large enough, $P_n(x) < 0$,
and there is an intermediate point where $P_n(x)$ vanishes.
There is therefore at least one 
singularity in the function $1/P_n(x)$.
Since as discussed in the previous section 
we can write $P_n(x) = - P_{n+1}^\prime (x)$,
where $n-1=2,4,...,$ is even, we know that 
the possible singularities are given 
by the points of zero of  $P_{n+1}^\prime (x)$,
represented already in Table~\ref{tabPNx}.

Note, however, that below the zero of $P_n(x)$, 
we are free of singularities for $1/P_n(x)$.
Therefore we can still use the function $P_n(x)$
with odd $n$, provided that $\frac{r_4}{(M_R-M)^2}$ 
is smaller than the value $x_0$ presented in Table~\ref{tabPNx}.

More specifically, in the case of the 
$\gamma^\ast N \to \Delta (1232)$ transition,
there is no danger of using 
the function $P_3(x)$, 
provided that $r_4  < \frac{1.596}{(M_R-M)^2}$ or $r_4 < 18.6$ GeV$^{-2}$.
Similarly, we can use  an expansion with $n=5$,
provided that $r_4 < \frac{2.18}{(M_R-M)^2} \simeq 25.4$ GeV$^{-2}$.
Other limits can be defined for higher powers $n$.

\subsection{Alternative expressions for $G_E$}
\label{appA3}

In the previous section, we concluded that, 
we can use the decomposition (\ref{eqFgen})
in the denominator of the functions $G_E$ and $G_C$,
under some conditions.
The functions $G_C$ for the 
MAID-SG1 and MAID-SG2 parametrizations,
given by Eqs.~(\ref{eqMAID-SG1b}) and (\ref{eqMAID-SG2b}) 
can be used for any positive values of $c_4$ (that replaces $r_4$).

As for the parametrizations for $G_E$ 
defined by  Eqs.~(\ref{eqMAID-SG1a}) and (\ref{eqMAID-SG2a}),
the values of $b_4$ have to be constrained 
respectively to the limits $b_4 < 18.6$  GeV$^{-2}$
and  $b_4 < 25.4$ GeV$^{-2}$, in order to avoid the 
singularities associated with the 
functions $P_3(x)$ and $P_5(x)$ respectively.

As discussed in the main text, 
the fits obtained in the present work are 
free from singularities, since  $b_4$ is below the critical value.
Nevertheless, in future, it may be more 
appropriate to define parametrizations 
for the function $G_E$, that are valid 
for all positive values of $b_4$.

We therefore propose alternative expressions for $G_E$, 
that replace the form Eq.~(\ref{eqMAID-SG1a})
for MAID-SG1 and Eq.~(\ref{eqMAID-SG2a})
for MAID-SG2.
For the parametrization MAID-SG1, we propose
\ba
G_E = \frac{C_0}{K} 
b_0 
\frac{1 + b_1 Q^2 + b_2 Q^4 + b_3 Q^6  + b_5 Q^8}{
1 + b_4 Q^2 + \frac{1}{2!}b_4^2 Q^4 + \frac{1}{3!}b_4^3 Q^6 +
\frac{1}{4!} b_4^4 Q^8 
}G_D, 
\nonumber \\
\ea
As for the parametrization MAID-SG2, we propose 
\ba
& &
G_E = \frac{C_0}{K}  b_0  \times
\nonumber \\
& &
\frac{1 + b_1 Q^2 + b_2 Q^4 + b_3 Q^6  + b_5 Q^8}{
1 + b_4 Q^2 + \frac{b_4^2}{2!}Q^4 + \frac{b_4^3}{3!}Q^6 +
\frac{b_4^4}{4!} Q^8  +  \frac{b_5^5}{5!} Q^{10} +
\frac{b_4^6}{6!} Q^{12} }. 
\nonumber \\
\ea

Compared to the expressions 
of Eqs.~(\ref{eqMAID-SG1a}) and (\ref{eqMAID-SG2a}),
we included an extra coefficient $b_5$,
adding some complexity to the parametrization.
Alternatively, we can drop 
the last two terms in the numerator (terms with $b_3$ and $b_5$)
and the last two terms in the denominator, 
generating simpler parametrizations 
of the function $G_E$, 
based on only 3 parameters ($b_1,b_2$ and $b_4$).
The last choice can also be  a good option, 
since the chi-squared associated 
with the $G_E$ data is smaller 
(more accurate description) than the 
chi-squared associated with the $G_C$ data.
In this case the quality of the global fit is not compromised.
As in the  MAID-SG1 and  MAID-SG2 cases,
the new parametrizations are consistent with the falloff 
$G_E \propto 1/Q^4$, for large $Q^2$.



\begin{thebibliography}{00}


\bibitem{NSTAR} 
  I.~G.~Aznauryan {\it et al.},
  Int.\ J.\ Mod.\ Phys.\ E {\bf 22}, 1330015 (2013)
  [arXiv:1212.4891 [nucl-th]].


\bibitem{Aznauryan12} 
  I.~G.~Aznauryan and V.~D.~Burkert,
  Prog.\ Part.\ Nucl.\ Phys.\  {\bf 67}, 1 (2012)
  [arXiv:1109.1720 [hep-ph]].




\bibitem{DeltaTL} 
  G.~Ramalho, M.~T.~Pe\~na, J.~Weil, H.~van Hees and U.~Mosel,
  Phys.\ Rev.\ D {\bf 93}, 033004 (2016)
  [arXiv:1512.03764 [hep-ph]];
  G.~Ramalho and M.~T.~Pe\~na,
  Phys.\ Rev.\ D {\bf 85}, 113014 (2012)
  [arXiv:1205.2575 [hep-ph]].



\bibitem{WhitePaper} 
  W.~J.~Briscoe, M.~Döring, H.~Haberzettl, D.~M.~Manley, M.~Naruki, I.~I.~Strakovsky and E.~S.~Swanson,
  Eur.\ Phys.\ J.\ A {\bf 51}, 129 (2015)
  [arXiv:1503.07763 [hep-ph]].



\bibitem{Buchmann98} 
  A.~J.~Buchmann, E.~Hernandez, U.~Meyer and A.~Faessler,
  Phys.\ Rev.\ C {\bf 58}, 2478 (1998).



\bibitem{Atti78} 
  C.~Ciofi Degli Atti,
  Prog.\ Part.\ Nucl.\ Phys.\  {\bf 3}, 163 (1978).


\bibitem{AmaldiBook}
   E.~Amaldi, S.~Fubini, and G.~Furlan,
   {\it Pion-Electroproduction
    Electroproduction at Low Energy and Hadron Form Factor},
   Springer Berlin Heidelberg (1979).





\bibitem{Drechsel92} 
  D.~Drechsel and L.~Tiator,
  J.\ Phys.\ G {\bf 18}, 449 (1992).



\bibitem{Tiator16}
   L.~Tiator, 
   Proceedings of the Workshop "Nucleon Resonances: From Photoproduction to High Photon Virtualities".
   October 2015, Trento, Italy.


\bibitem{Bjorken66} 
  J.~D.~Bjorken and J.~D.~Walecka,
  Annals Phys.\  {\bf 38}, 35 (1966).


\bibitem{Devenish76} 
  R.~C.~E.~Devenish, T.~S.~Eisenschitz and J.~G.~Korner,
  Phys.\ Rev.\ D {\bf 14}, 3063 (1976).


\bibitem{Jones73} 
  H.~F.~Jones and M.~D.~Scadron,
  Annals Phys.\  {\bf 81}, 1 (1973).



\bibitem{newPaper}
  G.~Ramalho,
  arXiv:1602.03444 [hep-ph].
  To appear in Phys.~Lett.~B.




\bibitem{Tiator2006} 
  L.~Tiator and S.~Kamalov,
  AIP Conf.\ Proc.\  {\bf 904}, 191 (2007)
  [nucl-th/0610113].


\bibitem{Drechsel2007} 
  D.~Drechsel, S.~S.~Kamalov and L.~Tiator,
  Eur.\ Phys.\ J.\ A {\bf 34}, 69 (2007)
  [arXiv:0710.0306 [nucl-th]].



\bibitem{MAID2009}
  L.~Tiator, D.~Drechsel, S.~S.~Kamalov and M.~Vanderhaeghen,
  Eur.\ Phys.\ J.\ ST {\bf 198}, 141 (2011)
  [arXiv:1109.6745 [nucl-th]];
  L.~Tiator, D.~Drechsel, S.~S.~Kamalov and M.~Vanderhaeghen,
  Chin.\ Phys.\ C {\bf 33}, 1069 (2009)
  [arXiv:0909.2335 [nucl-th]].





\bibitem{N1520} 
  G.~Ramalho and M.~T.~Pe\~na,
  Phys.\ Rev.\ D {\bf 89}, 094016 (2014)
  [arXiv:1309.0730 [hep-ph]].








\bibitem{Drechsel84} 
  D.~Drechsel and M.~M.~Giannini,
  Phys.\ Lett.\ B {\bf 143}, 329 (1984).



\bibitem{NDelta} 
  G.~Ramalho, M.~T.~Pe\~na and F.~Gross,
  Eur.\ Phys.\ J.\ A {\bf 36}, 329 (2008)
  [arXiv:0803.3034 [hep-ph]].


\bibitem{NDeltaD} 
  G.~Ramalho, M.~T.~Pe\~na and F.~Gross,
  Phys.\ Rev.\ D {\bf 78}, 114017 (2008)
  [arXiv:0810.4126 [hep-ph]].




\bibitem{Delta1600} 
  G.~Ramalho and K.~Tsushima,
  Phys.\ Rev.\ D {\bf 82}, 073007 (2010)
  [arXiv:1008.3822 [hep-ph]].




\bibitem{MokeevDatabase}
   V.~Mokeev,
   \url{https://userweb.jlab.org/~mokeev/} \\
   \url{resonance_ electrocouplings/}



\bibitem{Stave08} 
  S.~Stave {\it et al.} [A1 Collaboration],
  Phys.\ Rev.\ C {\bf 78}, 025209 (2008)
  [arXiv:0803.2476 [hep-ex]].






\bibitem{Data}
  N.~F.~Sparveris {\it et al.} [OOPS Collaboration],
  Phys.\ Rev.\ Lett.\  {\bf 94}, 022003 (2005)
  [nucl-ex/0408003];
  J.~J.~Kelly {\it et al.},
  Phys.\ Rev.\ C {\bf 75}, 025201 (2007)
  [nucl-ex/0509004].


\bibitem{Aznauryan09}
  I.~G.~Aznauryan {\it et al.} [CLAS Collaboration],
  Phys.\ Rev.\ C {\bf 80}, 055203 (2009)
  [arXiv:0909.2349 [nucl-ex]].




\bibitem{PDG} 
  K.~A.~Olive {\it et al.} [Particle Data Group Collaboration],
  Chin.\ Phys.\ C {\bf 38}, 090001 (2014).




\bibitem{LEGS01} 
  G.~Blanpied {\it et al.},
  Phys.\ Rev.\ C {\bf 64}, 025203 (2001).








\bibitem{Weyrauch86} 
  M.~Weyrauch and H.~J.~Weber,
  Phys.\ Lett.\ B {\bf 171}, 13 (1986)
  [Phys.\ Lett.\ B {\bf 181}, 415 (1986)].

\bibitem{Bourdeau87} 
  M.~Bourdeau and N.~C.~Mukhopadhyay,
  Phys.\ Rev.\ Lett.\  {\bf 58}, 976 (1987).


\bibitem{Capstick90} 
  S.~Capstick and G.~Karl,
  Phys.\ Rev.\ D {\bf 41}, 2767 (1990).





\bibitem{LatticeD} 
  G.~Ramalho and M.~T.~Pe\~na,
  Phys.\ Rev.\ D {\bf 80}, 013008 (2009)
  [arXiv:0901.4310 [hep-ph]].



\bibitem{Buchmann04} 
  A.~J.~Buchmann,
  Phys.\ Rev.\ Lett.\  {\bf 93}, 212301 (2004)
  [hep-ph/0412421].


\bibitem{Note1}
   In Ref.~\cite{Devenish76}, the Coulomb form factor 
$G_C$ is defined with the difference 
of a sign, compared to our convention~\cite{Aznauryan12,N1520}.


\bibitem{Note2}
We conclude, that, in this case,
the difference between two functions 
${\cal O}(1)$ leads to a function ${\cal O}(|{\bf q}|^2)$, 
as consequence of the relation between 
the amplitudes $A_{3/2}$ and $A_{1/2}$.
A similar result was also observed 
in the study of the $\gamma^\ast N \to N(1535)$ 
helicity amplitudes~\cite{newPaper}.
See Ref.~\cite{newPaper} for a more detailed 
explanation about the vanishment of the term ${\cal O}(|{\bf q}|)$.





\bibitem{Mokeev12} 
  V.~I.~Mokeev {\it et al.} [CLAS Collaboration],
  Phys.\ Rev.\ C {\bf 86}, 035203 (2012)
  [arXiv:1205.3948 [nucl-ex]].




\bibitem{Mokeev15} 
  V.~I.~Mokeev {\it et al.},
  arXiv:1509.05460 [nucl-ex].








\bibitem{Anisovich12} 
  A.~V.~Anisovich, R.~Beck, E.~Klempt, V.~A.~Nikonov, A.~V.~Sarantsev and U.~Thoma,
  Eur.\ Phys.\ J.\ A {\bf 48}, 15 (2012)
  [arXiv:1112.4937 [hep-ph]].

\bibitem{Workman12} 
  R.~L.~Workman, R.~A.~Arndt, W.~J.~Briscoe, M.~W.~Paris and I.~I.~Strakovsky,
  Phys.\ Rev.\ C {\bf 86}, 035202 (2012)
  [arXiv:1204.2277 [hep-ph]].






\bibitem{Diaz08} 
  B.~Julia-Diaz, T.-S.~H.~Lee, A.~Matsuyama, T.~Sato and L.~C.~Smith,
  Phys.\ Rev.\ C {\bf 77}, 045205 (2008)
  [arXiv:0712.2283 [nucl-th]].


\bibitem{Santopinto12} 
  E.~Santopinto and M.~M.~Giannini,
  Phys.\ Rev.\ C {\bf 86}, 065202 (2012)
  [arXiv:1506.01207 [nucl-th]].

\bibitem{Ronniger13} 
  M.~Ronniger and B.~C.~Metsch,
  Eur.\ Phys.\ J.\ A {\bf 49}, 8 (2013)
  [arXiv:1207.2640 [hep-ph]].

\bibitem{Golli13} 
  B.~Golli and S.~\v{S}irca,
  Eur.\ Phys.\ J.\ A {\bf 49}, 111 (2013)
  [arXiv:1306.3330 [nucl-th]].

\bibitem{Mokeev16} 
  V.~I.~Mokeev {\it et al.},
  Phys.\ Rev.\ C {\bf 93}, 025206 (2016)
  [arXiv:1509.05460 [nucl-ex]].



\bibitem{Meyer01} 
  U.~Meyer, E.~Hernandez and A.~J.~Buchmann,
  Phys.\ Rev.\ C {\bf 64}, 035203 (2001).




\bibitem{Carlson} 
  C.~E.~Carlson and N.~C.~Mukhopadhyay,
  Phys.\ Rev.\ Lett.\  {\bf 81}, 2646 (1998)
  [hep-ph/9804356];
  C.~E.~Carlson and J.~L.~Poor,
  Phys.\ Rev.\ D {\bf 38}, 2758 (1988);
  C.~E.~Carlson,
  Phys.\ Rev.\ D {\bf 34}, 2704 (1986).




\bibitem{Idilbi04} 
  A.~Idilbi, X.~Ji and J.~P.~Ma,
  Phys.\ Rev.\ D {\bf 69}, 014006 (2004)
  [hep-ph/0308018].




\bibitem{Pascalutsa07} 
  V.~Pascalutsa and M.~Vanderhaeghen,
  Phys.\ Rev.\ D {\bf 76}, 111501 (2007)
  [arXiv:0711.0147 [hep-ph]].




\bibitem{Buchmann02} 
  A.~J.~Buchmann, J.~A.~Hester and R.~F.~Lebed,
  Phys.\ Rev.\ D {\bf 66}, 056002 (2002)
  [hep-ph/0205108].




\bibitem{Jenkins02a} 
  E.~E.~Jenkins, X.~Ji and A.~V.~Manohar,
  Phys.\ Rev.\ Lett.\  {\bf 89}, 242001 (2002)
  [hep-ph/0207092].



\bibitem{Pascalutsa07b} 
  V.~Pascalutsa, M.~Vanderhaeghen and S.~N.~Yang,
  Phys.\ Rept.\  {\bf 437}, 125 (2007)
  [hep-ph/0609004].


\bibitem{Buchmann09a} 
  A.~J.~Buchmann,
  Can.\ J.\ Phys.\  {\bf 87}, 773 (2009)
  [arXiv:0910.4747 [physics.atom-ph]].


\bibitem{Grabmayr01} 
  P.~Grabmayr and A.~J.~Buchmann,
  Phys.\ Rev.\ Lett.\  {\bf 86}, 2237 (2001)
  [hep-ph/0104203].



\bibitem{Buchmann97a} 
  A.~J.~Buchmann, E.~Hernandez and A.~Faessler,
  Phys.\ Rev.\ C {\bf 55}, 448 (1997)
  [nucl-th/9610040].






\bibitem{Buchmann02b} 
  A.~J.~Buchmann and E.~M.~Henley,
  Phys.\ Rev.\ D {\bf 65}, 073017 (2002).





\bibitem{Galster71} 
  S.~Galster, H.~Klein, J.~Moritz, K.~H.~Schmidt, D.~Wegener and J.~Bleckwenn,
  Nucl.\ Phys.\ B {\bf 32}, 221 (1971).




\bibitem{Buchmann00a} 
  A.~J.~Buchmann and E.~M.~Henley,
  Phys.\ Rev.\ C {\bf 63}, 015202 (2000)
  [hep-ph/0101027].





\bibitem{Friedrich03} 
  J.~Friedrich and T.~Walcher,
  Eur.\ Phys.\ J.\ A {\bf 17}, 607 (2003)
  [hep-ph/0303054].


\bibitem{Kelly04} 
  J.~J.~Kelly,
  Phys.\ Rev.\ C {\bf 70}, 068202 (2004).



\bibitem{Forest66} 
  T.~De Forest, Jr. and J.~D.~Walecka,
  Adv.\ Phys.\  {\bf 15}, 1 (1966).


\end{thebibliography}
\end{document}